\documentclass[%
 reprint,
 amsmath,amssymb,
 aps,
]{revtex4-2}

\usepackage[margin=5pt,caption=false]{subfig}
\usepackage{comment}
\usepackage{threeparttable}
\usepackage{booktabs}
\usepackage[figuresright]{rotating}
\usepackage{multirow}
\newcommand{\be}{\begin{equation}}
\newcommand{\ee}{\end{equation}}
\newcommand{\bea}{\begin{eqnarray}}
\newcommand{\eea}{\end{eqnarray}}
\newcommand{\hunit}{$\rm{km \ s^{-1} \ Mpc^{-1}}$}
\newcommand{\lcdm}{$\Lambda$CDM}
\newcommand{\pcdm}{$\phi$CDM}
\usepackage{array}

\newcommand{\hiig}{H\,\textsc{ii}G}
\newcommand{\hii}{H\,\textsc{ii}}
\newcommand{\Om}{\Omega_{m0}}
\newcommand{\Ok}{\Omega_{k0}}
\newcommand{\om}{$\Omega_{m0}$}
\newcommand{\ok}{$\Omega_{k0}$}
\newcommand{\wx}{$w_{\rm X}$}
\newcommand{\wX}{w_{\rm X}}
\newcommand{\mq}{Mg\,\textsc{ii} QSO}
\newcommand{\mii}{Mg\,\textsc{ii}}
\newcommand{\cq}{C\,\textsc{iv} QSO}
\newcommand{\civ}{C\,\textsc{iv}}

\newcommand{\obh}{\Omega_{b}h^2}
\newcommand{\och}{\Omega_{c}h^2}
\newcommand{\onh}{\Omega_{\nu}h^2}
\newcommand{\obhs}{$\Omega_{b}h^2$}
\newcommand{\ochs}{$\Omega_{c}h^2$}
\newcommand{\plus}{\!\raisebox{0.2ex}{+}}

\usepackage[T1]{fontenc}
\usepackage{scalerel}
\usepackage{tikz}
\usetikzlibrary{svg.path}

\definecolor{orcidlogocol}{HTML}{A6CE39}
\tikzset{
  orcidlogo/.pic={
    \fill[orcidlogocol] svg{M256,128c0,70.7-57.3,128-128,128C57.3,256,0,198.7,0,128C0,57.3,57.3,0,128,0C198.7,0,256,57.3,256,128z};
    \fill[white] svg{M86.3,186.2H70.9V79.1h15.4v48.4V186.2z}
                 svg{M108.9,79.1h41.6c39.6,0,57,28.3,57,53.6c0,27.5-21.5,53.6-56.8,53.6h-41.8V79.1z M124.3,172.4h24.5c34.9,0,42.9-26.5,42.9-39.7c0-21.5-13.7-39.7-43.7-39.7h-23.7V172.4z}
                 svg{M88.7,56.8c0,5.5-4.5,10.1-10.1,10.1c-5.6,0-10.1-4.6-10.1-10.1c0-5.6,4.5-10.1,10.1-10.1C84.2,46.7,88.7,51.3,88.7,56.8z};
  }
}

\newcommand\orcidicon[1]{\href{https://orcid.org/#1}{\mbox{\scalerel*{
\begin{tikzpicture}[yscale=-1,transform shape]
\pic{orcidlogo};
\end{tikzpicture}
}{|}}}}

\usepackage{hyperref} 

\DeclareRobustCommand{\VAN}[3]{#2}
\let\VANthebibliography\thebibliography
\def\thebibliography{\DeclareRobustCommand{\VAN}[3]{##3}\VANthebibliography}
\usepackage{amsmath,bm}	
\usepackage{amssymb}	
\usepackage{fnpct}

\begin{document}

\preprint{APS/123-QED}

\title{\boldmath{$H_0=69.8\pm1.3$ $\rm{km \ s^{-1} \ Mpc^{-1}}$, $\Omega_{m0}=0.288\pm0.017$}, and other constraints from lower-redshift, non-CMB, expansion-rate data}

\author{Shulei Cao$^{\orcidicon{0000-0003-2421-7071}}$}
 \email{shulei@phys.ksu.edu}
\author{Bharat Ratra$^{\orcidicon{0000-0002-7307-0726}}$}%
 \email{ratra@phys.ksu.edu}
\affiliation{ 
Department of Physics, Kansas State University, 116 Cardwell Hall, Manhattan, KS 66506, USA
}%

\date{\today}

\begin{abstract}
We use updated Type Ia Pantheon\!\raisebox{0.2ex}{+} supernova, baryon acoustic oscillation, and Hubble parameter (now also accounting for correlations) data, as well as new reverberation-measured C\,\textsc{iv} quasar data, and quasar angular size, H\,\textsc{ii} starburst galaxy, reverberation-measured Mg\,\textsc{ii} quasar, and Amati-correlated gamma-ray burst data to constrain cosmological parameters. We show that these data sets result in mutually consistent constraints and jointly use them to constrain cosmological parameters in six different spatially-flat and non-flat cosmological models. Our analysis provides summary model-independent determinations of two key cosmological parameters: the Hubble constant, $H_0=69.8\pm1.3$ $\rm{km \ s^{-1} \ Mpc^{-1}}$, and the current non-relativistic matter density parameter, $\Omega_{m0}=0.288\pm0.017$. Our summary error bars are 2.4 and 2.3 times those obtained using the flat \lcdm\ model and \textit{Planck} TT,TE,EE+lowE+lensing cosmic microwave background (CMB) anisotropy data. Our $H_0$ value is very consistent with that from the local expansion rate based on the Tip of the Red Giant Branch and Type Ia supernova (SN Ia) data, is 2$\sigma$ lower than that from the local expansion rate based on Cepheid and SN Ia data, and is 2$\sigma$ higher than that in the flat \lcdm\ model based on \textit{Planck} TT,TE,EE+lowE+lensing CMB data. Our data compilation shows at most mild evidence for non-flat spatial hypersurfaces, but more significant evidence for dark energy dynamics, 2$\sigma$ or larger in the spatially-flat dynamical dark energy models we study.  
\end{abstract}

\maketitle

\section{Introduction} \label{sec:intro}

The Universe is currently expanding at an increasing rate, a finding supported by a number of observations. The most widely accepted explanation for this acceleration is dark energy, a hypothetical substance with negative pressure. In the spatially-flat \lcdm\ model, \citep{peeb84}, dark energy is taken to be a cosmological constant and contributes about 70\% of the total energy budget of the current Universe. However, recent observations may indicate potential discrepancies with this model, \citep{DiValentinoetal2021b,PerivolaropoulosSkara2021,Morescoetal2022,Abdallaetal2022,Hu:2023jqc}, and have led to the exploration of alternate models that allow for non-zero spatial curvature and dark energy dynamics. In our analyses here we also explore some of these alternatives. 

Cosmological models have been compared and cosmological parameter constraints have been determined using various observations, including cosmic microwave background (CMB) anisotropy data, \citep{planck2018b}, that probe the high-redshift Universe, and lower-redshift expansion-rate observations like those we use here. These lower-redshift data sets include better-established probes such as Hubble parameter [$H(z)$] data that reach to redshift $z\sim2$, and baryon acoustic oscillation (BAO) and type Ia supernova (SN Ia) measurements that reach to $z\sim2.3$, \cite{Yuetal2018, eBOSS_2020, Brout:2022vxf}, as well as emerging probes such as \hii\ starburst galaxy (\hiig) apparent magnitude data that reach to $z\sim2.5$, \citep{Mania_2012, Chavez_2014, GM2021, Johnsonetal2022, Mehrabietal2022, CaoRyanRatra2022}, quasar angular size (QSO-AS) measurements that reach to $z\sim2.7$, \citep{Cao_et_al2017b, Ryanetal2019, CaoRyanRatra2020, Zhengetal2021, Lian_etal_2021}, reverberation-measured (RM) \mii\ and \civ\ quasar (QSO) measurements that reach to $z\sim3.4$, \citep{Czernyetal2021, Zajaceketal2021, Yuetal2021, Khadkaetal_2021a, Khadka:2022ooh, Cao:2022pdv, Czerny:2022xfj}, and gamma-ray burst (GRB) data that reach to $z\sim8.2$, \citep{Wang_2016, Dirirsa2019, KhadkaRatra2020c, Caoetal_2021, Dainottetal2020, Huetal_2021, Daietal_2021, Demianskietal_2021, Khadkaetal_2021b, CaoDainottiRatra2022b, DainottiNielson2022}, of which only 118 Amati-correlated (A118) GRBs, with lower intrinsic dispersion, are suitable for cosmological purposes, \citep{Khadkaetal_2021b, LuongoMuccino2021, CaoKhadkaRatra2021, CaoDainottiRatra2022, Liuetal2022}. 

In our analyses here we also exclude RM $\mathrm{H}\beta$ QSO data that probe to $z \sim 0.9$, \citep{Czernyetal2021, Zajaceketal2021, Khadkaetal2021c}, because the resulting cosmological parameter constraints are in $\sim 2\sigma$ tension with those from more established probes. QSO flux observations that reach to $z \sim 7.5$ have been studied, \citep{RisalitiLusso2015, RisalitiLusso2019, KhadkaRatra2020a, Yangetal2020, KhadkaRatra2020b, Lussoetal2020, KhadkaRatra2021, KhadkaRatra2022, Rezaeietal2022, Luongoetal2021, DainottiBardiacchi2022}, however, we also exclude these QSOs from our analyses here since the latest QSO flux compilation, \cite{Lussoetal2020}, is not standardizable, \citep{KhadkaRatra2021, KhadkaRatra2022, Petrosian:2022tlp, Khadka:2022aeg}.

We use only the above-listed, not unreliable, lower-redshift ($z \leq 8.2$) expansion-rate data sets to derive cosmological parameter constraints. This is because we want to derive constraints that are independent of CMB aniostropy data that result in constraints that contradict some local distance-ladder measurements of the Hubble constant $H_0$, \citep{DiValentinoetal2021b,PerivolaropoulosSkara2021,Morescoetal2022,Abdallaetal2022,Hu:2023jqc}. We also do not use growth-rate data here, since use of such data requires assumption of a primordial inhomogeneity power spectrum and so additional freedom.

We emphasize, as discussed below, that we do use SN Ia that are also used in distance-ladder $H_0$ measurements, but unlike in the distance-ladder case we do not use, e.g., Cepheid or Tip of the Red Giant Branch (TRGB) distances to calibrate SN Ia data. We also note, as discussed below, that \hiig\ data we use assume a correlation relation that is calibrated by using a compilation of Cepheid, TRGB, and other data; in an Appendix we discuss cosmological constraints that do not make use of the \hiig\ measurements and so are independent of both CMB data and more conventional distance-ladder data.

In this paper we build and improve upon our earlier work in Ref.\ \cite{CaoRatra2022}. In particular we use new Pantheon\plus\ SN Ia (SNP\plus) data, update our BAO data compilation, and update as well as now account for the correlations between some of the $H(z)$ measurements. We now also include new RM \cq\ data and now also more correctly account for the asymmetric errors in RM \mq\ and \cq\ data. We show that the results from each of the individual data sets are mutually consistent and also show that the updated joint results do not differ significantly from the joint results of Ref.\ \cite{CaoRatra2022}. In particular, our joint analysis of new $H(z)$ + BAO + SNP\plus\ + QSO-AS + \hiig\ + \mii\ + \civ\ + A118 data here yields summary model-independent values of the non-relativisitic matter density parameter $\Om=0.288\pm0.017$ and $H_0=69.8\pm1.3$ \hunit, which are $0.29\sigma$ lower and $0.057\sigma$ higher than the summary joint constraints from Ref.\ \cite{CaoRatra2022}, $\Om=0.295\pm0.017$ and $H_0=69.7\pm1.2$ \hunit. 

Our $H_0 =69.8\pm1.3$ \hunit\ measurement is in better agreement with the median statistics $H_0 =68\pm2.8$ \hunit\ estimate of Ref.\ \cite{chenratmed} and the TRGB and SN Ia local expansion rate $H_0 =69.8\pm1.7$ \hunit\ estimate of Ref.\ \cite{Freedman2021} than with the Cepheids and SN Ia local expansion rate $H_0 =73.04\pm1.04$ \hunit\ estimate of Ref.\ \cite{Riessetal2022} and the flat \lcdm\ model $H_0 =67.36 \pm 0.54$ \hunit\ estimate from \textit{Planck} 2018 TT,TE,EE+lowE+lensing CMB anisotropy data \cite{planck2018b}, differing by $\sim2\sigma$ from the last two. (As discussed in the Appendix, excluding \hiig\ data results in $H_0$ values that are $\sim 0.9 \sigma$ higher than the \textit{Planck} flat \lcdm\ model value and $\sim 1.3-1.6\sigma$ lower than the Cepheids and SN Ia local expansion rate value of Ref.\ \cite{Riessetal2022}.) These data show at most mild evidence for non-flat spatial hypersurfaces, but more significant evidence for dark energy dynamics, 2$\sigma$ or larger in the spatially-flat dynamical dark energy models we study. Based on the deviance information criterion, flat \pcdm\ is the most favored cosmological model.

This paper is organized as follows. In Section \ref{sec:model} we summarize the cosmological models and parametrizations used in our analyses. Section \ref{sec:data} provides a detailed description of the data sets used in our analyses. The methods we employ are summarized in Section \ref{sec:analysis}. In Section \ref{sec:results} we present our findings on the constraints of cosmological parameters. We summarize our conclusions in Section \ref{sec:conclusion}. Constraints that do not make use of \hiig\ data are discussed in the Appendix.

\section{Cosmological models}
\label{sec:model}

We use six cosmological models to study the data sets we consider. These data are used to constrain the parameters of the six cosmological models that apply for different combinations of flat or non-flat spatial geometry and a constant cosmological constant or a dynamical dark energy density. For recent determinations of constraints on spatial curvature see Refs.\ \citep{Ranaetal2017, Oobaetal2018a, Oobaetal2018b, ParkRatra2019a, ParkRatra2019b, DESCollaboration2019, Lietal2020, EfstathiouGratton2020, DiValentinoetal2021a, KiDSCollaboration2021, Vagnozzietal2021, ArjonaNesseris2021, Dhawanetal2021, Renzietal2021, Gengetal2022, WeiMelia2022, MukherjeeBanerjee2022, Glanvilleetal2022, Wuetal2022, deCruzPerezetal2022, DahiyaJain2022, Stevensetal2022, Favale:2023lnp} and references therein. We compare the goodness of fit of these models. We use multiple, potentially very different, cosmological models in order to be able to determine which cosmological parameters are constrained in a model-independent manner by the data sets we use.

To compute cosmological parameter constraints in these models we use the Hubble parameter, $H(z, \textbf{p})$, which is a function of redshift $z$ and the cosmological parameters $\textbf{p}$ in the given model. The Hubble parameter is related to the expansion rate function and the Hubble constant as $H(z, \textbf{p}) = H_0E(z, \textbf{p})$. In these models we assume the presence of one massive and two massless neutrino species, with the total neutrino mass ($\sum m_{\nu}$) being $0.06$ eV and an effective number of relativistic neutrino species of $N_{\rm eff} = 3.046$. This allows us to compute the current non-relativistic matter density parameter value, \om, from the current values of the physical energy density parameters for non-relativistic neutrinos ($\onh$), baryons (\obhs), and cold dark matter (\ochs), and the Hubble constant $h$ in units of 100 \hunit, as $\Om = (\onh + \obh + \och)/{h^2}$, where $\onh=\sum m_{\nu}/(93.14\ \rm eV)$ is a constant.

In the flat and non-flat \lcdm\ models the expansion rate function is
\be
\label{eq:EzL}
    E(z, \textbf{p}) = \sqrt{\Om\left(1 + z\right)^3 + \Ok\left(1 + z\right)^2 + \Omega_{\Lambda}},
\ee
where the current value of the spatial curvature energy density parameter $\Ok = 0$ in flat \lcdm\ and the cosmological constant dark energy density parameter $\Omega_{\Lambda} = 1 - \Om - \Ok$. The cosmological parameters being constrained are $\textbf{p}=\{H_0, \obh\!, \och\}$ and $\textbf{p}=\{H_0, \obh\!, \och\!, \Ok\}$ in the flat and non-flat \lcdm\ models, respectively. 

In the flat and non-flat XCDM parametrizations, 
\be
\label{eq:EzX}
\resizebox{0.475\textwidth}{!}{%
    $E(z, \textbf{p}) = \sqrt{\Om\left(1 + z\right)^3 + \Ok\left(1 + z\right)^2 + \Omega_{{\rm X}0}\left(1 + z\right)^{3\left(1 + \wX\right)}},$%
    }
\ee
where the current value of the X-fluid dynamical dark energy density parameter $\Omega_{{\rm X}0} = 1 - \Om - \Ok$ and the X-fluid (dark energy) equation of state parameter $w_{\rm X} = p_{\rm X}/\rho_{\rm X}$ is allowed to take values different from $-1$ (which corresponds to a cosmological constant), where $p_{\rm X}$ and $\rho_{\rm X}$ are the pressure and energy density of the X-fluid, respectively. The cosmological parameters being constrained are $\textbf{p}=\{H_0, \obh\!, \och\!, \wX\}$ and $\textbf{p}=\{H_0, \obh\!, \och\!, \wX, \Ok\}$ in the flat and non-flat XCDM parametrizations, respectively. 
Discussions of parameterizations of dynamical dark energy may be traced through Refs.\ \citep{SolaPeracaula:2021gxi, deCruzPerez:2023wzd,2023arXiv230411157S}.

In the flat and non-flat \pcdm\ models, \citep{peebrat88,ratpeeb88,pavlov13},
\be
\label{eq:Ezp}
    E(z, \textbf{p}) = \sqrt{\Om\left(1 + z\right)^3 + \Ok\left(1 + z\right)^2 + \Omega_{\phi}(z,\alpha)},
\ee
where the scalar field ($\phi$) dynamical dark energy density parameter is
\be
\label{Op}
\Omega_{\phi}(z,\alpha)=\frac{1}{6H_0^2}\bigg[\frac{1}{2}\dot{\phi}^2+V(\phi)\bigg],
\ee
which is determined by numerically solving the Friedmann equation \eqref{eq:Ezp} and the equation of motion of the scalar field
\be
\label{em}
\ddot{\phi}+3H\dot{\phi}+V'(\phi)=0,
\ee 
with an overdot and a prime denoting a derivative with respect to time and $\phi$, respectively. Here we assume an inverse power-law scalar field potential energy density
\be
\label{PE}
V(\phi)=\frac{1}{2}\kappa m_p^2\phi^{-\alpha},
\ee
where $m_p$ is the Planck mass, $\alpha$ is a positive constant ($\alpha=0$ corresponds to a cosmological constant), and $\kappa$ is a constant that is determined by the shooting method in the Cosmic Linear Anisotropy Solving System (\textsc{class}) code \citep{class}. The cosmological parameters being constrained are $\textbf{p}=\{H_0, \obh\!, \och\!, \alpha\}$ and $\textbf{p}=\{H_0, \obh\!, \och\!, \alpha, \Ok\}$ in the flat and non-flat \pcdm\ models, respectively. For recent studies on constraints on \pcdm\ see Refs.\ \citep{Sola:2016hnq, Zhaietal2017, ooba_etal_2018b, ooba_etal_2019, park_ratra_2018, park_ratra_2019b, park_ratra_2020, SolaPercaulaetal2019, Singhetal2019, UrenaLopezRoy2020, SinhaBanerjee2021, Xuetal2021, deCruzetal2021, Jesusetal2022, Adiletal2022} and related references within these papers.

Note that in analyses of some of the data sets we use, we set $H_0=70$ \hunit\ and $\Omega_{b}=0.05$, with $\textbf{p}$ changing accordingly, because these data are unable to constrain $H_0$ and $\Omega_{b}$.

\section{Data}
\label{sec:data}

\begin{table}
\centering
\begin{threeparttable}
\caption{32 $H(z)$ data.}\label{tab:hz}
\setlength{\tabcolsep}{7.5mm}{
\begin{tabular}{lcc}
\hline
$z$ & $H(z)$\tnote{a} & Reference\\
\hline
0.07 & $69.0\pm19.6$ & \cite{73}\\
0.09 & $69.0\pm12.0$ & \cite{69}\\
0.12 & $68.6\pm26.2$ & \cite{73}\\
0.17 & $83.0\pm8.0$ & \cite{69}\\
0.2 & $72.9\pm29.6$ & \cite{73}\\
0.27 & $77.0\pm14.0$ & \cite{69}\\
0.28 & $88.8\pm36.6$ & \cite{73}\\
0.4 & $95.0\pm17.0$ & \cite{69}\\
0.47 & $89.0\pm50.0$ & \cite{15}\\
0.48 & $97.0\pm62.0$ & \cite{71}\\
0.75 & $98.8\pm33.6$ & \cite{Borghi_etal_2021}\\
0.88 & $90.0\pm40.0$ & \cite{71}\\
0.9 & $117.0\pm23.0$ & \cite{69}\\
1.3 & $168.0\pm17.0$ & \cite{69}\\
1.43 & $177.0\pm18.0$ & \cite{69}\\
1.53 & $140.0\pm14.0$ & \cite{69}\\
1.75 & $202.0\pm40.0$ & \cite{69}\\
0.1791 & 74.91 & \cite{Morescoetal2020}\tnote{b}\\
0.1993 & 74.96 & \cite{Morescoetal2020}\tnote{b}\\
0.3519 & 82.78 & \cite{Morescoetal2020}\tnote{b}\\
0.3802 & 83.0 &  \cite{Morescoetal2020}\tnote{b}\\
0.4004 & 76.97 &  \cite{Morescoetal2020}\tnote{b}\\
0.4247 & 87.08 &  \cite{Morescoetal2020}\tnote{b}\\
0.4497 & 92.78 &  \cite{Morescoetal2020}\tnote{b}\\
0.4783 & 80.91 &  \cite{Morescoetal2020}\tnote{b}\\
0.5929 & 103.8 & \cite{Morescoetal2020}\tnote{b}\\
0.6797 & 91.6 & \cite{Morescoetal2020}\tnote{b}\\
0.7812 & 104.5 & \cite{Morescoetal2020}\tnote{b}\\
0.8754 & 125.1 & \cite{Morescoetal2020}\tnote{b}\\
1.037 & 153.7 & \cite{Morescoetal2020}\tnote{b}\\
1.363 & 160.0 & \cite{Morescoetal2020}\tnote{b}\\
1.965 & 186.5 & \cite{Morescoetal2020}\tnote{b}\\
\hline
\end{tabular}}
\begin{tablenotes}[flushleft]
\item[a] \hunit.
\item[b] These 15 measurements are correlated and used in our analyses with a full covariance matrix as noted in Sec. \ref{sec:data}.
\end{tablenotes}
\end{threeparttable}
\end{table}

\begin{table}
\centering
\begin{threeparttable}
\caption{12 BAO data.}\label{tab:bao}
\setlength{\tabcolsep}{2.3mm}{
\begin{tabular}{lccc}
\toprule
$z$ & Measurement\tnote{a} & Value & Reference\\
\midrule
$0.122$ & $D_V\left(r_{s,{\rm fid}}/r_s\right)$ & $539\pm17$ & \cite{Carter_2018}\\
$0.38$ & $D_M/r_s$ & 10.23406 & \cite{eBOSSG_2020}\tnote{b}\\
$0.38$ & $D_H/r_s$ & 24.98058 & \cite{eBOSSG_2020}\tnote{b}\\
$0.51$ & $D_M/r_s$ & 13.36595 & \cite{eBOSSG_2020}\tnote{b}\\
$0.51$ & $D_H/r_s$ & 22.31656 & \cite{eBOSSG_2020}\tnote{b}\\
$0.698$ & $D_M/r_s$ & 17.85823691865007 & \cite{eBOSSG_2020, eBOSSL_2021}\tnote{c}\\
$0.698$ & $D_H/r_s$ & 19.32575373059217 & \cite{eBOSSG_2020, eBOSSL_2021}\tnote{c}\\
$0.835$ & $D_M/r_s$ & $18.92\pm0.51$ & \cite{DES_2022}\tnote{d}\\
$1.48$ & $D_M/r_s$ & 30.6876 & \cite{eBOSSQ_2020, eBOSSQ_2021}\tnote{e}\\
$1.48$ & $D_H/r_s$ & 13.2609 & \cite{eBOSSQ_2020, eBOSSQ_2021}\tnote{e}\\
$2.334$ & $D_M/r_s$ & 37.5 & \cite{duMas2020}\tnote{f}\\
$2.334$ & $D_H/r_s$ & 8.99 & \cite{duMas2020}\tnote{f}\\
\bottomrule
\end{tabular}}
\begin{tablenotes}[flushleft]
\item[a] $D_V$, $r_s$, $r_{s, {\rm fid}}$, $D_M$, $D_H$, and $D_A$ have units of Mpc.
\item[b] The four measurements from Ref.\ \cite{eBOSSG_2020} are correlated; see equation \eqref{CovM2} below for their correlation matrix.
\item[c] The two measurements from Refs.\ \cite{eBOSSG_2020} and \cite{eBOSSL_2021} are correlated; see equation \eqref{CovM3} below for their correlation matrix.
\item[d] This measurement is updated relative to the one from Ref.\ \cite{DES_2019b} used in Ref.\ \cite{CaoRatra2022}.
\item[e] The two measurements from Refs.\ \cite{eBOSSQ_2020} and \cite{eBOSSQ_2021} are correlated; see equation \eqref{CovM4} below for their correlation matrix.
\item[f] The two measurements from Ref.\ \cite{duMas2020} are correlated; see equation \eqref{CovM1} below for their correlation matrix.
\end{tablenotes}
\end{threeparttable}
\end{table}

In this paper, compared to data we used in Ref.\ \citep{CaoRatra2022}, we now use updated $H(z)$ data and an improved $H(z)$ analysis that now includes the covariance matrix for a subset of these data from Refs.\ \cite{Morescoetal2020}, updated BAO data, new Pantheon\plus\ SN Ia data, an improved analysis of reverberation measured \mii\ QSO data, and new reverberation measured \civ\ QSO data, as well as other data sets, to constrain cosmological parameters. We also correct an error in one GRB measurement used in Ref.\ \citep{CaoRatra2022}, as discussed below. These data are summarized next.

\begin{itemize}

\item[]{$\textbf{\emph{H(z)}}$ \bf data.} The 32 $H(z)$ measurements listed in Table \ref{tab:hz} have a redshift range of $0.07 \leq z \leq 1.965$. The covariance matrix of the 15 correlated measurements originally from Refs.\ \cite{70,72,moresco_et_al_2016}, discussed in Ref.\ \cite{Morescoetal2020}, can be found at \url{https://gitlab.com/mmoresco/CCcovariance/}. In the following we refer to the $H(z)$ data set used in Ref.\ \citep{CaoRatra2022} as Old $H(z)$ data.

\item[]{\bf BAO data}. The 12 BAO measurements listed in Table \ref{tab:bao} cover the redshift range $0.122 \leq z \leq 2.334$. The quantities listed in Table \ref{tab:bao} are described as follows:
\begin{itemize}
    \item $D_V(z)$: Spherically averaged BAO distance, $D_V(z)=[cz(1+z)^2H(z)^{-1}D^2_A(z)]^{1/3}$, where $c$ is the speed of light and the angular diameter distance $D_A(z) = D_M(z)/(1+z)$ with $D_M(z)$ defined in the following
    \item $D_H(z)$: Hubble distance, $D_H(z)=c/H(z)$
    \item $r_s$: Sound horizon at the drag epoch, $r_{s, {\rm fid}}=147.5$ Mpc in Ref. \cite{Carter_2018}
    \item $D_M(z)$: Transverse comoving distance,
    \begin{equation}
      \label{eq:DM}
        D_M(z) = 
        \begin{cases}
        \frac{c}{H_0\sqrt{\Omega_{k0}}}\sinh\left[\frac{H_0\sqrt{\Omega_{k0}}}{c}D_C(z)\right] & \text{if}\ \Omega_{k0} > 0, \\
        D_C(z) & \text{if}\ \Omega_{k0} = 0,\\
        \frac{c}{H_0\sqrt{|\Omega_{k0}|}}\sin\left[\frac{H_0\sqrt{|\Omega_{k0}|}}{c}D_C(z)\right] & \text{if}\ \Omega_{k0} < 0,
        \end{cases}
    \end{equation}
    where the comoving distance
    \begin{equation}
    \label{eq:gz}
       D_C(z) = c\int^z_0 \frac{dz'}{H(z')}.
    \end{equation}
\end{itemize}

The covariance matrices for given BAO data are as follows. The covariance matrix $\textbf{C}$ for BAO data from Ref.\ \cite{duMas2020} is
\be
\label{CovM1}
    \begin{bmatrix}
    1.3225 & -0.1009 \\
    -0.1009 & 0.0380
    \end{bmatrix},
\ee
for BAO data from Ref.\ \cite{eBOSSG_2020} it is
\be
\label{CovM2}
    \resizebox{\columnwidth}{!}{%
    $\begin{bmatrix}
    0.02860520 & -0.04939281 & 0.01489688 & -0.01387079\\
    -0.04939281 & 0.5307187 & -0.02423513 & 0.1767087\\
    0.01489688 & -0.02423513 & 0.04147534 & -0.04873962\\
    -0.01387079 & 0.1767087 & -0.04873962 & 0.3268589
    \end{bmatrix},$%
    }
\ee
for BAO data from Refs.\ \cite{eBOSSG_2020} and \cite{eBOSSL_2021} it is
\be
\label{CovM3}
    \begin{bmatrix}
    0.1076634008565565 & -0.05831820341302727\\
    -0.05831820341302727 & 0.2838176386340292 
    \end{bmatrix},
\ee
and for BAO data from Refs.\ \cite{eBOSSQ_2020} and \cite{eBOSSQ_2021} it is
\be
\label{CovM4}
    \begin{bmatrix}
    0.63731604 & 0.1706891\\
    0.1706891 & 0.30468415
    \end{bmatrix}.
\ee

In the following we refer to the BAO data compilation used in Ref.\ \citep{CaoRatra2022} as Old BAO data. As discussed, e.g., below eq.\ (45) of Ref.\ \citep{eBOSSL_2021}, BAO measurements are robust to the choice of fiducial cosmology, close to the assumed one.

\item[]{\bf SN Ia data.} We used two sets of SN Ia data jointly in the analyses of Ref.\ \cite{CaoRatra2022}: 1048 Pantheon (abbreviated as ``SNP'') measurements from Ref.\ \cite{scolnic_et_al_2018} that span a redshift range of $0.01 < z < 2.3$ and 20 binned DES 3yr (abbreviated as ``SND'') measurements from Ref.\ \cite{DES_2019d} that span a redshift range of $0.015 \leq z \leq 0.7026$. In the following we refer to the SN Ia data compilation used in Ref.\ \citep{CaoRatra2022} as SNP + SND data. In this paper we use 1590 of 1701 Pantheon\plus\ SN Ia (abbreviated as ``SNP\plus'') measurements from Ref.\ \cite{Brout:2022vxf} that span a redshift range of $0.01016 \leq z \leq 2.26137$, with a minimum redshift of $z>0.01$ to avoid dependence on peculiar velocity models. We note that SNP\plus\ data includes updated SNP and SND data.

\item[]{\bf QSO angular size (QSO-AS) data.} The 120 QSO angular size measurements are listed in table 1 of Ref.\ \cite{Cao_et_al2017b} and span the redshift range $0.462 \leq z \leq 2.73$. The angular size of a QSO can be predicted in a given cosmological model by using the formula $\theta(z)=l_{\rm m}/D_A(z)$, where $l_{\rm m}$ is the characteristic linear size of the QSOs in the sample and $D_A(z)$ is the angular diameter distance at redshift $z$. 

\item[]{\bf H\,\textsc{ii}G data.} The 181 \hiig\ measurements listed in table A3 of Ref.\ \cite{GM2021} include 107 low-$z$ ones from Ref.\ \cite{Chavez_2014} recalibrated in Ref.\ \cite{G-M_2019}, spanning the redshift range $0.0088 \leq z \leq 0.16417$, and 74 high-$z$ ones spanning the redshift range $0.63427 \leq z \leq 2.545$. These sources follow a correlation between the observed luminosity ($L$) of the Balmer emission lines and the velocity dispersion ($\sigma$) of the ionized gas, represented by the equation $\log L = \beta \log \sigma + \gamma$, where $\log = \log_{10}$. In Ref.\ \cite{GM2021}, 107 low-$z$ \hiig\ and 36 Giant Extragalactic \hii\ Regions (GH\,\textsc{ii}R) data are used to determine the intercept and slope parameters, $\beta$ and $\gamma$, which are found to be $5.022 \pm 0.058$ and $33.268 \pm 0.083$, respectively. To infer the distances for GH\,\textsc{ii}R data, they relied on primary indicators such as Cepheids, TRGB, and theoretical model calibrations, \cite{FernandezArenas}. Using this relation the observed distance modulus of an \hiig\ can be computed as $\mu_{\rm obs} = 2.5\log L - 2.5\log f - 100.2$, where $f(z)$ is the measured flux at redshift $z$ corrected for extinction using the Gordon extinction law, \cite{Gordon_2003}. The theoretical distance modulus in a given cosmological model is $\mu_{\rm th}(z) = 5\log D_{L}(z) + 25$, where $D_L(z)=(1+z)D_M(z)$ is the luminosity distance.

\item[]{\bf \mq\ and \cq\ sample}. The sample of 78 \mq\ and 38 \cq\ measurements, listed in tables A1 of Refs.\ \cite{Khadkaetal_2021a} and \cite{Cao:2022pdv}, respectively, span a wide range of redshifts, from $0.0033 \leq z \leq 1.89$ for \mii\ QSOs and from $0.001064 \leq z \leq 3.368$ for \civ\ QSOs. Both \mq\ and \cq\ sources follow the radius-luminosity ($R-L$) relation $\log \tau=\beta+\gamma(\log L-44)$. Here $\tau$ (days) is the QSO time-lag, $\beta$ is the intercept parameter, and $\gamma$ is the slope parameter. For \mii\ QSOs and \civ\ QSOs we denote $\beta$ and $\gamma$ as $\beta_{\rm\textsc{m}}$ and $\gamma_{\rm\textsc{m}}$, and $\beta_{\rm\textsc{c}}$ and $\gamma_{\rm\textsc{c}}$, respectively. The monochromatic luminosity $L=4\pi D_L^2F$ where $F$ is the QSO flux measured in units of $\rm erg\ s^{-1}\ cm^{-2}$ at 1350\,\AA\ and 3000\,\AA\ for \mii\ QSOs and \civ\ QSOs, respectively. As described in Refs.\ \cite{Khadkaetal_2021a} and \cite{Cao:2022pdv}, in our analysis we must simultaneously determine both $R-L$ relation and cosmological parameters and verify that the $R-L$ relation parameters are independent of the assumed cosmological model for the \mii\ QSOs and \civ\ QSOs to be standardizable. In addition to what we used in Ref.\ \cite{CaoRatra2022}, here we also use the 38 \civ\ QSOs and improve on the analyses of these QSOs by now accounting for the asymmetry in the data error bars, as described in Ref.\ \cite{Cao:2022pdv}. 

\item[]{\bf A118 GRB sample.} The A118 sample, which is listed in table 7 of Ref.\ \cite{Khadkaetal_2021b}, consists of 118 long GRBs and spans a wide redshift range, from $0.3399$ to $8.2$. The isotropic radiated energy $E_{\rm iso}$ of a GRB source in its rest frame is $E_{\rm iso}=4\pi D_L^2S_{\rm bolo}/(1+z)$, where $S_{\rm bolo}$ is the bolometric fluence computed in the standard rest-frame energy band $1-10^4$ keV. The peak energy of a source is $E_{\rm p} = (1+z)E_{\rm p, obs}$, where $E_{\rm p, obs}$ is the observed peak energy. There is a correlation between $E_{\rm iso}$ and $E_{\rm p}$ known as the Amati correlation, \citep{Amati2008, Amati2009}, which is given by $\log E_{\rm iso} = \beta_{\rm\textsc{a}} + \gamma_{\rm\textsc{a}}\log E_{\rm p}$. As described in Ref.\ \cite{Khadkaetal_2021b}, we must simultaneously determine both Amati relation and cosmological parameters and verify that the Amati relation parameters are independent of the assumed cosmological model for the A118 GRBs to be  standardizable. Note that here we use the correct value of $E_{\rm p}=871\pm123$ keV for GRB081121, as discussed in Ref.\ \cite{Liuetal2022}, rather than the value listed in table 7 of Ref.\ \cite{Khadkaetal_2021b} and used in Ref.\ \citep{CaoRatra2022}.

\end{itemize}

\section{Data Analysis Methodology}
\label{sec:analysis}

The natural log of the likelihood function for the \civ, \mii, and A118 data sets (denoted with subscript ``\textsc{s}'' that is either \civ, \mii, or A118) with measured quantity \textbf{Q}, \citep{D'Agostini_2005}, is
\be
\label{eq:LF_s1}
    \ln\mathcal{L}_{\rm\textsc{s}}= -\frac{1}{2}\Bigg[\chi^2_{\rm\textsc{s}}+\sum^{N}_{i=1}\ln\left(2\pi\sigma^2_{\mathrm{tot,\textsc{s}},i}\right)\Bigg],
\ee
where
\be
\label{eq:chi2_s1}
    \chi^2_{\rm\textsc{s}} = \sum^{N}_{i=1}\bigg[\frac{(\mathbf{Q}_{\mathrm{obs,\textsc{s}},i} - \mathbf{Q}_{\mathrm{th,\textsc{s}},i})^2}{\sigma^2_{\mathrm{tot,\textsc{s}},i}}\bigg]
\ee
with total uncertainty
\be
\label{eq:sigma_s1}
\sigma^2_{\mathrm{tot,\textsc{s}},i}=\sigma_{\rm int,\,\textsc{s}}^2+\sigma_{{\mathbf{Q}_{\mathrm{obs,\textsc{s}},i}}}^2+\sigma_{{\mathbf{Q}_{\mathrm{th,\textsc{s}},i}}}^2.
\ee
$\sigma_{\rm int,\,\textsc{s}}$ represents the intrinsic scatter parameter for data \textsc{s}, which also accounts for unknown systematic uncertainties. $N$ is the total number of data points.

The natural log of the likelihood function for some $H(z)$ and BAO data and for QSO-AS and \hiig\ data (also denoted ``\textsc{s}'') with measured quantity \textbf{Q} is
\be
\label{eq:LF_s2}
    \ln\mathcal{L}_{\rm\textsc{s}}= -\frac{1}{2}\chi^2_{\rm\textsc{s}},
\ee
where
\be
\label{eq:chi2_s2}
    \chi^2_{\rm\textsc{s}} = \sum^{N}_{i=1}\bigg[\frac{(\mathbf{Q}_{\mathrm{obs,\textsc{s}},i} - \mathbf{Q}_{\mathrm{th,\textsc{s}},i})^2}{\sigma^2_{\mathrm{tot,\textsc{s}},i}}\bigg]
\ee
with total uncertainty
\be
\label{eq:sigma_s2}
\sigma^2_{\mathrm{tot,\textsc{s}},i}=\sigma_{\mathrm{sys,\textsc{s}},i}^2+\sigma_{{\mathbf{Q}_{\mathrm{obs,\textsc{s}},i}}}^2+\sigma_{{\mathbf{Q}_{\mathrm{th,\textsc{s}},i}}}^2,
\ee
with $\sigma_{\mathrm{sys,\textsc{s}},i}$ being the systematic uncertainty at redshift $z_i$. It is important to note that we have ignored the systematic uncertainties for \hiig\ data because they are not yet properly quantified. Following Ref.\ \cite{Cao_et_al2017b}, the total uncertainty for QSO-AS data is calculated as $\sigma_{\mathrm{tot},i}=\sigma_{\theta_{\mathrm{obs},i}} + 0.1\theta_{\mathrm{obs},i}$, taking into consideration a 10\% margin for both observational and intrinsic errors in the observed angular sizes $\theta_{\mathrm{obs},i}$.

For those $H(z)$ and BAO data (also denoted ``\textsc{s}'') with covariance matrix $\textbf{C}_{\textsc{s}}$,
\be
\label{eq:chi2_s3}
    \chi^2_{\textsc{s}} = [\mathbf{Q}_{\mathrm{th,\textsc{s}},i} - \mathbf{Q}_{\mathrm{obs,\textsc{s}},i}]^T\textbf{C}_{\textsc{s}}^{-1}[\mathbf{Q}_{\mathrm{th,\textsc{s}},i} - \mathbf{Q}_{\mathrm{obs,\textsc{s}},i}],
\ee
in which superscripts $T$ and $-1$ represent transpose and inverse of the matrix, respectively.

For SN Ia data, $\chi^2_{\rm SN}$ is calculated using equation (C1) in appendix C of Ref.\ \cite{Conley_et_al_2011}. In this equation the variable $\mathcal{M}$, that includes the SN Ia absolute magnitude M and the Hubble constant $H_0$, is marginalized, therefore here SN Ia data cannot constrain $H_0$. However, when we allow \obhs\ and \ochs\ to be free parameters $H_0$ is derived from the \obhs\ and \ochs\ constraints.

\begin{table}
\centering
\resizebox{\columnwidth}{!}{%
\begin{threeparttable}
\caption{Flat priors of the constrained parameters.}
\label{tab:priors}
\begin{tabular}{lcc}
\toprule
Parameter & & Prior\\
\midrule
 & Cosmological Parameters & \\
\midrule
$h$\tnote{a} &  & [None, None]\\
\obhs\,\tnote{b} &  & [0, 1]\\
\ochs\,\tnote{c} &  & [0, 1]\\
\ok &  & [$-2$, 2]\\
$\alpha$ &  & [0, 10]\\
\wx &  & [$-5$, 0.33]\\
\midrule
 & Non-Cosmological Parameters & \\
\midrule
$\gamma_{\mathrm{\textsc{m}}}$ &  & [0, 5]\\
$\beta_{\mathrm{\textsc{m}}}$ &  & [0, 10]\\
$\sigma_{\rm int,\,\textsc{m}}$ &  & [0, 5]\\
$\gamma_{\mathrm{\textsc{c}}}$ &  & [0, 5]\\
$\beta_{\mathrm{\textsc{c}}}$ &  & [0, 10]\\
$\sigma_{\rm int,\,\textsc{c}}$ &  & [0, 5]\\
$\gamma_{\mathrm{\textsc{a}}}$ &  & [0, 5]\\
$\beta_{\mathrm{\textsc{a}}}$ &  & [0, 300]\\
$\sigma_{\rm int,\,\textsc{a}}$ &  & [0, 5]\\
$l_{\rm m}$ &  & [None, None]\\
\bottomrule
\end{tabular}
\begin{tablenotes}[flushleft]
\item [a] $H_0$ in unit of 100 \hunit. In the \mii\ + \civ\ and A118 cases $h=0.7$, in the SNP + SND and SNP\plus\ cases $0.2\leq h\leq 1$, and in other cases the $h$ prior range is irrelevant (unbounded).
\item [b] In the \mii\ + \civ\ and A118 cases \obhs\ is set to be 0.0245, i.e. $\Omega_{b}=0.05$.
\item [c] In the \mii\ + \civ\ and A118 cases $\Om\in[0,1]$ is ensured.
\end{tablenotes}
\end{threeparttable}%
}
\end{table}

In this paper, we do not use CMB data. Instead, in our analyses of BAO data we determine the sound horizon $r_s$ by also constraining \obhs\ and \ochs\ from these data. Consequently our cosmological parameter constraints do not depend on CMB data. 

The QSO-AS + \hiig\ and \mii\ + \civ\ constraints used in our paper are taken from Refs.\ \cite{CaoRatra2022} and \cite{Cao:2022pdv}, respectively. The analyses of Ref.\ \cite{Cao:2022pdv} account for the asymmetric errors of the \mii\ + \civ\ measurements.

The flat priors for the free cosmological and non-cosmological parameters are listed in Table \ref{tab:priors}. Some of the individual data sets (or smaller combinations of individual data sets) are unable to constrain some cosmological parameters. In these cases we fix these cosmological parameter values, see discussion in Table \ref{tab:priors} footnotes. Note that we do not need to do this for larger combinations of individual data sets; these data combinations determine cosmological parameter values in a prior-independent manner.

The Markov chain Monte Carlo (MCMC) code \textsc{MontePython}, \citep{Audrenetal2013,Brinckmann2019}, is utilized to maximize the likelihood functions and determine the posterior distributions of all free parameters. The \textsc{python} package \textsc{getdist}, \citep{Lewis_2019}, is used to analyze the MCMC results and create plots.

The Akaike Information Criterion (AIC), the Bayesian Information Criterion (BIC), and the Deviance Information Criterion (DIC) are
\be
\label{AIC}
    \mathrm{AIC}=-2\ln \mathcal{L}_{\rm max} + 2n,
\ee
\be
\label{BIC}
    \mathrm{BIC}=-2\ln \mathcal{L}_{\rm max} + n\ln N,
\ee
and
\be
\label{DIC}
    \mathrm{DIC}=-2\ln \mathcal{L}_{\rm max} + 2n_{\rm eff},
\ee
where $n$ is the number of parameters in the given model, and $n_{\rm eff}=\langle-2\ln \mathcal{L}\rangle+2\ln \mathcal{L}_{\rm max}$ represents the number of effectively constrained parameters. Here, the angular brackets indicate an average over the posterior distribution and $\mathcal{L}_{\rm max}$ is the maximum value of the likelihood function.

$\Delta \mathrm{A/B/DIC}$ values are computed by subtracting the A/B/DIC value of the flat \lcdm\ reference model from the values of the other five cosmological dark energy models. Negative values of $\Delta \mathrm{A/B/DIC}$ indicate that the model being evaluated fits the data set better than does the flat \lcdm\ reference model, while positive values indicate a worse fit. In comparison to the model with the lowest A/B/DIC value, a $\Delta \mathrm{A/B/DIC}$ value within the range $(0, 2]$ indicates weak evidence against the model being evaluated, a value within $(2, 6]$ indicates positive evidence against the model, a value within $(6, 10]$ indicates strong evidence against the model, and a value greater than 10 indicates very strong evidence against the model.

\section{Results}
\label{sec:results}

\subsection{Comparison of constraints obtained from Old $H(z)$ data and $H(z)$ data, Old $H(z)$ + Old BAO data and $H(z)$ + BAO data, and SNP + SND data and SNP\plus\ data}
 \label{subsec:comp1}

\begin{table*}
\centering
\resizebox*{2.05\columnwidth}{2.35\columnwidth}{%
\begin{threeparttable}
\caption{Unmarginalized best-fitting parameter values for all models from various combinations of BAO, $H(z)$, and SN Ia data.}\label{tab:BFP}
\begin{tabular}{lcccccccccccccc}
\toprule
Model & Data set & $\Omega_{b}h^2$ & $\Omega_{c}h^2$ & \om & \ok & $w_{\mathrm{X}}$/$\alpha$\tnote{a} & $H_0$\tnote{b} & $-2\ln\mathcal{L}_{\mathrm{max}}$ & AIC & BIC & DIC & $\Delta \mathrm{AIC}$ & $\Delta \mathrm{BIC}$ & $\Delta \mathrm{DIC}$ \\
\midrule
\multirow{8}{*}{Flat \lcdm} & Old $H(z)$ & 0.0273 & 0.1201 & 0.319 & -- & -- & 68.16 & 14.54 & 20.54 & 24.94 & 18.87 & 0.00 & 0.00 & 0.00\\
 & $H(z)$ & 0.0120 & 0.1366 & 0.309 & -- & -- & 69.43 & 14.50 & 20.50 & 24.90 & 18.78 & 0.00 & 0.00 & 0.00\\
 & Old $H(z)$ + Old BAO & 0.0244 & 0.1181 & 0.301 & -- & -- & 68.98 & 25.64 & 31.64 & 36.99 & 32.32 & 0.00 & 0.00 & 0.00\\
 & $H(z)$ + BAO & 0.0254 & 0.1200 & 0.297 & -- & -- & 70.12 & 30.56 & 36.56 & 41.91 & 37.32 & 0.00 & 0.00 & 0.00\\
 & SNP + SND & 0.0102 & 0.0133 & 0.309 & -- & -- & 27.93 & 1056.64 & 1062.64 & 1077.56 & 1058.68 & 0.00 & 0.00 & 0.00\\
 & SNP\plus & 0.0139 & 0.1031 & 0.331 & -- & -- & 59.59 & 1406.97 & 1412.97 & 1429.08 & 1409.17 & 0.00 & 0.00 & 0.00\\
 & Old $H(z)$ + Old BAO + SNP + SND & 0.0242 & 0.1191 & 0.304 & -- & -- & 68.86 & 1082.39 & 1088.39 & 1103.44 & 1089.92 & 0.00 & 0.00 & 0.00\\
 & $H(z)$ + BAO + SNP\plus\ & 0.0239 & 0.1256 & 0.312 & -- & -- & 69.35 & 1439.59 & 1445.59 & 1461.79 & 1446.05 & 0.00 & 0.00 & 0.00\\
\\
\multirow{8}{*}{Non-flat \lcdm} & Old $H(z)$ & 0.0205 & 0.1515 & 0.362 & $-0.136$ & -- & 69.09 & 14.49 & 22.49 & 28.36 & 20.19 & 1.96 & 3.42 & 1.32\\
 & $H(z)$ & 0.0180 & 0.1328 & 0.314 & $-0.012$ & -- & 69.47 & 14.50 & 22.50 & 28.37 & 20.09 & 2.00 & 3.47 & 1.31\\
 & Old $H(z)$ + Old BAO & 0.0260 & 0.1098 & 0.292 & 0.048 & -- & 68.35 & 25.30 & 33.30 & 40.43 & 33.87 & 1.66 & 3.44 & 1.54\\
 & $H(z)$ + BAO & 0.0269 & 0.1128 & 0.289 & 0.041 & -- & 69.61 & 30.34 & 38.34 & 45.48 & 38.80 & 1.78 & 3.56 & 1.48\\
 & SNP + SND & 0.0335 & 0.1292 & 0.326 & $-0.043$ & -- & 70.84 & 1056.58 & 1064.58 & 1084.47 & 1060.96 & 1.94 & 6.91 & 2.28\\
 & SNP\plus & 0.0336 & 0.0663 & 0.295 & 0.095 & -- & 58.43 & 1406.46 & 1414.46 & 1435.95 & 1410.81 & 1.49 & 6.87 & 1.65\\
 & Old $H(z)$ + Old BAO + SNP + SND & 0.0255 & 0.1121 & 0.295 & 0.035 & -- & 68.53 & 1082.11 & 1090.11 & 1110.16 & 1091.17 & 1.72 & 6.72 & 1.24\\
 & $H(z)$ + BAO + SNP\plus\ & 0.0276 & 0.1078 & 0.288 & 0.084 & -- & 68.69 & 1437.61 & 1445.61 & 1467.21 & 1446.04 & 0.02 & 5.42 & $-0.01$\\
\\
\multirow{8}{*}{Flat XCDM} & Old $H(z)$ & 0.0376 & 0.1236 & 0.321 & -- & $-1.261$ & 70.95 & 14.39 & 22.39 & 28.25 & 22.17 & 1.85 & 3.32 & 3.30\\
 & $H(z)$ & 0.0106 & 0.1464 & 0.316 & -- & $-1.140$ & 70.63 & 14.47 & 22.47 & 28.33 & 22.28 & 1.97 & 3.43 & 3.49\\
 & Old $H(z)$ + Old BAO & 0.0296 & 0.0951 & 0.290 & -- & $-0.754$ & 65.79 & 22.39 & 30.39 & 37.52 & 30.63 & $-1.25$ & 0.53 & $-1.69$\\
 & $H(z)$ + BAO & 0.0318 & 0.0938 & 0.283 & -- & $-0.734$ & 66.67 & 26.58 & 34.58 & 41.71 & 34.83 & $-1.98$ & $-0.20$ & $-2.49$\\
 & SNP + SND & 0.0162 & 0.1648 & 0.319 & -- & $-1.028$ & 75.45 & 1056.62 & 1064.62 & 1084.52 & 1061.01 & 1.98 & 6.96 & 2.33\\
 & SNP\plus & 0.0243 & 0.0745 & 0.288 & -- & $-0.895$ & 58.73 & 1406.52 & 1414.52 & 1436.00 & 1410.84 & 1.55 & 6.92 & 1.67\\
 & Old $H(z)$ + Old BAO + SNP + SND & 0.0258 & 0.1115 & 0.295 & -- & $-0.940$ & 68.37 & 1081.34 & 1089.34 & 1109.40 & 1090.43 & 0.95 & 5.96 & 0.50\\
 & $H(z)$ + BAO + SNP\plus\ & 0.0283 & 0.1092 & 0.290 & -- & $-0.883$ & 68.96 & 1434.63 & 1442.63 & 1464.22 & 1443.28 & $-2.96$ & 2.43 & $-2.77$\\
\\
\multirow{8}{*}{Non-flat XCDM} & Old $H(z)$ & 0.0223 & 0.0736 & 0.172 & 0.324 & $-2.272$ & 75.05 & 14.14 & 24.14 & 31.47 & 20.73 & 3.60 & 6.53 & 1.86\\
 & $H(z)$ & 0.0316 & 0.0530 & 0.151 & 0.378 & $-2.278$ & 75.06 & 14.21 & 24.21 & 31.54 & 21.46 & 3.71 & 6.64 & 2.68\\
 & Old $H(z)$ + Old BAO & 0.0289 & 0.0985 & 0.296 & $-0.053$ & $-0.730$ & 65.76 & 22.13 & 32.13 & 41.05 & 32.51 & 0.49 & 4.06 & 0.19\\
 & $H(z)$ + BAO & 0.0305 & 0.0998 & 0.293 & $-0.084$ & $-0.703$ & 66.79 & 26.00 & 36.00 & 44.92 & 36.11 & $-0.56$ & 3.01 & $-1.21$\\
 & SNP + SND & 0.0057 & 0.0965 & 0.145 & $-0.576$ & $-0.596$ & 84.24 & 1056.24 & 1066.24 & 1091.11 & 1064.60 & 3.60 & 13.55 & 5.92\\
 & SNP\plus & 0.0334 & 0.0649 & 0.295 & 0.194 & $-1.155$ & 57.96 & 1406.43 & 1416.43 & 1443.29 & 1413.23 & 3.46 & 14.21 & 4.06\\
 & Old $H(z)$ + Old BAO + SNP + SND & 0.0255 & 0.1155 & 0.300 & $-0.030$ & $-0.922$ & 68.68 & 1081.28 & 1091.28 & 1116.35 & 1092.47 & 2.89 & 12.91 & 2.55\\
 & $H(z)$ + BAO + SNP\plus\ & 0.0278 & 0.1118 & 0.294 & $-0.032$ & $-0.865$ & 69.07 & 1434.46 & 1444.46 & 1471.45 & 1445.42 & $-1.13$ & 9.66 & $-0.63$\\
\\
\multirow{8}{*}{Flat \pcdm} & Old $H(z)$ & 0.0140 & 0.1341 & 0.321 & -- & 0.000 & 68.04 & 14.54 & 22.54 & 28.40 & 20.81 & 2.00 & 3.47 & 1.94\\
 & $H(z)$ & 0.0158 & 0.1335 & 0.312 & -- & 0.001 & 69.31 & 14.51 & 22.51 & 28.37 & 20.05 & 2.00 & 3.47 & 1.27\\
 & Old $H(z)$ + Old BAO & 0.0330 & 0.0911 & 0.278 & -- & 1.018 & 66.98 & 22.14 & 30.14 & 37.27 & 29.56 & $-1.33$ & 0.46 & $-2.42$\\
 & $H(z)$ + BAO & 0.0336 & 0.0866 & 0.271 & -- & 1.157 & 66.80 & 26.50 & 34.50 & 41.64 & 34.15 & $-2.05$ & $-0.27$ & $-3.17$\\
 & SNP + SND & 0.0198 & 0.0089 & 0.308 & -- & 0.001 & 30.81 & 1056.64 & 1064.64 & 1084.53 & 1062.54 & 2.00 & 6.98 & 3.86\\
 & SNP\plus & 0.0132 & 0.2553 & 0.279 & -- & 0.399 & 98.21 & 1406.50 & 1414.50 & 1435.98 & 1411.49 & 1.53 & 6.90 & 2.33\\
 & Old $H(z)$ + Old BAO + SNP + SND & 0.0263 & 0.1097 & 0.292 & -- & 0.203 & 68.39 & 1081.22 & 1089.22 & 1109.28 & 1089.91 & 0.83 & 5.84 & $-0.01$\\
 & $H(z)$ + BAO + SNP\plus\ & 0.0288 & 0.1060 & 0.286 & -- & 0.402 & 68.84 & 1434.43 & 1442.43 & 1464.02 & 1442.92 & $-3.16$ & 2.23 & $-3.14$\\
\\
\multirow{8}{*}{Non-flat \pcdm} & Old $H(z)$ & 0.0213 & 0.1514 & 0.365 & $-0.144$ & 0.036 & 68.91 & 14.50 & 24.50 & 31.83 & 21.42 & 3.96 & 6.89 & 2.55\\
 & $H(z)$ & 0.0358 & 0.1120 & 0.310 & 0.003 & 0.011 & 69.18 & 14.51 & 24.51 & 31.84 & 20.63 & 4.00 & 6.94 & 1.85\\
 & Old $H(z)$ + Old BAO & 0.0306 & 0.0920 & 0.284 & $-0.058$ & 1.200 & 65.91 & 22.05 & 32.05 & 40.97 & 31.30 & 0.41 & 3.98 & $-1.02$\\
 & $H(z)$ + BAO & 0.0337 & 0.0894 & 0.275 & $-0.074$ & 1.393 & 67.16 & 25.92 & 35.92 & 44.84 & 35.29 & $-0.64$ & 2.93 & $-2.03$\\
 & SNP + SND & 0.0283 & 0.1387 & 0.251 & $-0.251$ & 1.107 & 81.71 & 1056.40 & 1066.40 & 1091.26 & 1062.52 & 3.76 & 13.71 & 3.84\\
 & SNP\plus & 0.0140 & 0.0525 & 0.297 & 0.085 & 0.005 & 47.56 & 1406.47 & 1416.47 & 1443.32 & 1411.50 & 3.50 & 14.24 & 2.33\\
 & Old $H(z)$ + Old BAO + SNP + SND & 0.0261 & 0.1119 & 0.295 & $-0.023$ & 0.253 & 68.56 & 1081.12 & 1091.12 & 1116.19 & 1091.27 & 2.73 & 12.75 & 1.35\\
 & $H(z)$ + BAO + SNP\plus\ & 0.0282 & 0.1104 & 0.291 & $-0.045$ & 0.480 & 69.13 & 1434.23 & 1444.23 & 1471.22 & 1444.26 & $-1.36$ & 9.43 & $-1.79$\\
\bottomrule
\end{tabular}
\begin{tablenotes}[flushleft]
\item [a] \wx\ corresponds to flat/non-flat XCDM and $\alpha$ corresponds to flat/non-flat \pcdm.
\item [b] \hunit.
\end{tablenotes}
\end{threeparttable}%
}
\end{table*}

\begin{table*}
\centering
\resizebox*{2.05\columnwidth}{2.35\columnwidth}{%
\begin{threeparttable}
\caption{One-dimensional posterior mean parameter values and uncertainties ($\pm 1\sigma$ error bars or $2\sigma$ limits) for all models from various combinations of BAO, $H(z)$, and SN Ia data.}\label{tab:1d_BFP}
\begin{tabular}{lccccccc}
\toprule
Model & Data set & $\Omega_{b}h^2$ & $\Omega_{c}h^2$ & \om & \ok & $w_{\mathrm{X}}$/$\alpha$\tnote{a} & $H_0$\tnote{b}\\
\midrule
\multirow{8}{*}{Flat \lcdm} & Old $H(z)$ & $0.0225\pm0.0108$ & $0.1264\pm0.0207$ & $0.328^{+0.052}_{-0.073}$ & -- & -- & $67.98\pm3.24$ \\
 & $H(z)$ & $0.0225\pm0.0107$ & $0.1275\pm0.0208$ & $0.319^{+0.050}_{-0.074}$ & -- & -- & $69.31\pm4.25$ \\
 & Old $H(z)$ + Old BAO & $0.0247\pm0.0030$ & $0.1186^{+0.0076}_{-0.0083}$ & $0.301^{+0.016}_{-0.018}$ & -- & -- & $69.14\pm1.85$ \\
 & $H(z)$ + BAO & $0.0260\pm0.0040$ & $0.1212^{+0.0091}_{-0.0101}$ & $0.297^{+0.015}_{-0.017}$ & -- & -- & $70.49\pm2.74$ \\
 & SNP + SND & $0.0224\pm0.0109$ & $0.1658^{+0.0927}_{-0.0598}$ & $0.310^{+0.021}_{-0.023}$ & -- & -- & $>45.87$ \\
 & SNP\plus & $0.0224\pm0.0109$ & $0.1785^{+0.1038}_{-0.0623}$ & $0.332\pm0.020$ & -- & -- & $>44.39$ \\
 & Old $H(z)$ + Old BAO + SNP + SND & $0.0244\pm0.0027$ & $0.1199\pm0.0076$ & $0.304^{+0.014}_{-0.015}$ & -- & -- & $69.04\pm1.77$ \\
 & $H(z)$ + BAO + SNP\plus\ & $0.0243\pm0.0034$ & $0.1267^{+0.0080}_{-0.0089}$ & $0.312\pm0.013$ & -- & -- & $69.65\pm2.48$ \\
\\
\multirow{8}{*}{Non-flat \lcdm} & Old $H(z)$ & $0.0223^{+0.0109}_{-0.0108}$ & $0.1685^{+0.0736}_{-0.1139}$ & $0.390^{+0.167}_{-0.172}$ & $-0.174^{+0.501}_{-0.491}$ & -- & $69.09^{+4.70}_{-4.67}$ \\
 & $H(z)$ & $0.0222\pm0.0108$ & $0.1612^{+0.0691}_{-0.1207}$ & $0.374^{+0.151}_{-0.210}$ & $-0.136^{+0.564}_{-0.457}$ & -- & $69.56^{+4.89}_{-4.88}$ \\
 & Old $H(z)$ + Old BAO & $0.0266^{+0.0039}_{-0.0045}$ & $0.1088\pm0.0166$ & $0.291\pm0.023$ & $0.059^{+0.081}_{-0.091}$ & -- & $68.37\pm2.10$ \\
 & $H(z)$ + BAO & $0.0275^{+0.0046}_{-0.0051}$ & $0.1131^{+0.0180}_{-0.0181}$ & $0.289\pm0.023$ & $0.047^{+0.082}_{-0.089}$ & -- & $69.81\pm2.80$ \\
 & SNP + SND & $0.0224\pm0.0107$ & $0.1698^{+0.0766}_{-0.0911}$ & $0.317\pm0.068$ & $-0.017^{+0.172}_{-0.174}$ & -- & $>46.83$ \\
 & SNP\plus & $0.0224\pm0.0107$ & $0.1745^{+0.0750}_{-0.0752}$ & $0.298\pm0.056$ & $0.089\pm0.132$ & -- & $>53.95$ \\
 & Old $H(z)$ + Old BAO + SNP + SND & $0.0260^{+0.0037}_{-0.0043}$ & $0.1119\pm0.0157$ & $0.294\pm0.022$ & $0.040\pm0.070$ & -- & $68.62\pm1.90$ \\
 & $H(z)$ + BAO + SNP\plus\ & $0.0282^{+0.0046}_{-0.0050}$ & $0.1082\pm0.0152$ & $0.288\pm0.021$ & $0.087\pm0.063$ & -- & $68.89\pm2.44$ \\
\\
\multirow{8}{*}{Flat XCDM} & Old $H(z)$ & $0.0225\pm0.0107$ & $0.1505^{+0.0337}_{-0.0206}$ & $0.285^{+0.061}_{-0.075}$ & -- & $-1.972^{+1.164}_{-0.588}$ & $79.55^{+7.01}_{-15.05}$ \\
 & $H(z)$ & $0.0225\pm0.0108$ & $0.1505^{+0.0303}_{-0.0200}$ & $0.278^{+0.065}_{-0.081}$ & -- & $-2.127^{+1.335}_{-0.629}$ & $80.96^{+7.59}_{-16.10}$ \\
 & Old $H(z)$ + Old BAO & $0.0295^{+0.0042}_{-0.0050}$ & $0.0969^{+0.0178}_{-0.0152}$ & $0.289\pm0.020$ & -- & $-0.784^{+0.140}_{-0.107}$ & $66.22^{+2.31}_{-2.54}$ \\
 & $H(z)$ + BAO & $0.0308^{+0.0053}_{-0.0046}$ & $0.0978^{+0.0184}_{-0.0164}$ & $0.285\pm0.019$ & -- & $-0.776^{+0.130}_{-0.103}$ & $67.18\pm3.18$ \\
 & SNP + SND & $0.0224\pm0.0108$ & $0.1750^{+0.0824}_{-0.0980}$ & $0.315^{+0.083}_{-0.057}$ & -- & $-1.054^{+0.237}_{-0.171}$ & $>48.66$ \\
 & SNP\plus & $0.0224\pm0.0107$ & $0.1552^{+0.0718}_{-0.0909}$ & $0.281^{+0.079}_{-0.061}$ & -- & $-0.900^{+0.166}_{-0.124}$ & $>49.26$ \\
 & Old $H(z)$ + Old BAO + SNP + SND & $0.0262^{+0.0033}_{-0.0037}$ & $0.1120\pm0.0110$ & $0.295\pm0.016$ & -- & $-0.941\pm0.064$ & $68.55\pm1.85$ \\
 & $H(z)$ + BAO + SNP\plus\ & $0.0287\pm0.0044$ & $0.1097\pm0.0117$ & $0.290\pm0.016$ & -- & $-0.886\pm0.053$ & $69.15\pm2.52$\\
\\
\multirow{8}{*}{Non-flat XCDM} & Old $H(z)$ & $0.0218^{+0.0075}_{-0.0144}$ & $0.0927^{+0.0217}_{-0.0890}$ & $0.228^{+0.055}_{-0.168}$ & $0.241^{+0.451}_{-0.261}$ & $-2.148^{+1.682}_{-0.776}$ & $71.98^{+5.86}_{-11.09}$ \\
 & $H(z)$ & $0.0218^{+0.0093}_{-0.0119}$ & $<0.2364$ & $0.228^{+0.054}_{-0.175}$ & $0.228^{+0.456}_{-0.267}$ & $-2.149^{+1.673}_{-0.772}$ & $73.06^{+6.61}_{-11.48}$ \\
 & Old $H(z)$ + Old BAO & $0.0294^{+0.0047}_{-0.0050}$ & $0.0980^{+0.0186}_{-0.0187}$ & $0.292\pm0.025$ & $-0.027\pm0.109$ & $-0.770^{+0.149}_{-0.098}$ & $66.13^{+2.35}_{-2.36}$ \\
 & $H(z)$ + BAO & $0.0303^{+0.0054}_{-0.0048}$ & $0.1021\pm0.0193$ & $0.292\pm0.024$ & $-0.054\pm0.103$ & $-0.757^{+0.135}_{-0.093}$ & $67.33\pm2.96$ \\
 & SNP + SND & $0.0224\pm0.0107$ & $0.1455^{+0.0644}_{-0.0987}$ & $0.271\pm0.085$ & $0.130^{+0.426}_{-0.249}$ & $-1.499^{+0.901}_{-0.237}$ & $>48.72$ \\
 & SNP\plus & $0.0224\pm0.0107$ & $0.1404^{+0.0654}_{-0.0845}$ & $0.259^{+0.071}_{-0.063}$ & $0.215^{+0.350}_{-0.202}$ & $-1.424^{+0.798}_{-0.251}$ & $>49.14$ \\
 & Old $H(z)$ + Old BAO + SNP + SND & $0.0262^{+0.0037}_{-0.0043}$ & $0.1119\pm0.0157$ & $0.295\pm0.022$ & $-0.001\pm0.098$ & $-0.948^{+0.098}_{-0.068}$ & $68.53\pm1.90$ \\
 & $H(z)$ + BAO + SNP\plus\ & $0.0284\pm0.0047$ & $0.1115^{+0.0151}_{-0.0165}$ & $0.293\pm0.021$ & $-0.017\pm0.095$ & $-0.884^{+0.082}_{-0.058}$ & $69.23\pm2.53$ \\
\\
\multirow{8}{*}{Flat \pcdm} & Old $H(z)$ & $0.0222\pm0.0107$ & $0.0642^{+0.0288}_{-0.0432}$ & $0.215^{+0.066}_{-0.108}$ & -- & $3.959^{+1.353}_{-3.789}$ & $63.95^{+3.01}_{-3.29}$ \\
 & $H(z)$ & $0.0221\pm0.0108$ & $0.0620^{+0.0273}_{-0.0440}$ & $0.199^{+0.059}_{-0.106}$ & -- & $3.972^{+1.394}_{-3.739}$ & $65.80^{+4.12}_{-4.10}$ \\
 & Old $H(z)$ + Old BAO & $0.0320^{+0.0054}_{-0.0041}$ & $0.0855^{+0.0175}_{-0.0174}$ & $0.275\pm0.023$ & -- & $1.267^{+0.536}_{-0.807}$ & $65.47^{+2.22}_{-2.21}$ \\
 & $H(z)$ + BAO & $0.0326^{+0.0061}_{-0.0030}$ & $0.0866^{+0.0197}_{-0.0180}$ & $0.272^{+0.024}_{-0.022}$ & -- & $1.271^{+0.507}_{-0.836}$ & $66.19^{+2.89}_{-2.88}$ \\
 & SNP + SND & $0.0221\pm0.0107$ & $0.0875^{+0.0333}_{-0.0733}$ & $0.181^{+0.075}_{-0.076}$ & -- & $<4.052$ & $>50.07$ \\%
 & SNP\plus & $0.0220^{+0.0100}_{-0.0118}$ & $0.0844^{+0.0288}_{-0.0757}$ & $0.175^{+0.065}_{-0.092}$ & -- & $1.966^{+0.479}_{-1.907}$ & $>50.58$ \\%
 & Old $H(z)$ + Old BAO + SNP + SND & $0.0278^{+0.0032}_{-0.0039}$ & $0.1054^{+0.0117}_{-0.0100}$ & $0.287\pm0.017$ & -- & $0.324^{+0.122}_{-0.264}$ & $68.29\pm1.78$ \\
 & $H(z)$ + BAO + SNP\plus\ & $0.0300^{+0.0047}_{-0.0046}$ & $0.1040\pm0.0129$ & $0.282\pm0.018$ & -- & $0.475^{+0.189}_{-0.265}$ & $69.01\pm2.43$ \\
\\
\multirow{8}{*}{Non-flat \pcdm} & Old $H(z)$ & $0.0215^{+0.0090}_{-0.0120}$ & $0.0536^{+0.0160}_{-0.0495}$ & $0.193^{+0.056}_{-0.117}$ & $0.262^{+0.265}_{-0.337}$ & $3.911^{+1.188}_{-3.871}$ & $62.81^{+2.65}_{-3.18}$ \\
 & $H(z)$ & $0.0213^{+0.0083}_{-0.0124}$ & $0.0481^{+0.0135}_{-0.0452}$ & $0.172^{+0.048}_{-0.104}$ & $0.277^{+0.249}_{-0.331}$ & $4.087^{+1.393}_{-3.890}$ & $64.32^{+3.59}_{-4.04}$ \\
 & Old $H(z)$ + Old BAO & $0.0320^{+0.0057}_{-0.0038}$ & $0.0865^{+0.0172}_{-0.0198}$ & $0.277^{+0.023}_{-0.026}$ & $-0.034^{+0.087}_{-0.098}$ & $1.360^{+0.584}_{-0.819}$ & $65.53\pm2.19$ \\
 & $H(z)$ + BAO & $0.0325^{+0.0064}_{-0.0029}$ & $0.0881^{+0.0199}_{-0.0201}$ & $0.275\pm0.025$ & $-0.052^{+0.093}_{-0.087}$ & $1.427^{+0.572}_{-0.830}$ & $66.24\pm2.88$ \\
 & SNP + SND & $0.0223\pm0.0107$ & $0.1098^{+0.0426}_{-0.0861}$ & $0.212^{+0.080}_{-0.082}$ & $-0.026^{+0.126}_{-0.150}$ & $<3.067$ & $>50.82$ \\%
 & SNP\plus & $0.0224\pm0.0107$ & $0.1113^{+0.0464}_{-0.0758}$ & $0.210^{+0.068}_{-0.069}$ & $-0.001^{+0.108}_{-0.132}$ & $1.282^{+0.290}_{-1.255}$ & $>52.64$ \\%
 & Old $H(z)$ + Old BAO + SNP + SND & $0.0271^{+0.0038}_{-0.0043}$ & $0.1095\pm0.0152$ & $0.292\pm0.022$ & $-0.038^{+0.071}_{-0.085}$ & $0.382^{+0.151}_{-0.299}$ & $68.48\pm1.85$ \\
 & $H(z)$ + BAO + SNP\plus\ & $0.0296^{+0.0048}_{-0.0047}$ & $0.1067^{+0.0153}_{-0.0154}$ & $0.286\pm0.021$ & $-0.035^{+0.071}_{-0.085}$ & $0.550^{+0.231}_{-0.314}$ & $69.15\pm2.53$ \\
\bottomrule
\end{tabular}
\begin{tablenotes}[flushleft]
\item [a] \wx\ corresponds to flat/non-flat XCDM and $\alpha$ corresponds to flat/non-flat \pcdm.
\item [b] \hunit.
\end{tablenotes}
\end{threeparttable}%
}
\end{table*}

\begin{figure*}
\centering
 \subfloat[]{%
    \includegraphics[width=0.45\textwidth,height=0.35\textwidth]{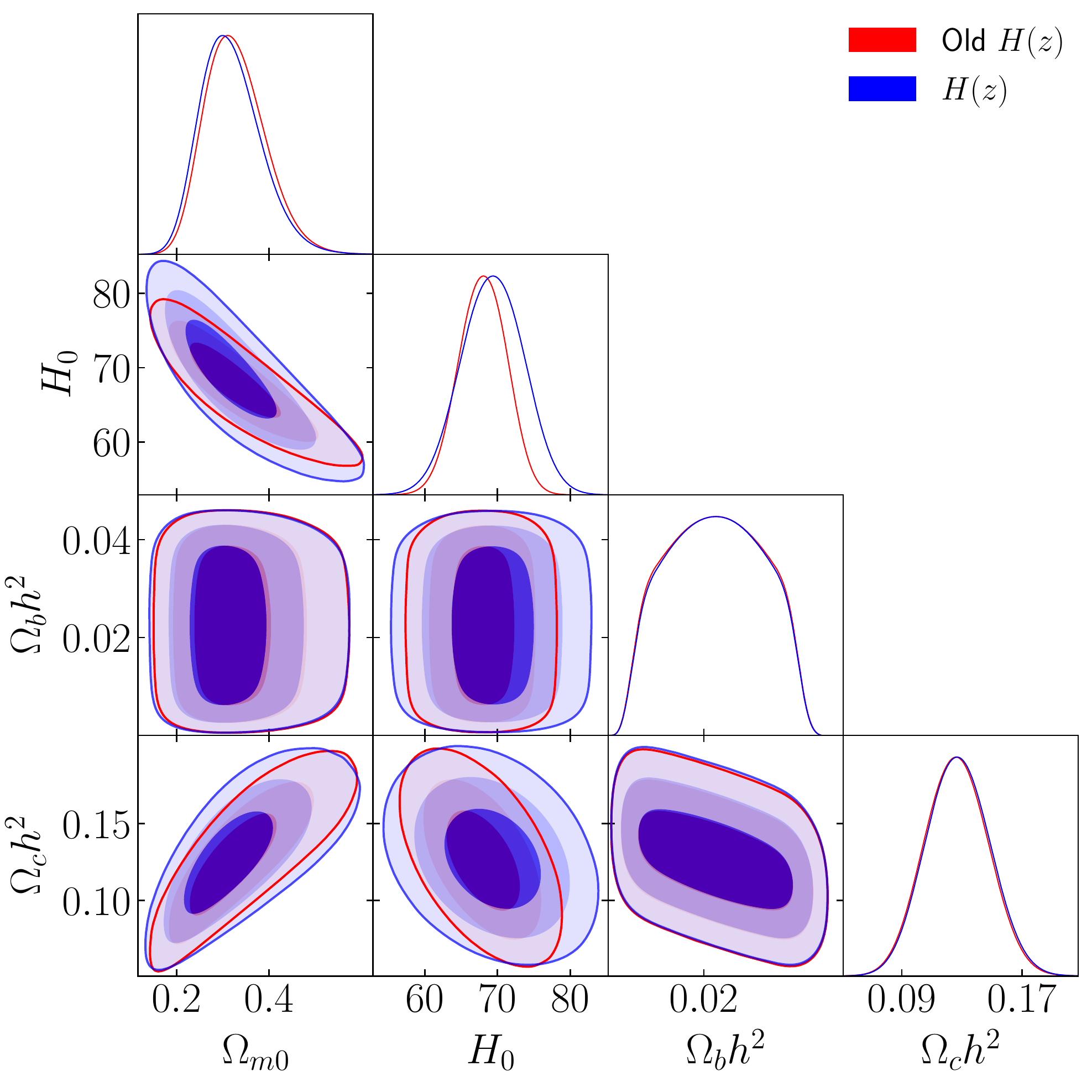}}
 \subfloat[]{%
    \includegraphics[width=0.45\textwidth,height=0.35\textwidth]{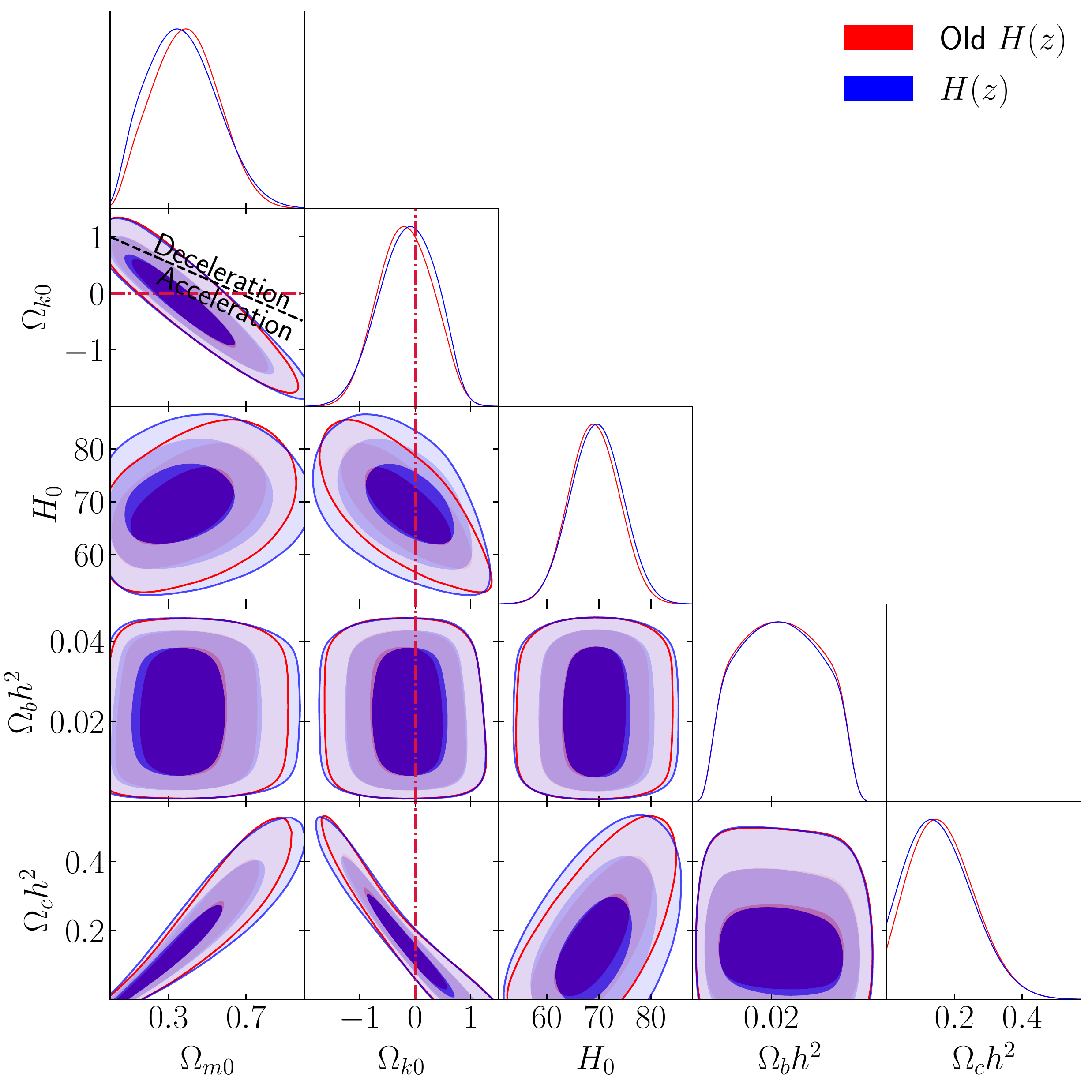}}\\
 \subfloat[]{%
    \includegraphics[width=0.45\textwidth,height=0.35\textwidth]{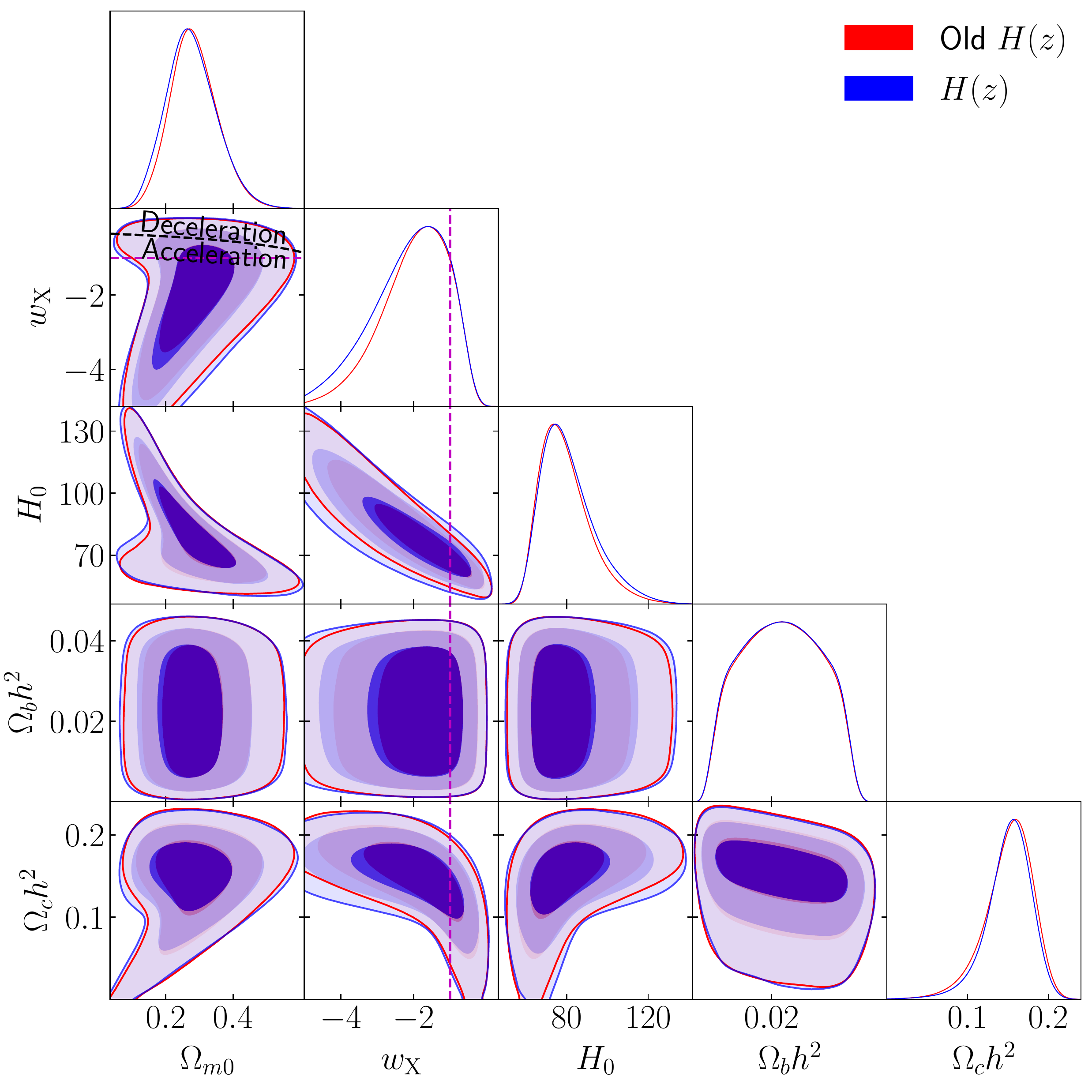}}
 \subfloat[]{%
    \includegraphics[width=0.45\textwidth,height=0.35\textwidth]{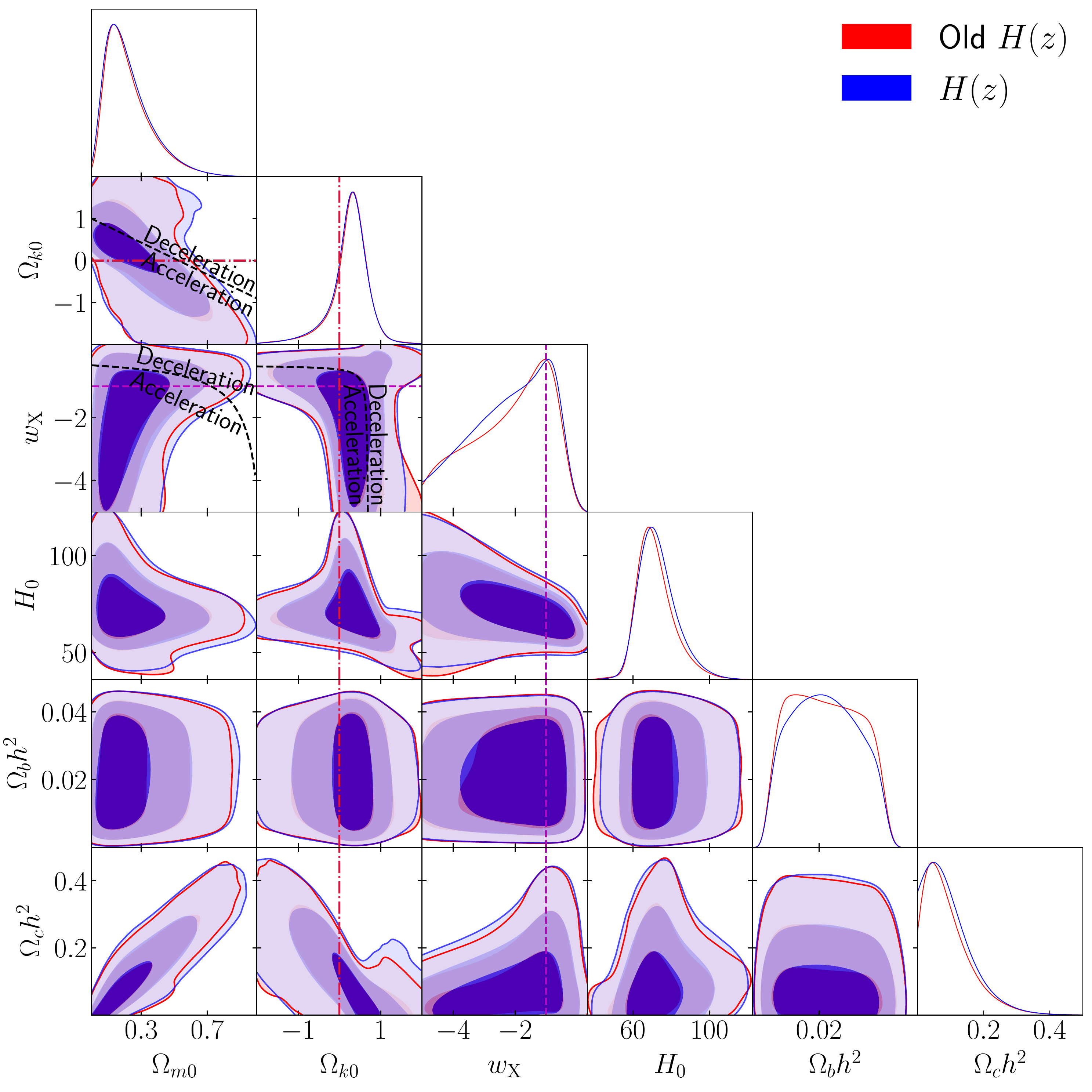}}\\
 \subfloat[]{%
    \includegraphics[width=0.45\textwidth,height=0.35\textwidth]{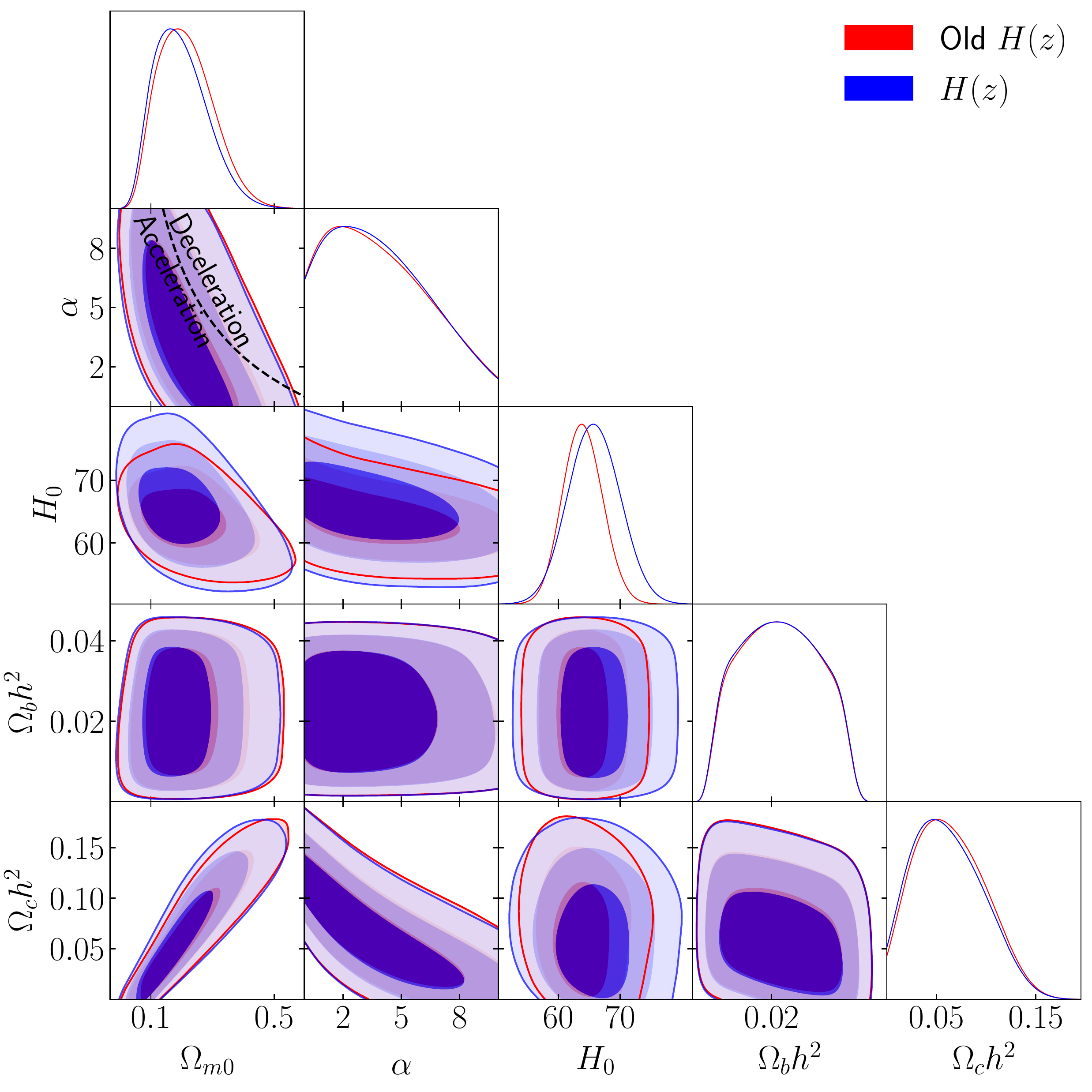}}
 \subfloat[]{%
    \includegraphics[width=0.45\textwidth,height=0.35\textwidth]{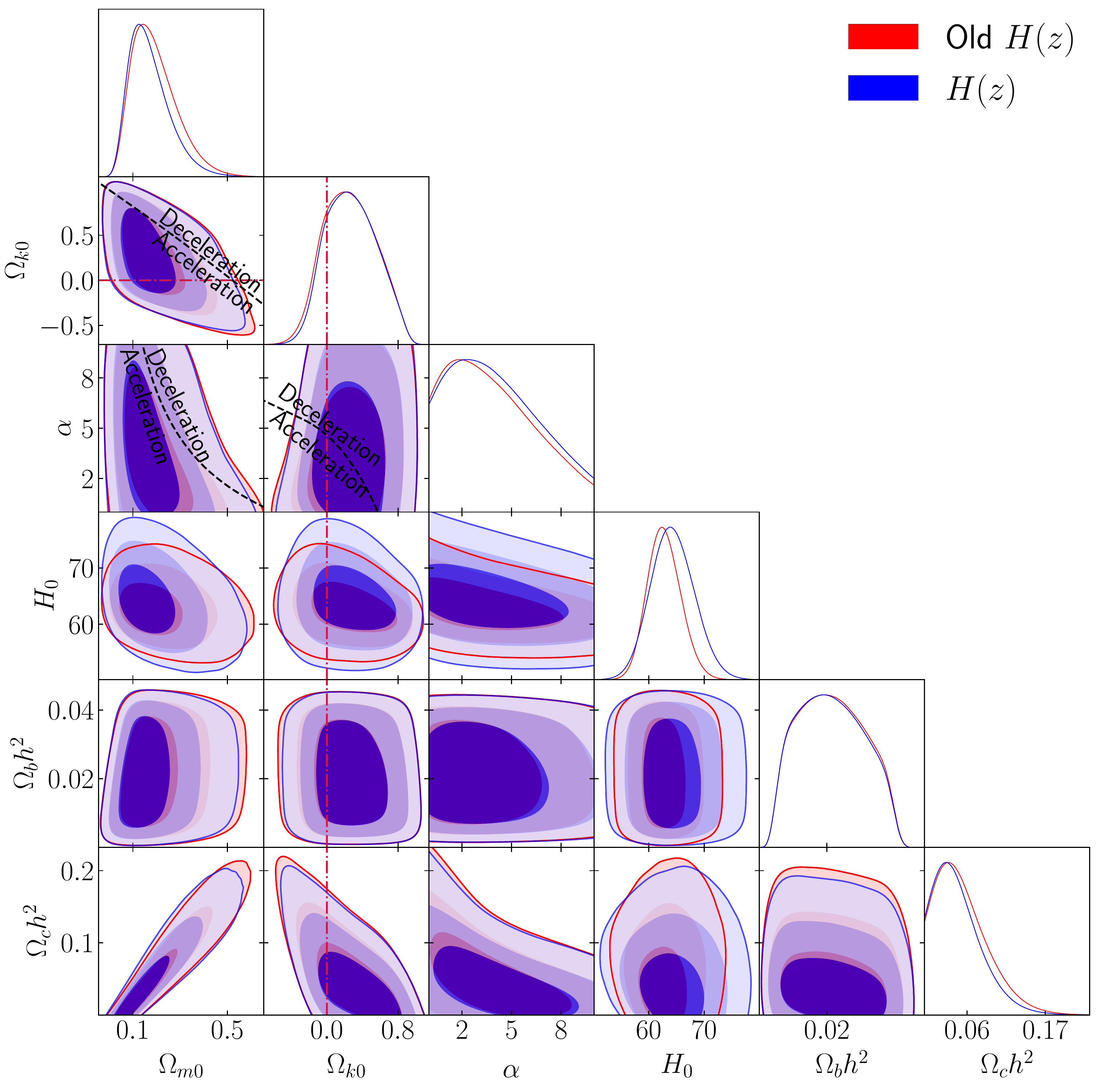}}\\
\caption{One-dimensional likelihoods and 1$\sigma$, 2$\sigma$, and 3$\sigma$ two-dimensional likelihood confidence contours from Old $H(z)$ (red) and $H(z)$ (blue) data for six different models. The black dashed zero-acceleration lines in panels (b)--(f), computed for the third cosmological parameter set to the $H(z)$ + BAO data best-fitting values listed in Table \ref{tab:BFP} in panels (d) and (f), divides the parameter space into regions associated with currently-accelerating (below or below left) and currently-decelerating (above or above right) cosmological expansion. The crimson dash-dot lines represent flat hypersurfaces, with closed spatial hypersurfaces either below or to the left. The magenta lines represent $w_{\rm X}=-1$, i.e.\ flat or non-flat \lcdm\ models. The $\alpha = 0$ axes correspond to flat and non-flat \lcdm\ models in panels (e) and (f), respectively.}
\label{fig1}
\end{figure*}

\begin{figure*}
\centering
 \subfloat[]{%
    \includegraphics[width=0.45\textwidth,height=0.35\textwidth]{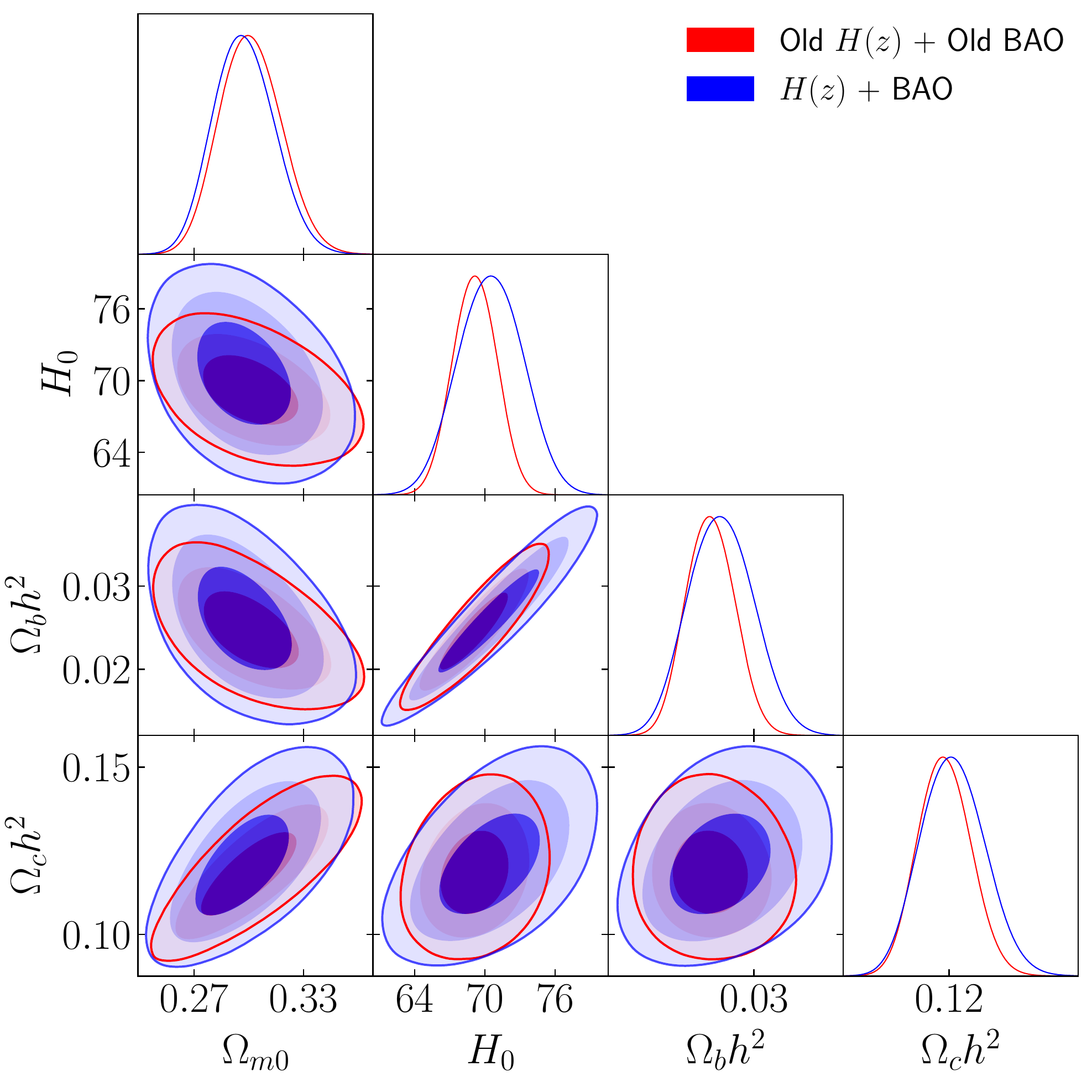}}
 \subfloat[]{%
    \includegraphics[width=0.45\textwidth,height=0.35\textwidth]{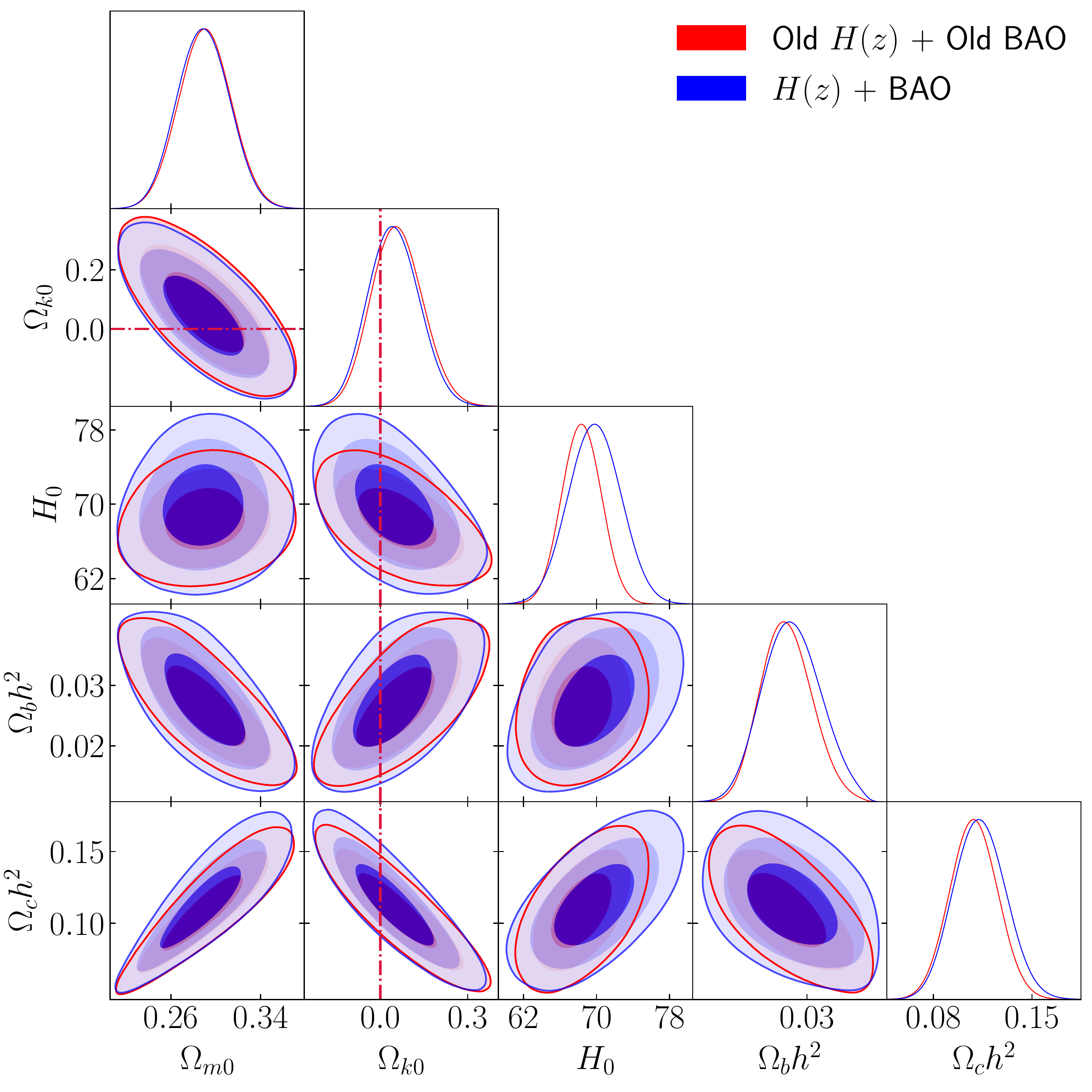}}\\
 \subfloat[]{%
    \includegraphics[width=0.45\textwidth,height=0.35\textwidth]{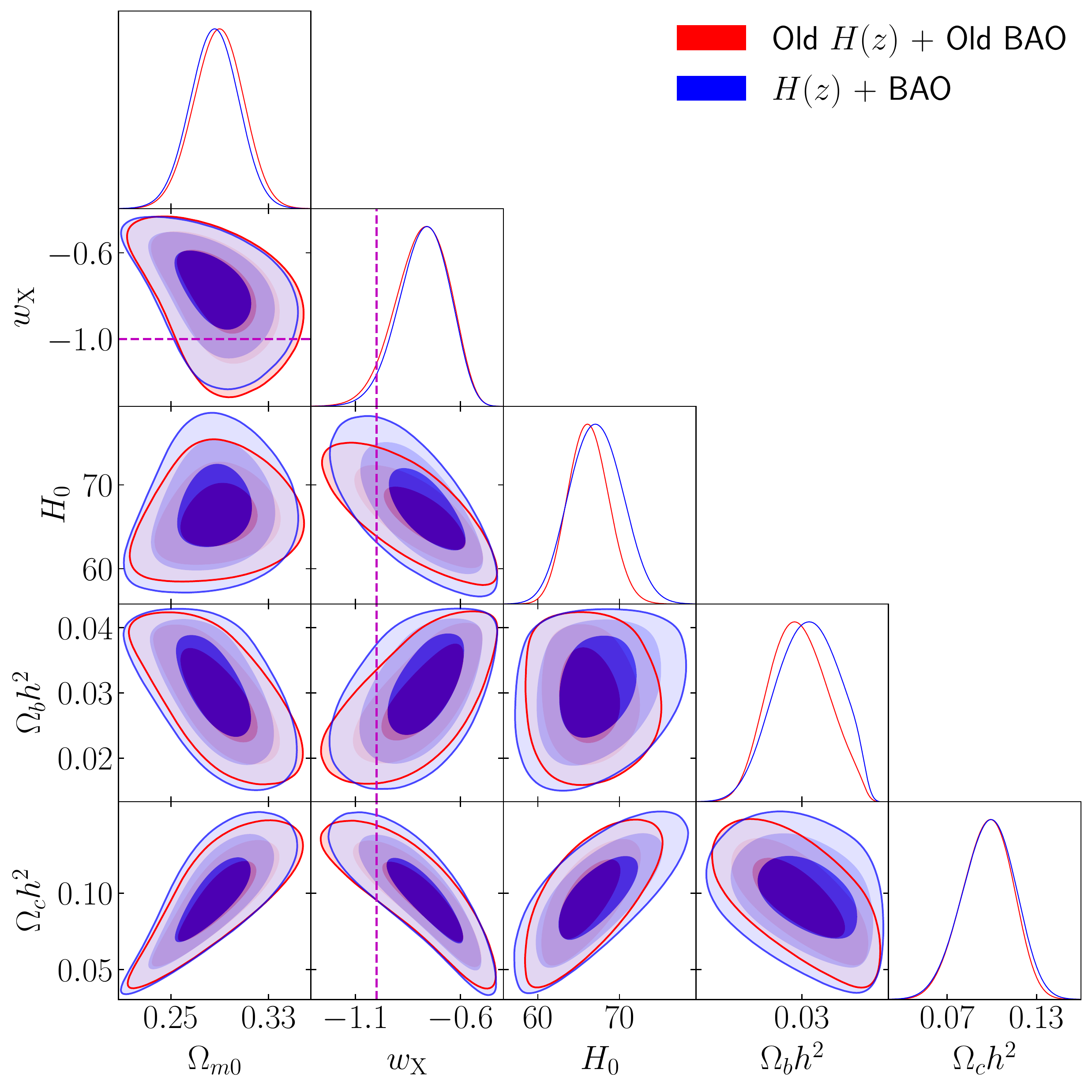}}
 \subfloat[]{%
    \includegraphics[width=0.45\textwidth,height=0.35\textwidth]{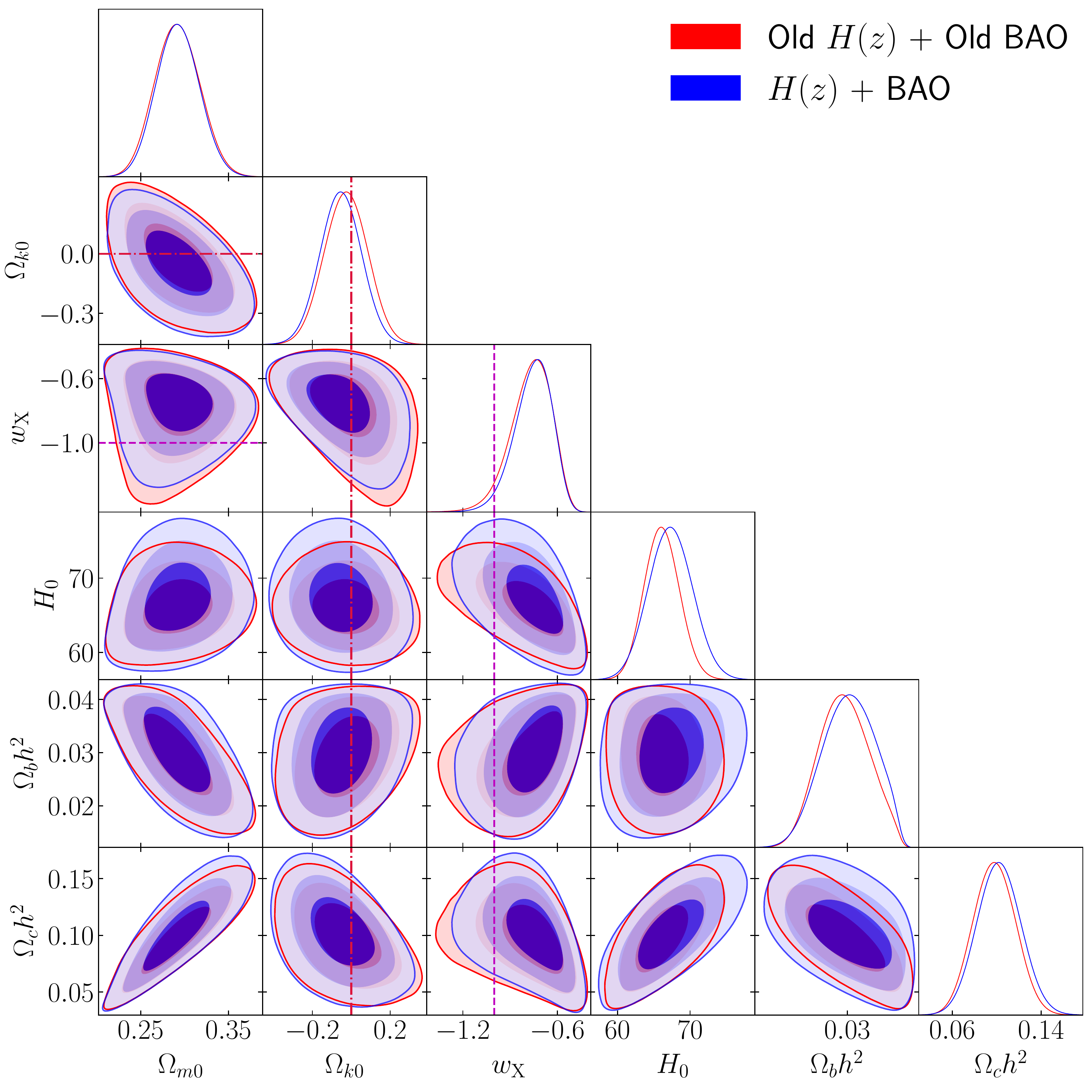}}\\
 \subfloat[]{%
    \includegraphics[width=0.45\textwidth,height=0.35\textwidth]{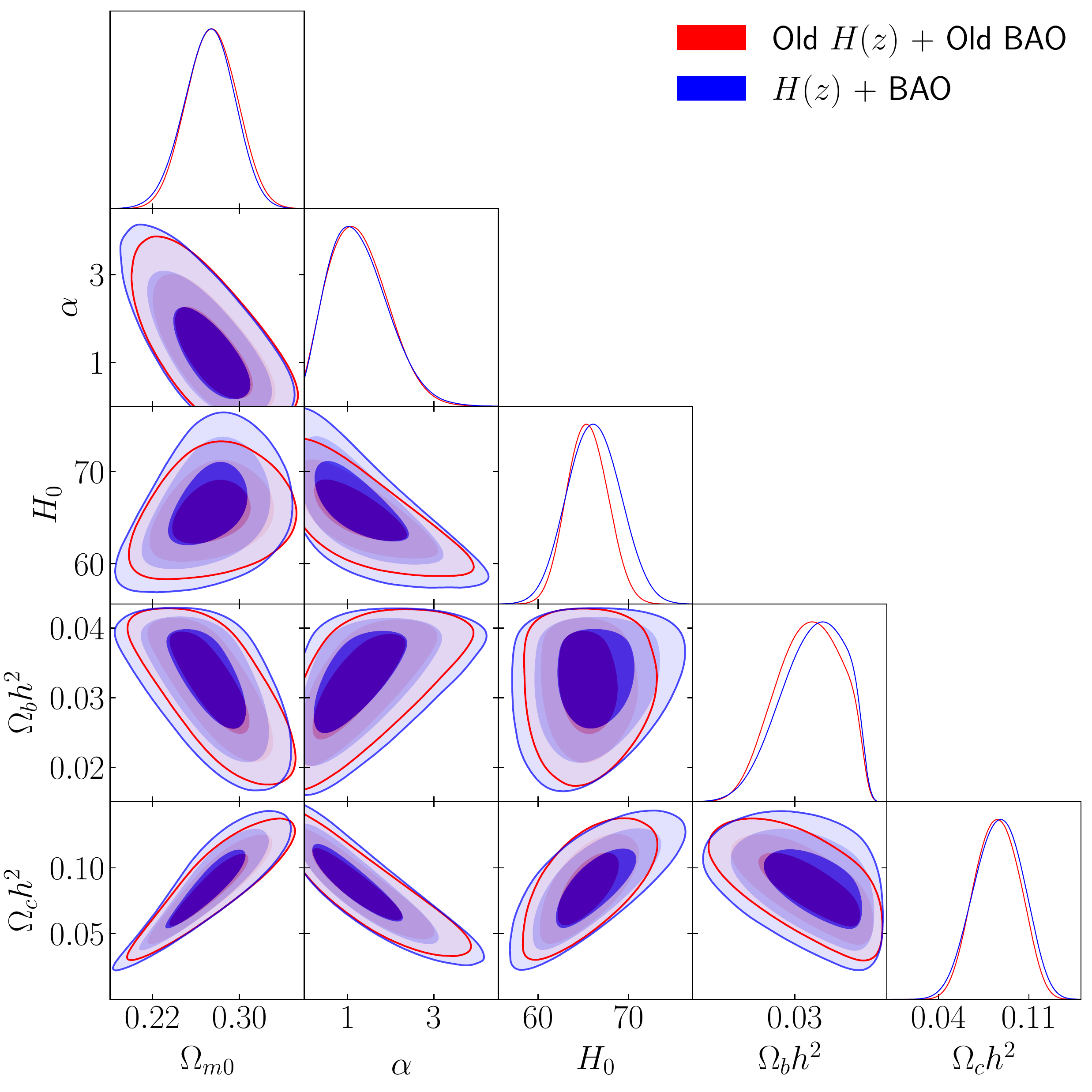}}
 \subfloat[]{%
    \includegraphics[width=0.5\textwidth,height=0.35\textwidth]{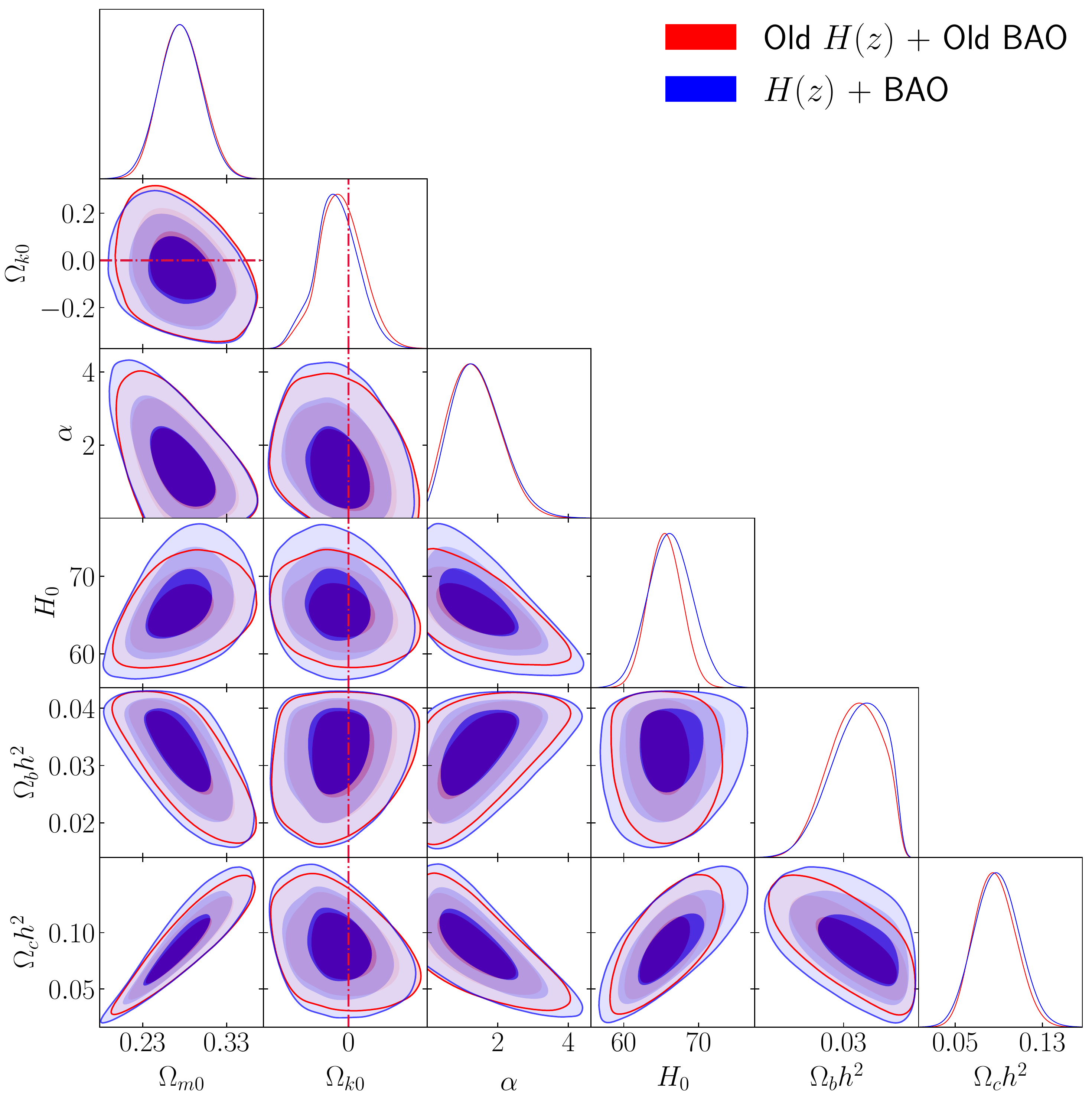}}\\
\caption{Same as Fig. \ref{fig1} but for Old $H(z)$ + Old BAO (red) and $H(z)$ + BAO (blue) data.}
\label{fig2}
\end{figure*}

\begin{figure*}
\centering
 \subfloat[]{%
    \includegraphics[width=0.45\textwidth,height=0.35\textwidth]{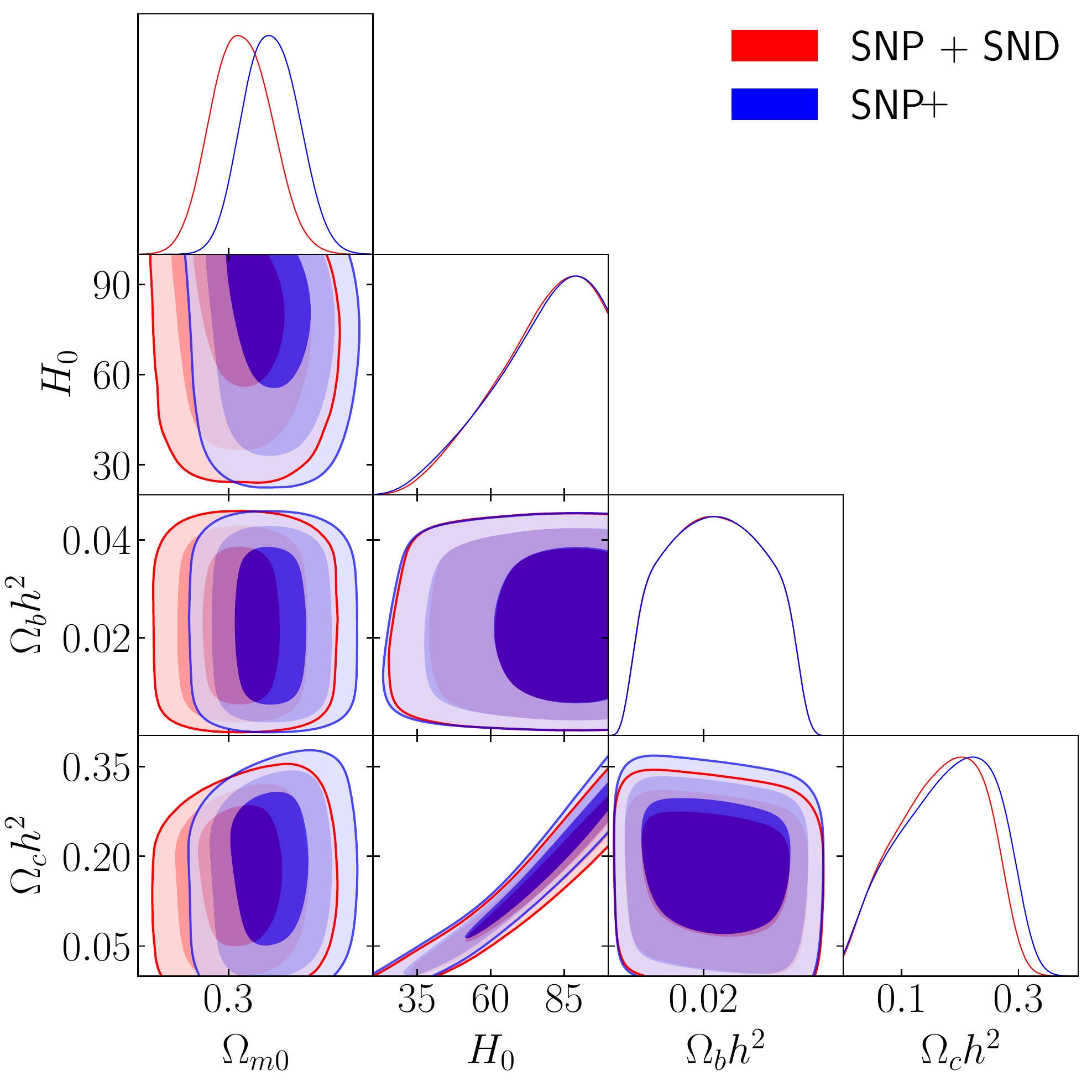}}
 \subfloat[]{%
    \includegraphics[width=0.45\textwidth,height=0.35\textwidth]{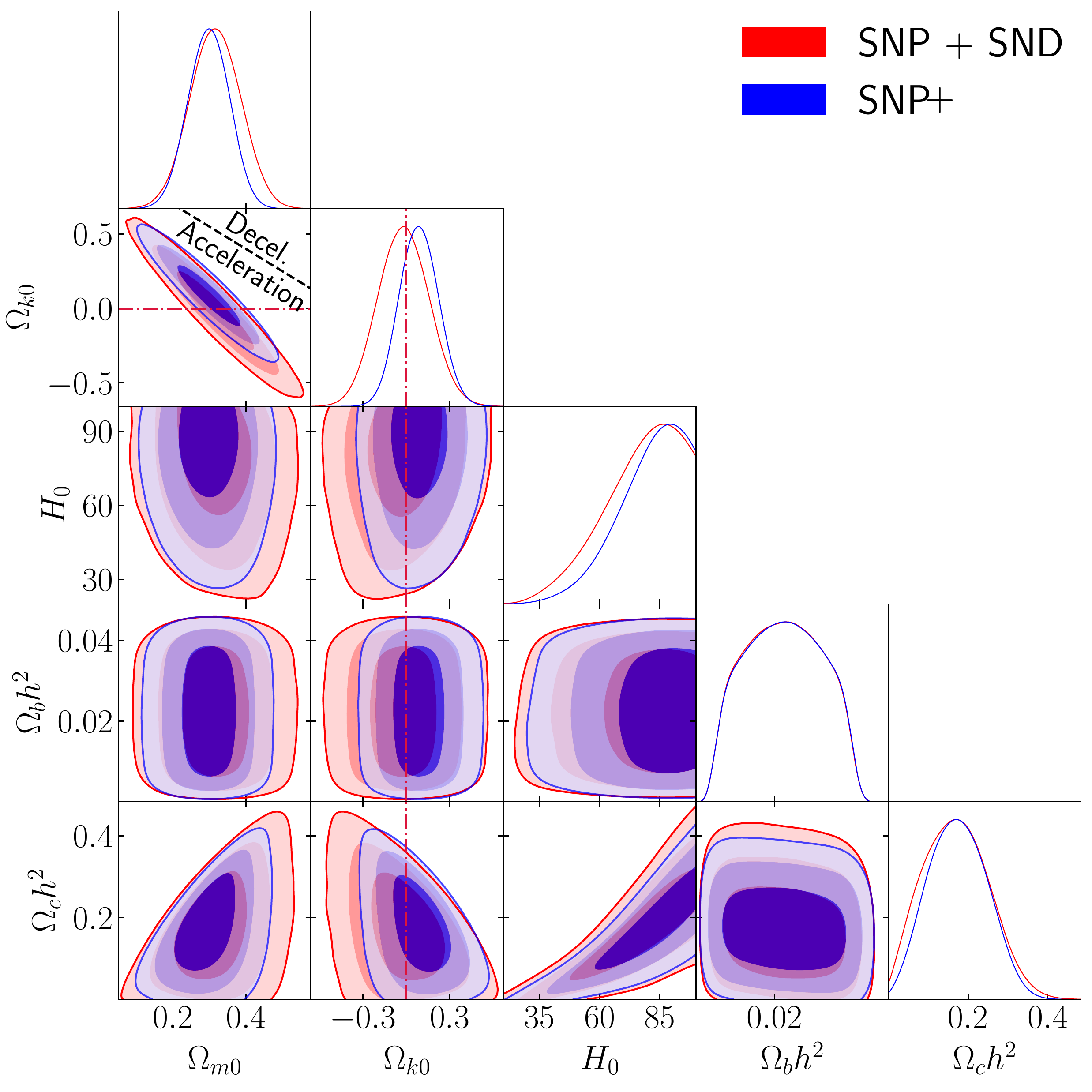}}\\
 \subfloat[]{%
    \includegraphics[width=0.45\textwidth,height=0.35\textwidth]{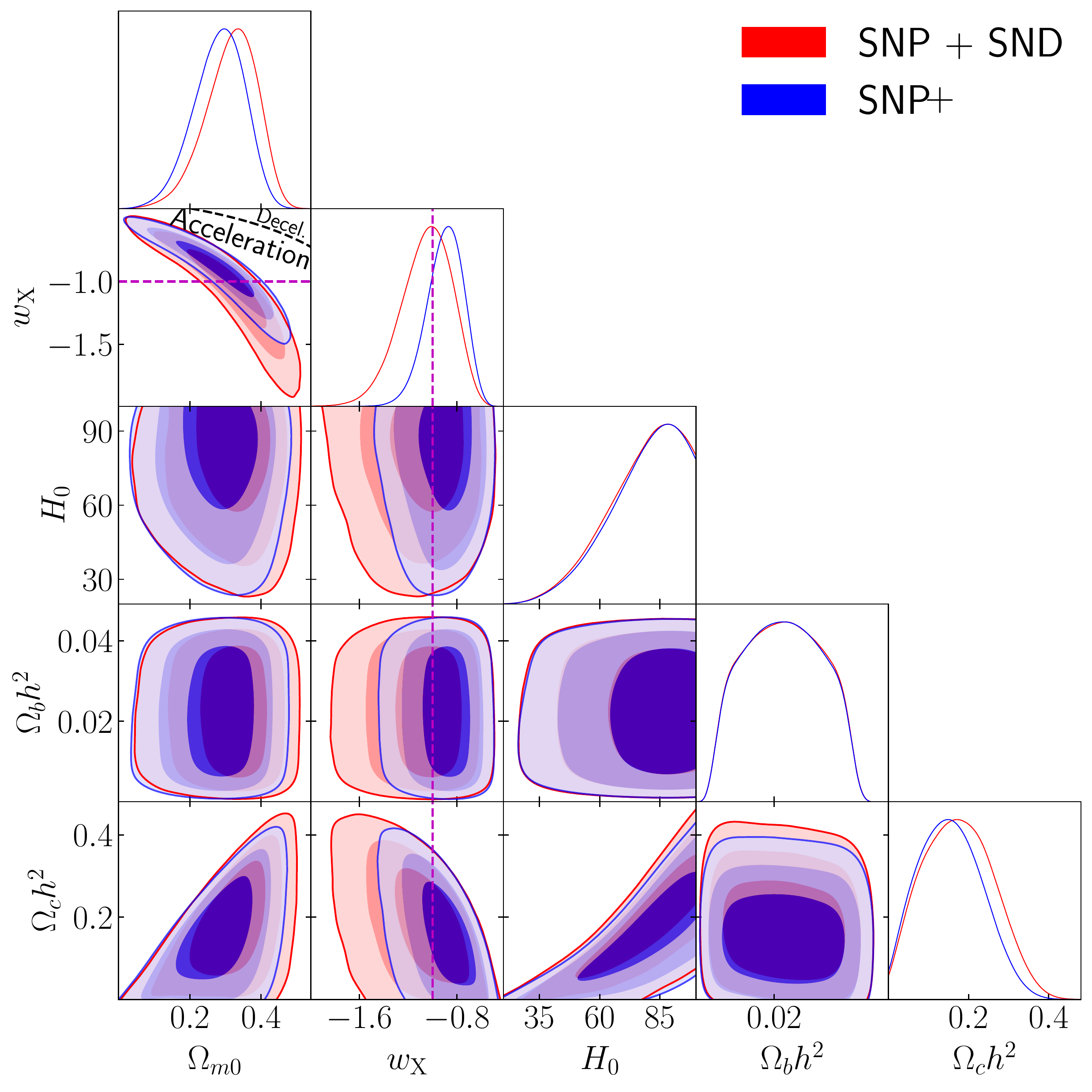}}
 \subfloat[]{%
    \includegraphics[width=0.45\textwidth,height=0.35\textwidth]{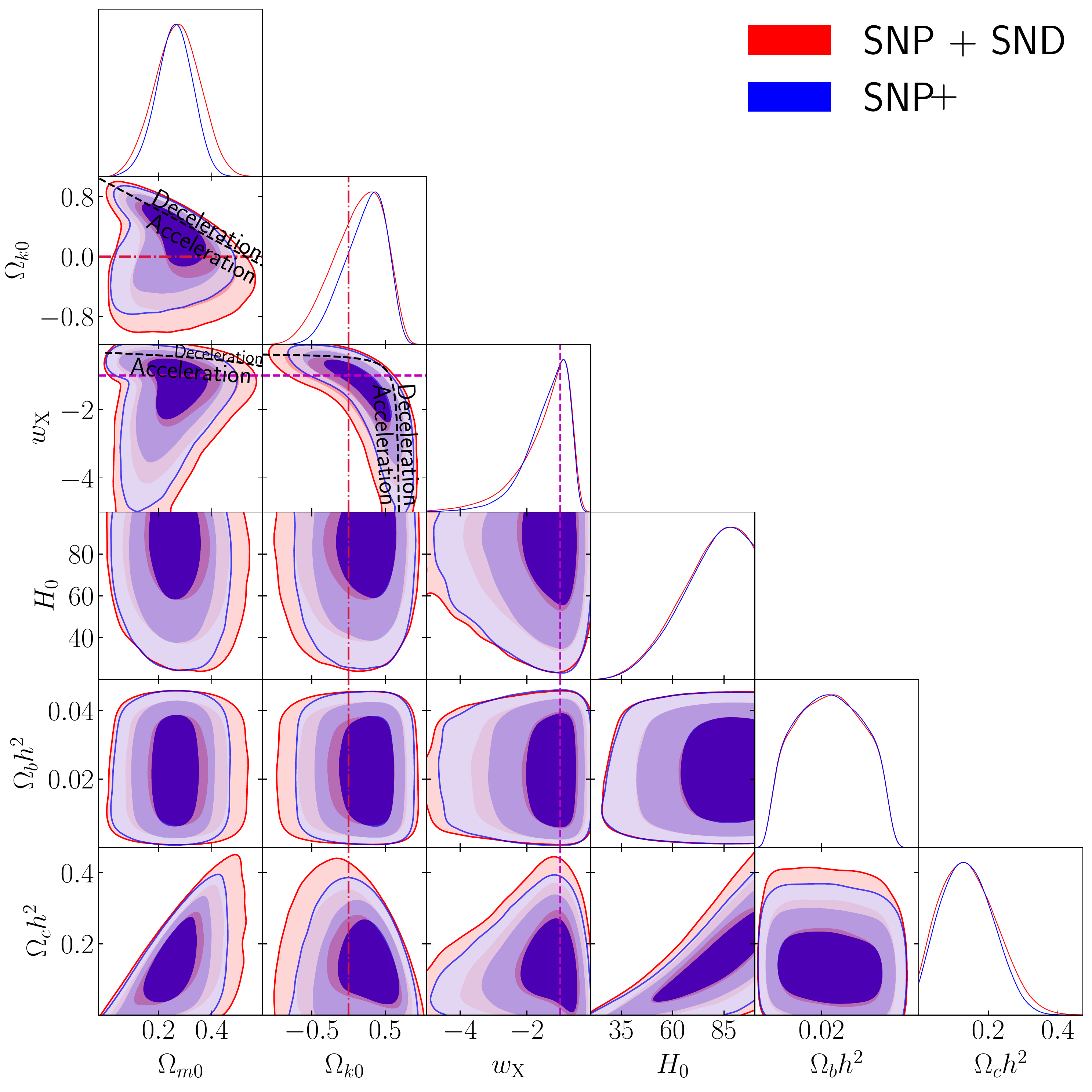}}\\
 \subfloat[]{%
    \includegraphics[width=0.45\textwidth,height=0.35\textwidth]{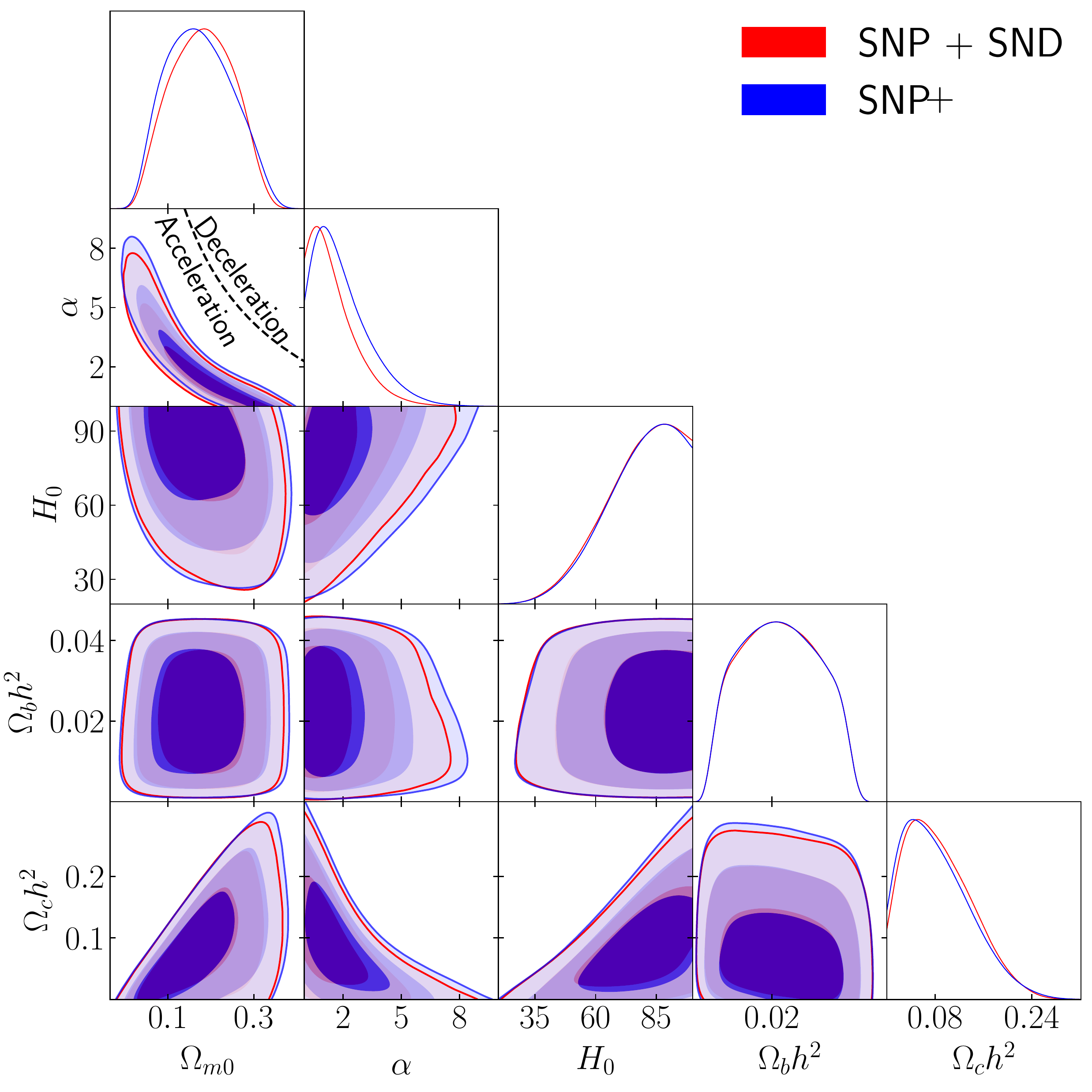}}
 \subfloat[]{%
    \includegraphics[width=0.45\textwidth,height=0.35\textwidth]{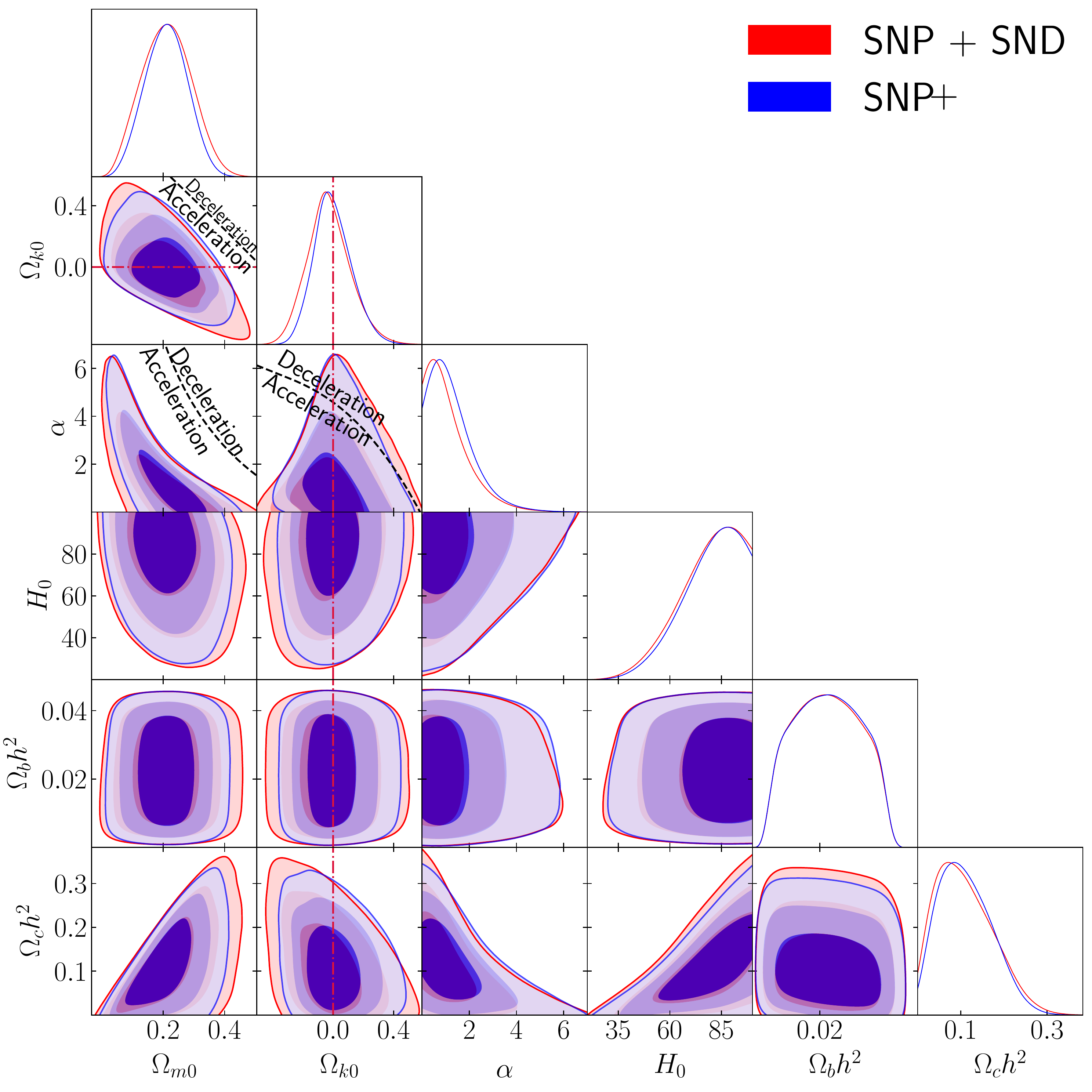}}\\
\caption{Same as Fig.\ \ref{fig1} but for SNP + SND (red) and SNP\plus\ (blue) data. The black dashed zero-acceleration lines in panels (b)--(f), computed for the third cosmological parameter set to the $H(z)$ + BAO data best-fitting values listed in Table \ref{tab:BFP} in panels (d) and (f), divides the parameter space into regions associated with currently-accelerating (below or below left) and currently-decelerating (above or above right) cosmological expansion.}
\label{fig3}
\end{figure*}

For Old $H(z)$, $H(z)$, Old $H(z)$ + Old BAO, $H(z)$ + BAO, SNP + SND, and SNP\plus\ data, the best-fitting parameter values, likelihood values, and information criteria values for all models are given in Table \ref{tab:BFP} and the marginalized posterior mean parameter values and uncertainties for all models are listed in Table \ref{tab:1d_BFP}. Figures \ref{fig1}--\ref{fig3} show the probability distributions and confidence regions of cosmological parameters, obtained from Old $H(z)$ and $H(z)$ data, from Old $H(z)$ + Old BAO and $H(z)$ + BAO data, and from SNP + SND and SNP\plus\ data, respectively.

Old $H(z)$ data constraints on \om\ range from $0.193^{+0.056}_{-0.117}$ (non-flat \pcdm) to $0.390^{+0.167}_{-0.172}$ (non-flat \lcdm), with a difference of $1.1\sigma$. In contrast, $H(z)$ data favor values of \om\ lower by $\lesssim0.17\sigma$ and ranging from $0.172^{+0.048}_{-0.104}$ (non-flat \pcdm) to $0.374^{+0.151}_{-0.210}$ (non-flat \lcdm), with a difference of $0.94\sigma$.

Old $H(z)$ data constraints on $H_0$ range from $62.81^{+2.65}_{-3.18}$ \hunit\ (non-flat \pcdm) to $79.55^{+7.01}_{-15.05}$ \hunit\ (flat XCDM), with a difference of $1.1\sigma$. In contrast, $H(z)$ data favor values of $H_0$ higher by $\lesssim0.36\sigma$ (and with larger error bars), ranging from $64.32^{+3.59}_{-4.04}$ \hunit\ (non-flat \pcdm) to $80.96^{+7.59}_{-16.10}$ \hunit\ (flat XCDM), with a difference of $1.0\sigma$.

Old $H(z)$ data constraints on \ok\ are $-0.174^{+0.501}_{-0.491}$, $0.241^{+0.451}_{-0.261}$, and $0.262^{+0.265}_{-0.337}$ for non-flat \lcdm, XCDM, and \pcdm, respectively. In contrast, $H(z)$ data constraints on \ok\ are $-0.136^{+0.564}_{-0.457}$, $0.228^{+0.456}_{-0.267}$, and $0.277^{+0.249}_{-0.331}$ for non-flat \lcdm, XCDM, and \pcdm, which are $0.056\sigma$ higher, $0.025\sigma$ lower, and $0.035\sigma$ higher than those from Old $H(z)$ data, respectively. Both Old $H(z)$ and $H(z)$ data indicate that closed spatial geometry is mildly favored by non-flat \lcdm, and that open spatial geometry is mildly favored by non-flat XCDM and non-flat \pcdm, but flat hypersurfaces are still within 1$\sigma$.

Both Old $H(z)$ data and $H(z)$ data indicate a slight preference for dark energy dynamics, with approximately similar evidence for deviation from the \lcdm\ models. For $H(z)$ data, for the flat and non-flat XCDM models, the $w_{\rm X}$ parameter is found to be $0.84\sigma$ and $0.68\sigma$ lower than $-1$ respectively. Similarly, for both the flat and non-flat \pcdm\ models, the $\alpha$ parameter is found to be $1.1\sigma$ away from $0$.

Old $H(z)$ + Old BAO data constraints on \om\ range from $0.275\pm0.023$ (flat \pcdm) to $0.301^{+0.016}_{-0.018}$ (flat \lcdm), with a difference of $0.89\sigma$. In contrast, $H(z)$ + BAO data favor values of \om\ lower by $\lesssim0.17\sigma$ and ranging from $0.272^{+0.024}_{-0.022}$ (flat \pcdm) to $0.297^{+0.015}_{-0.017}$ (flat \lcdm), with a difference of $0.85\sigma$.

Old $H(z)$ + Old BAO data constraints on $H_0$ range from $65.47^{+2.22}_{-2.21}$ \hunit\ (flat \pcdm) to $69.14\pm1.85$ \hunit\ (flat \lcdm), with a difference of $1.3\sigma$. In contrast, $H(z)$ + BAO data favor values of $H_0$ higher by $\lesssim0.41\sigma$ (and with larger error bars), ranging from $66.19^{+2.89}_{-2.88}$ \hunit\ (flat \pcdm) to $70.49\pm2.74$ \hunit\ (flat \lcdm), with a difference of $1.1\sigma$.

Old $H(z)$ + Old BAO data constraints on \ok\ are $0.059^{+0.081}_{-0.091}$, $-0.027\pm0.109$, and $-0.034^{+0.087}_{-0.098}$ for non-flat \lcdm, XCDM, and \pcdm, respectively. In contrast, $H(z)$ + BAO data constraints on \ok\ are $0.047^{+0.082}_{-0.089}$, $-0.054\pm0.103$, and $-0.052^{+0.093}_{-0.087}$ for non-flat \lcdm, XCDM, and \pcdm, which are $0.10\sigma$, $0.18\sigma$, and $0.13\sigma$ lower than those from Old $H(z)$ + Old BAO data, respectively. In contrast to Old $H(z)$ and $H(z)$ data, both Old $H(z)$ + Old BAO and $H(z)$ + BAO data indicate that open spatial geometry is mildly favored by non-flat \lcdm, and closed spatial geometry is mildly favored by non-flat XCDM and non-flat \pcdm, but flat hypersurfaces are still within 1$\sigma$.

Both Old $H(z)$ + Old BAO data and $H(z)$ + BAO data show strong evidence for dark energy dynamics. In the Old $H(z)$ + Old BAO data case, for flat (non-flat) XCDM ($1\sigma$ and $2\sigma$), $w_{\rm X}=-0.784^{+0.140}_{-0.107}{}^{+0.230}_{-0.243}$ ($w_{\rm X}=-0.770^{+0.149}_{-0.098}{}^{+0.233}_{-0.256}$), with central values being $<2\sigma$ higher than $w_{\rm X}=-1$ (\lcdm); and for flat (non-flat) \pcdm\ ($1\sigma$ and $2\sigma$), $\alpha=1.267^{+0.536}_{-0.807}{}^{+1.240}_{-1.221}$ ($\alpha=1.360^{+0.584}_{-0.819}{}^{+1.289}_{-1.300}$), with central values being $>2\sigma$ away from $\alpha=0$ (\lcdm). In the $H(z)$ + BAO data case, for flat (non-flat) XCDM ($1\sigma$ and $2\sigma$), $w_{\rm X}=-0.776^{+0.130}_{-0.103}{}^{+0.221}_{-0.232}$ ($w_{\rm X}=-0.757^{+0.135}_{-0.093}{}^{+0.215}_{-0.236}$), with central values being $<2\sigma$ ($>2\sigma$) higher than $w_{\rm X}=-1$ (\lcdm); and for flat (non-flat) \pcdm\ ($1\sigma$ and $2\sigma$), $\alpha=1.271^{+0.507}_{-0.836}{}^{+1.294}_{-1.228}$ ($\alpha=1.427^{+0.572}_{-0.830}{}^{+1.365}_{-1.317}$), with central values being $>2\sigma$ away from $\alpha=0$ (\lcdm).

Since SN Ia data alone cannot constrain $H_0$ we choose a narrower prior of $H_0\in[20,100]$ \hunit\ for these data. The constraints on $H_0$ derived from those on \obhs\ and \ochs\ for both SN Ia data sets provide $2\sigma$ lower limits that are in agreement with $H_0$ constraints derived from other data.

SNP + SND data constraints on \om\ range from $0.181^{+0.075}_{-0.076}$ (flat \pcdm) to $0.317\pm0.068$ (non-flat \lcdm), with a difference of $1.3\sigma$. In contrast, SNP\plus\ data constraints on \om\ differ from the SNP + SND ones by $-0.35\sigma$ to $0.76\sigma$, ranging from $0.175^{+0.065}_{-0.092}$ (flat \pcdm) to $0.332\pm0.020$ (flat \lcdm), with a difference of $2.3\sigma$.

SNP + SND data constraints on \ok\ are $-0.017^{+0.172}_{-0.174}$, $0.130^{+0.426}_{-0.249}$, and $-0.026^{+0.126}_{-0.150}$ for non-flat \lcdm, XCDM, and \pcdm, respectively. In contrast, SNP\plus\ data constraints on \ok\ are $0.089\pm0.132$, $0.215^{+0.350}_{-0.202}$, and $-0.001^{+0.108}_{-0.132}$ for non-flat \lcdm, XCDM, and \pcdm, which are $0.49\sigma$, $0.18\sigma$, and $0.14\sigma$ higher than those from SNP + SND data, respectively. SNP + SND data show that open spatial geometry is mildly favored by non-flat XCDM, and closed spatial geometry is mildly favored by non-flat \lcdm\ and \pcdm; while SNP\plus\ data show that open geometry is mildly favored by non-flat \lcdm, XCDM, and \pcdm. Both sets of data indicate that flat hypersurfaces remain within the 1$\sigma$ confidence interval, with the exception of the non-flat XCDM scenario, where SNP\plus\ data deviates by $1.1\sigma$ from flatness.

Both SNP + SND and SNP\plus\ data show a slight preference for dark energy dynamics, but deviations from the \lcdm\ models are within $1\sigma$. In the SNP + SND case, for flat (non-flat) XCDM, $w_{\rm X}=-1.054^{+0.237}_{-0.171}$ ($w_{\rm X}=-1.499^{+0.901}_{-0.237}$), with central values being $0.23\sigma$ ($0.55\sigma$) lower than $w_{\rm X}=-1$ (\lcdm); and for flat (non-flat) \pcdm, $2\sigma$ upper limits of $\alpha<4.052$ and $\alpha<3.067$ suggest that $\alpha=0$ (\lcdm) is within $1\sigma$. In the SNP\plus\ case, for flat (non-flat) XCDM, $w_{\rm X}=-0.900^{+0.166}_{-0.124}$ ($w_{\rm X}=-1.424^{+0.798}_{-0.251}$), with central values being $0.81\sigma$ ($0.53\sigma$) higher (lower) than $w_{\rm X}=-1$ (\lcdm); and for flat (non-flat) \pcdm, $\alpha=1.966^{+0.479}_{-1.907}$ ($\alpha=1.282^{+0.290}_{-1.255}$), with both central values being $1.0\sigma$ away from $\alpha=0$ (\lcdm).

\subsection{Constraints from Old $H(z)$ + Old BAO + SNP + SND data and $H(z)$ + BAO + SNP\plus\ data}
 \label{subsec:comp2}

\begin{figure*}
\centering
 \subfloat[]{%
    \includegraphics[width=0.45\textwidth,height=0.35\textwidth]{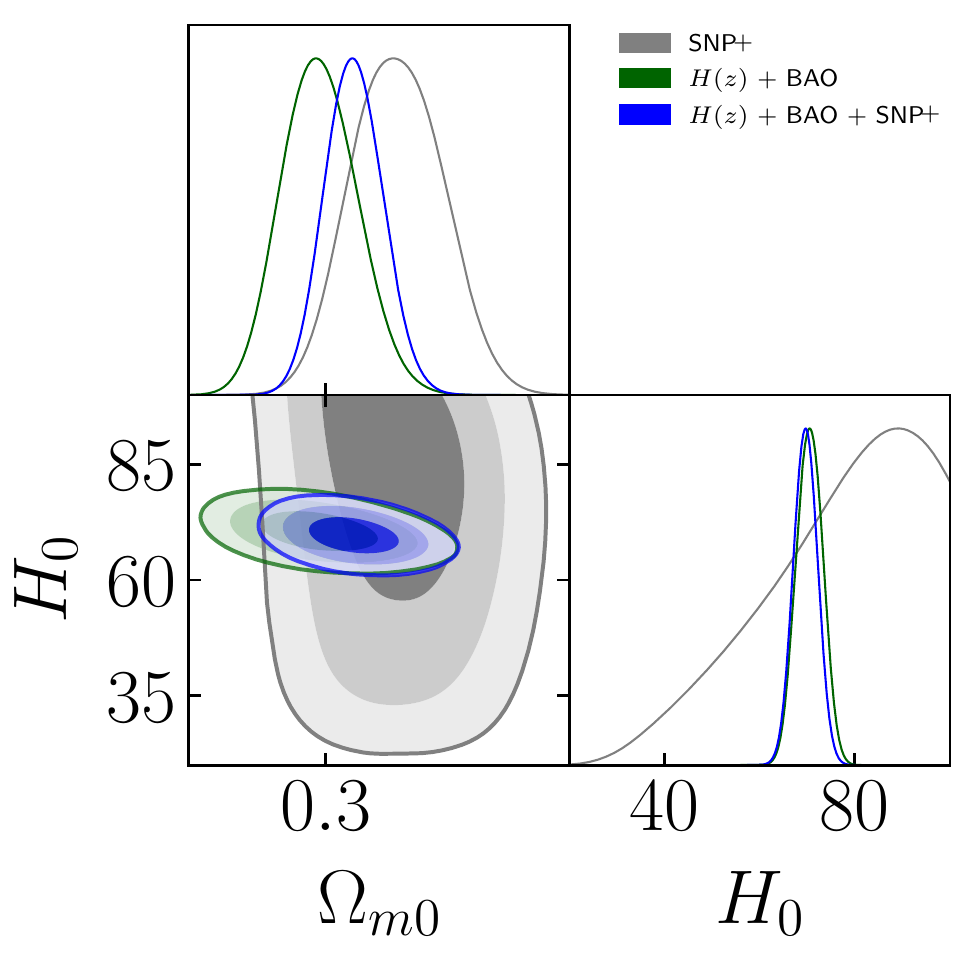}}
 \subfloat[]{%
    \includegraphics[width=0.45\textwidth,height=0.35\textwidth]{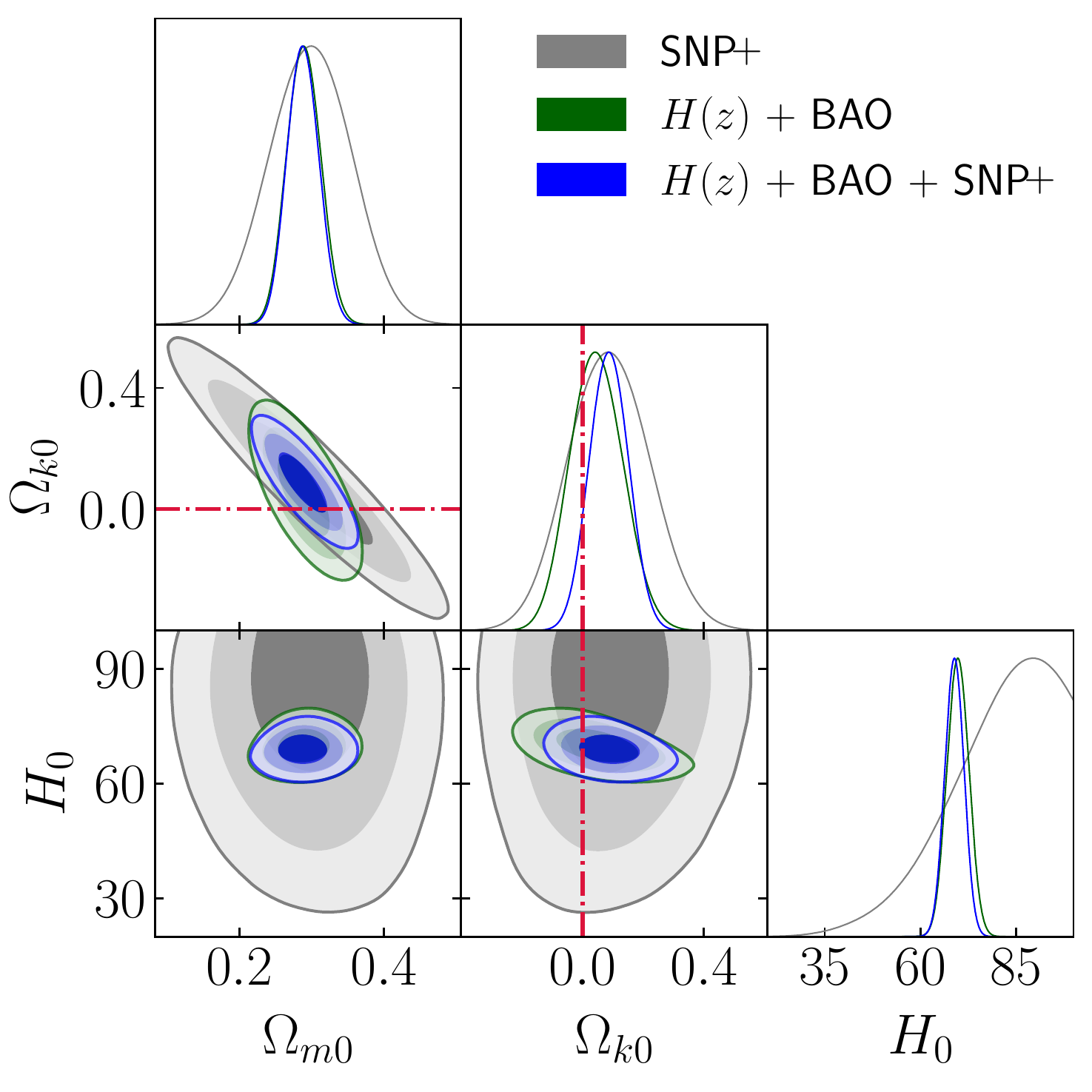}}\\
 \subfloat[]{%
    \includegraphics[width=0.45\textwidth,height=0.35\textwidth]{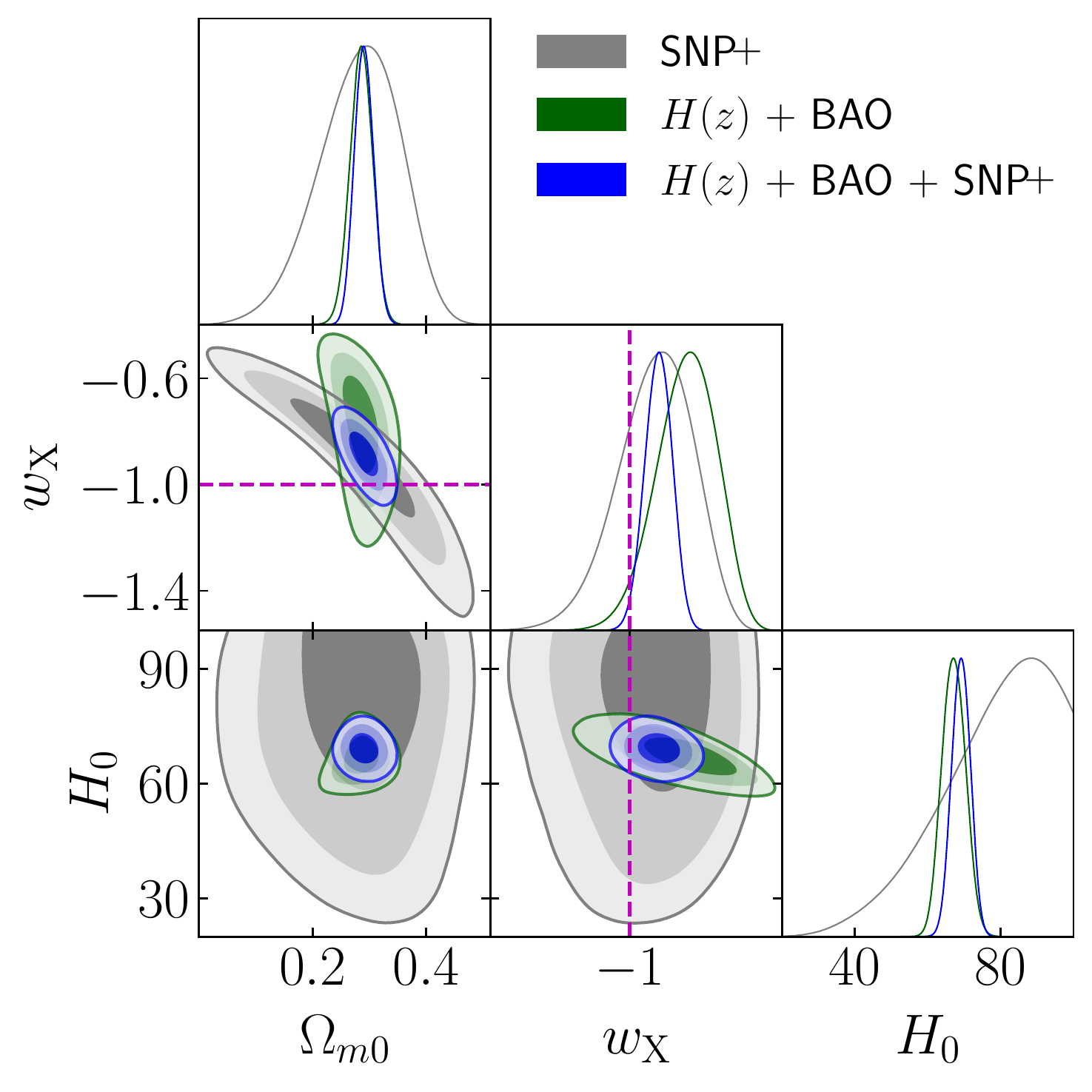}}
 \subfloat[]{%
    \includegraphics[width=0.45\textwidth,height=0.35\textwidth]{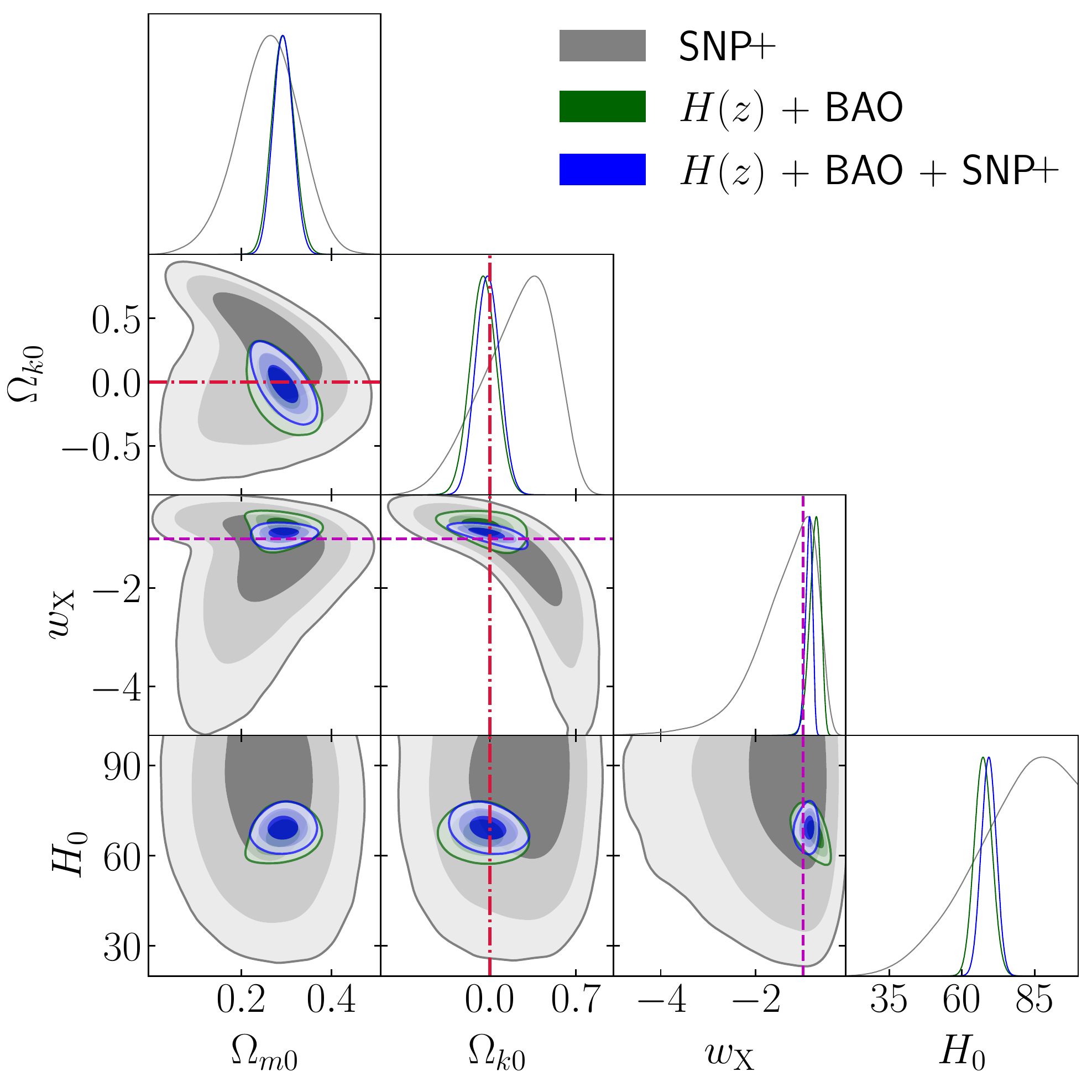}}\\
 \subfloat[]{%
    \includegraphics[width=0.45\textwidth,height=0.35\textwidth]{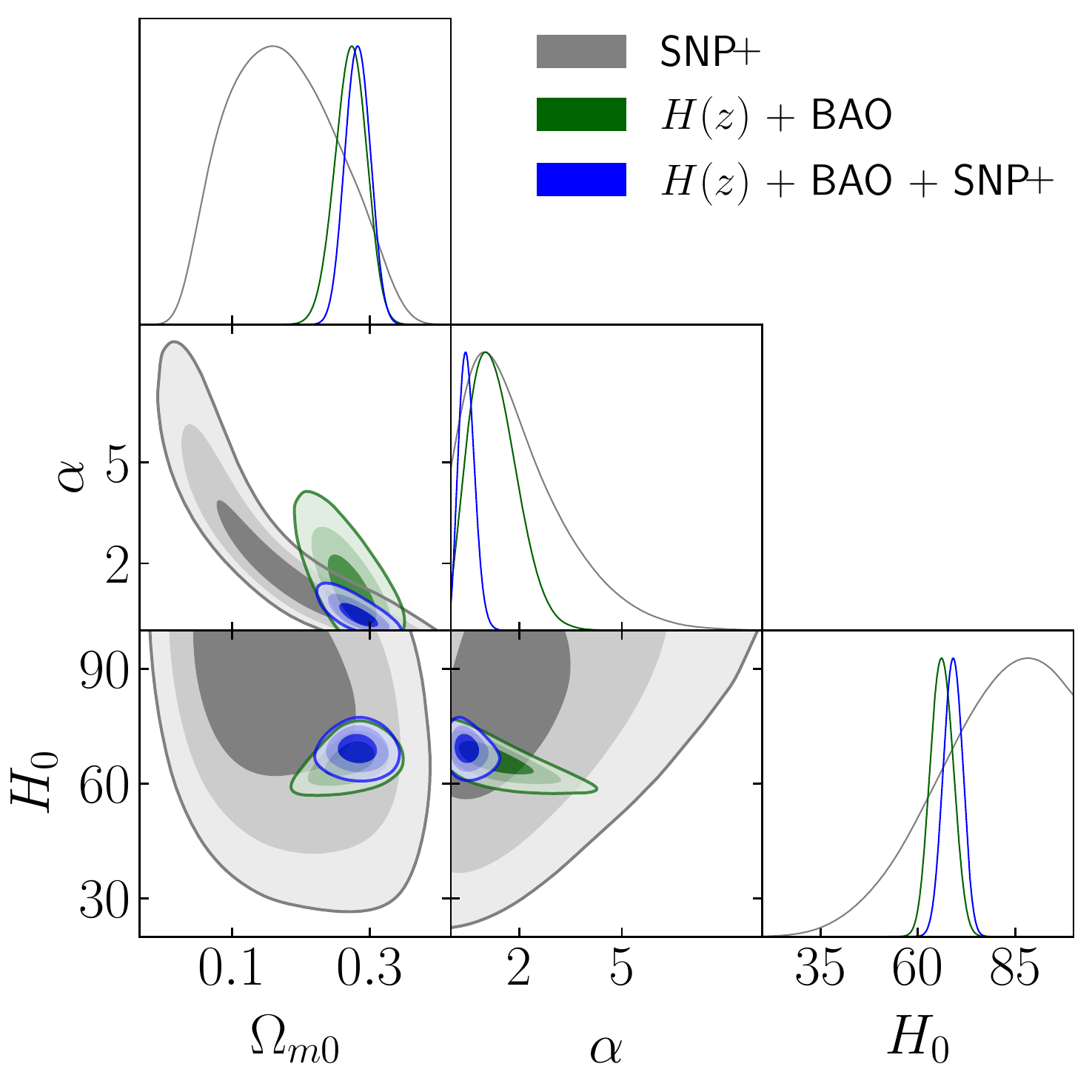}}
 \subfloat[]{%
    \includegraphics[width=0.45\textwidth,height=0.35\textwidth]{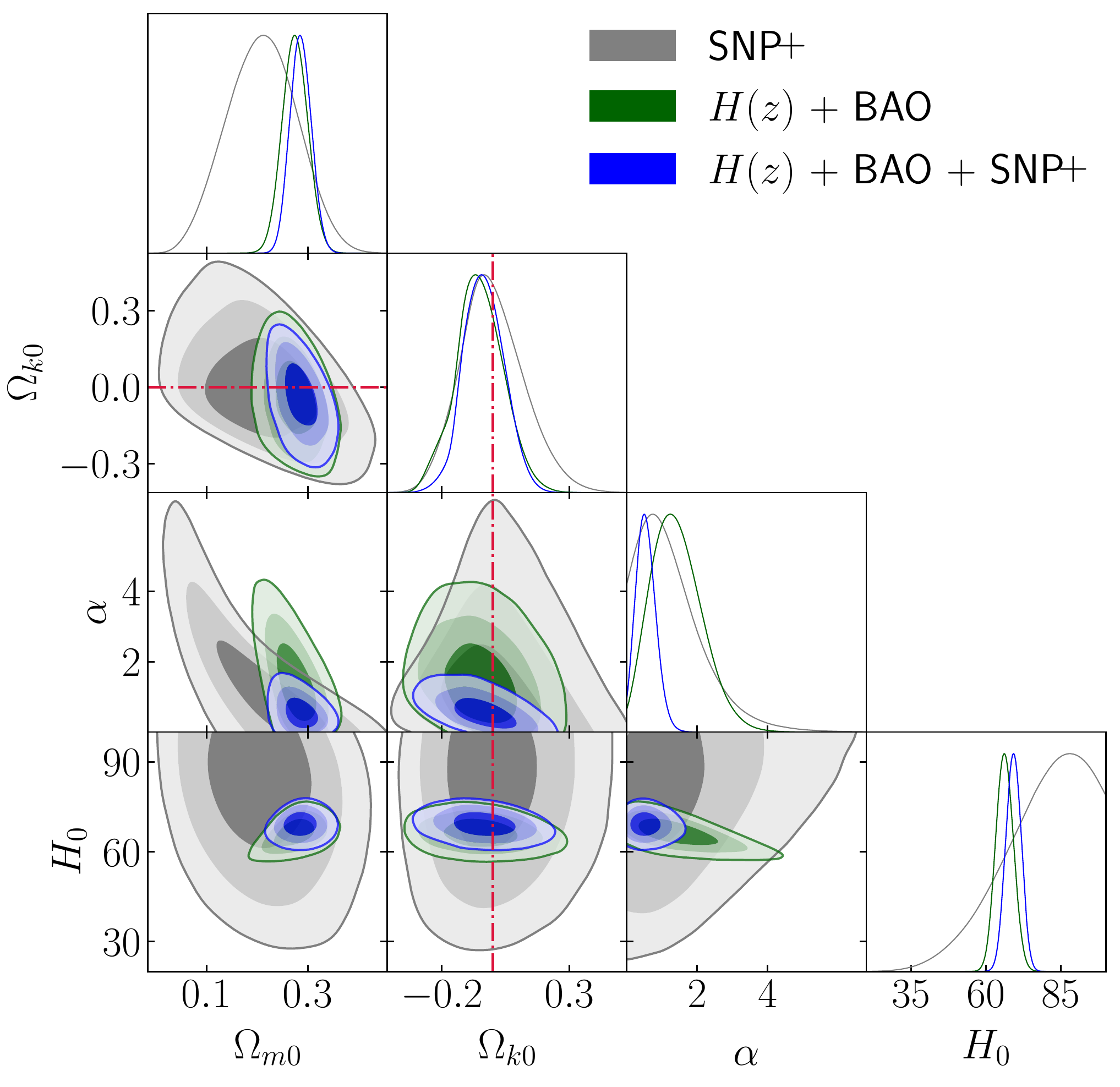}}\\
\caption{Same as Fig.\ \ref{fig1} but for SNP\plus\ (gray), $H(z)$ + BAO (green), and $H(z)$ + BAO + SNP\plus\ (blue) data.}
\label{fig4}
\end{figure*}

\begin{figure*}
\centering
 \subfloat[]{%
    \includegraphics[width=0.45\textwidth,height=0.35\textwidth]{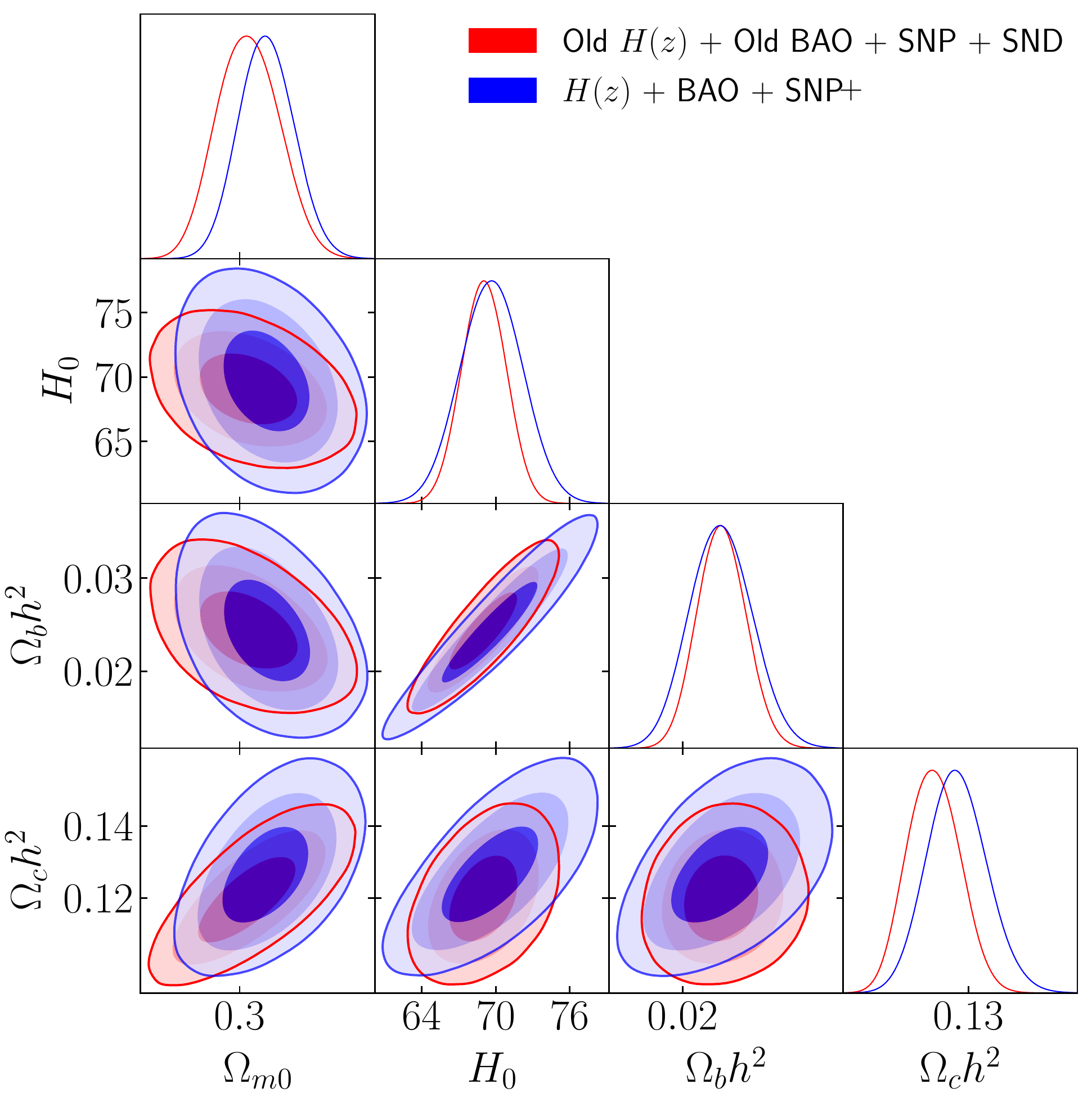}}
 \subfloat[]{%
    \includegraphics[width=0.45\textwidth,height=0.35\textwidth]{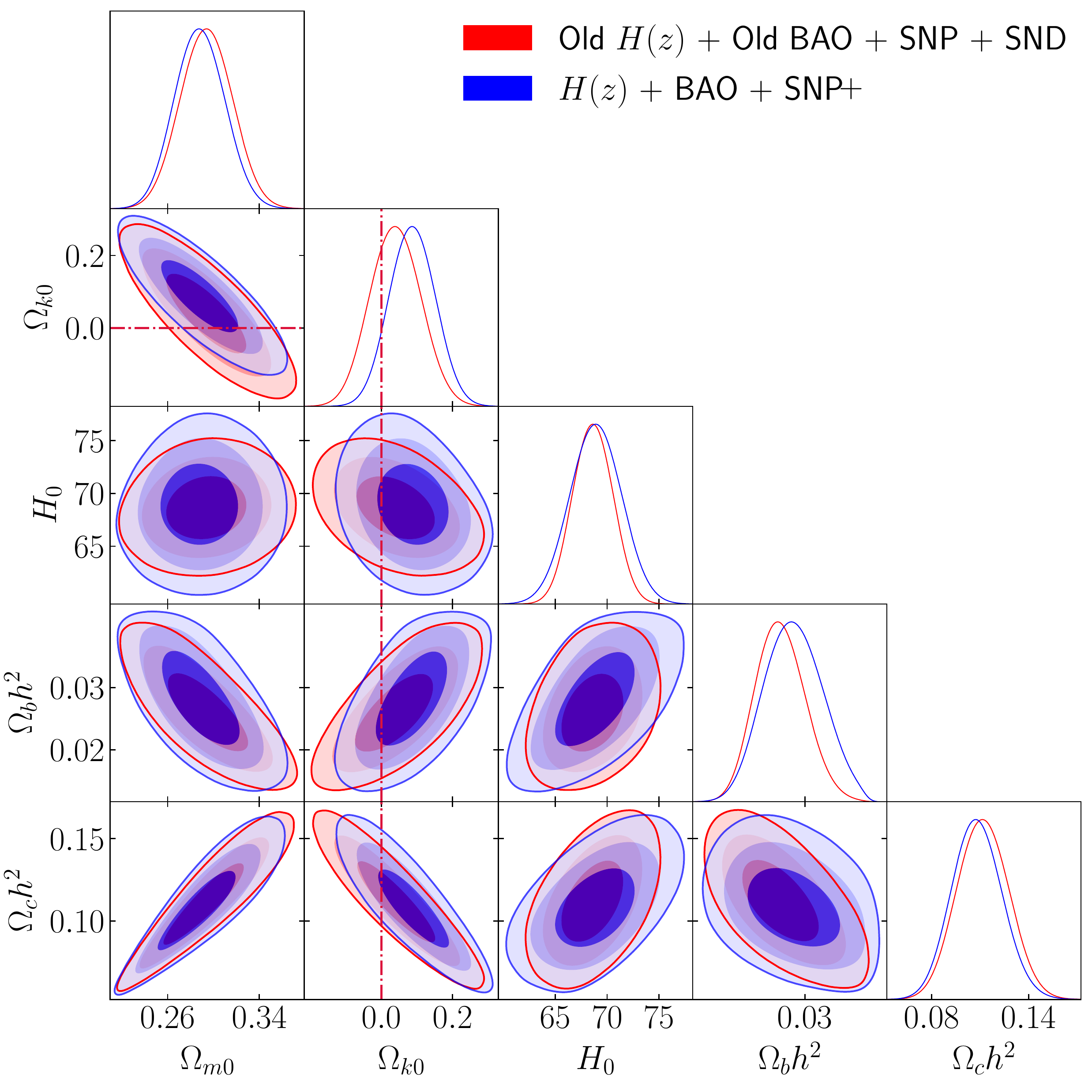}}\\
 \subfloat[]{%
    \includegraphics[width=0.45\textwidth,height=0.35\textwidth]{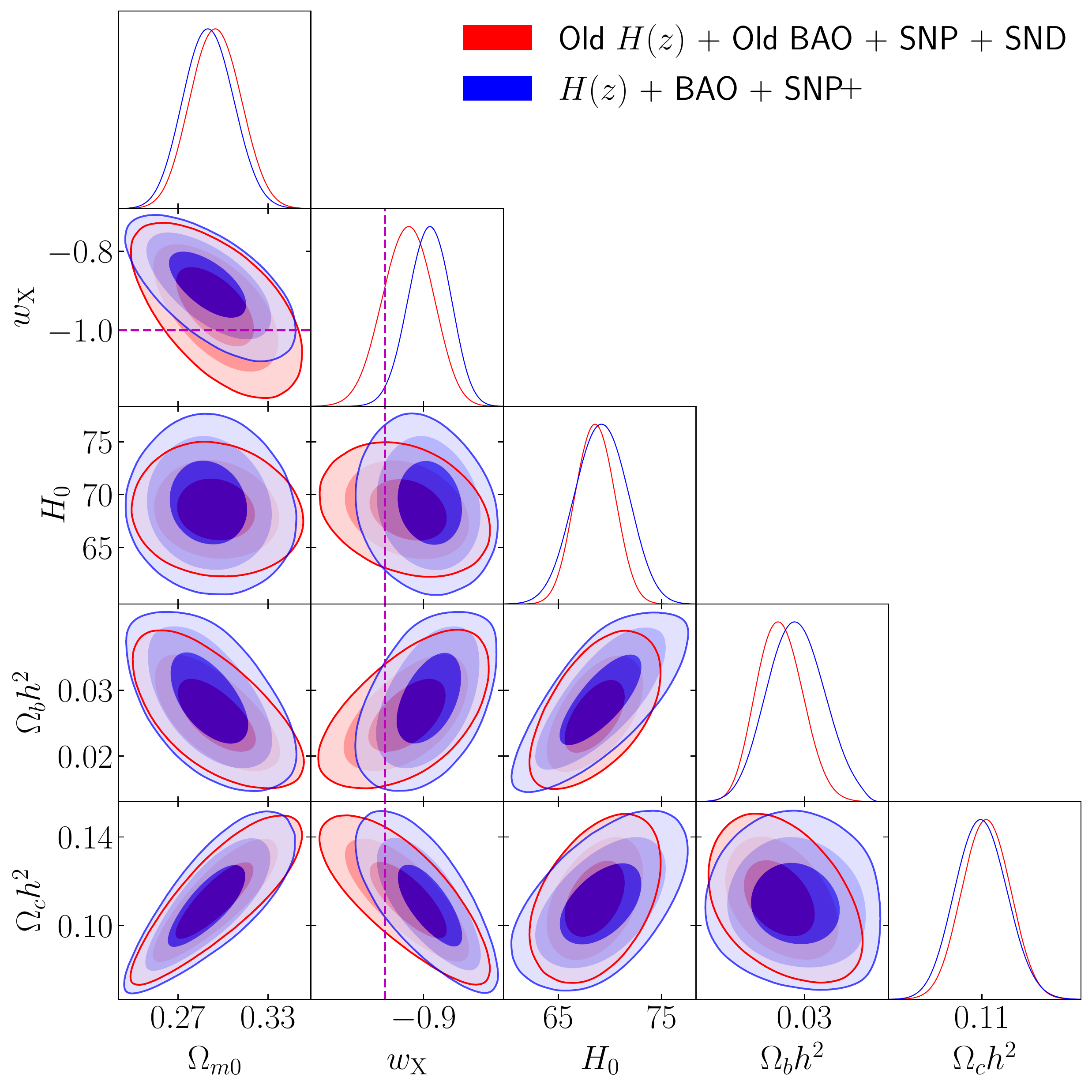}}
 \subfloat[]{%
    \includegraphics[width=0.45\textwidth,height=0.35\textwidth]{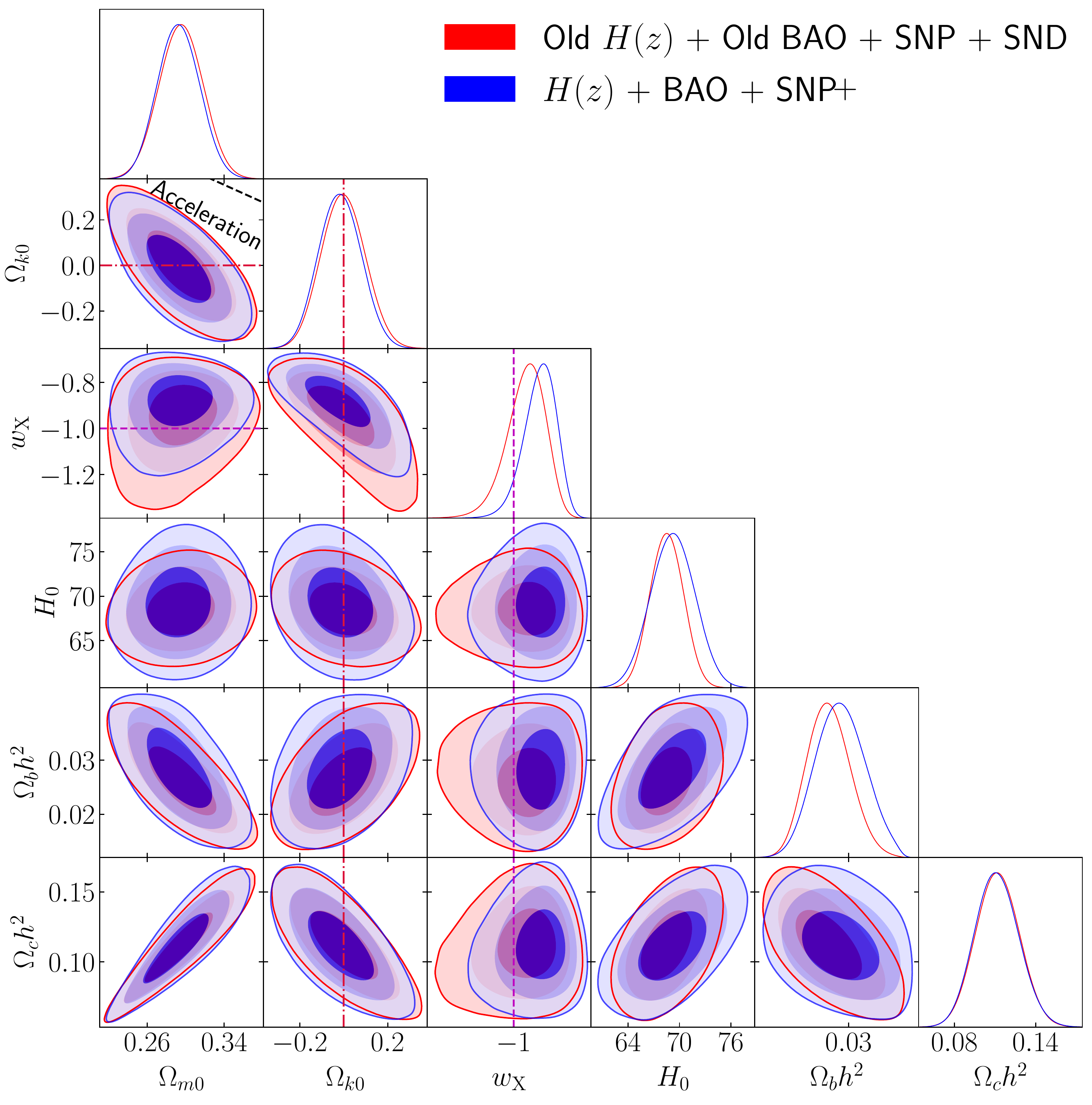}}\\
 \subfloat[]{%
    \includegraphics[width=0.45\textwidth,height=0.35\textwidth]{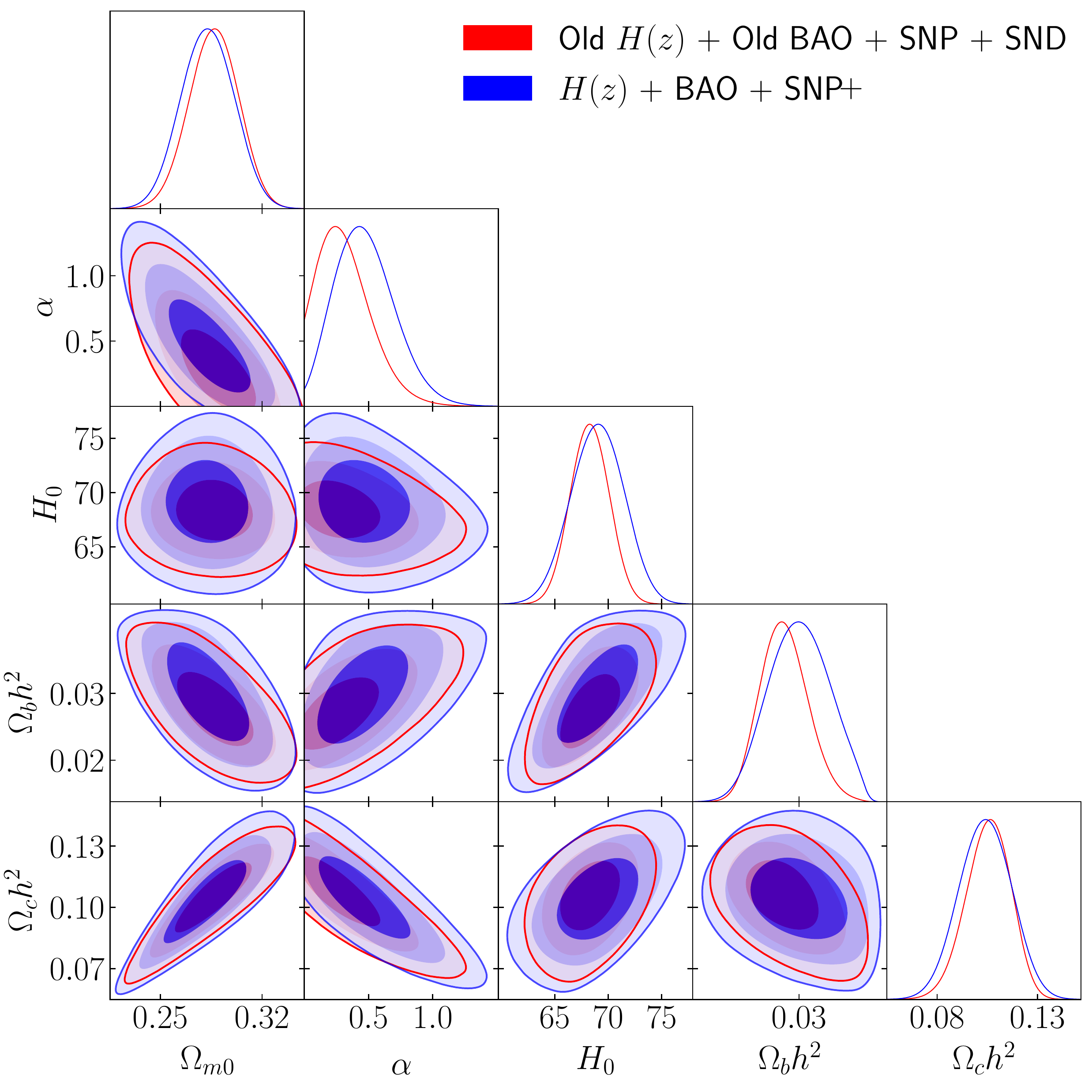}}
 \subfloat[]{%
    \includegraphics[width=0.5\textwidth,height=0.35\textwidth]{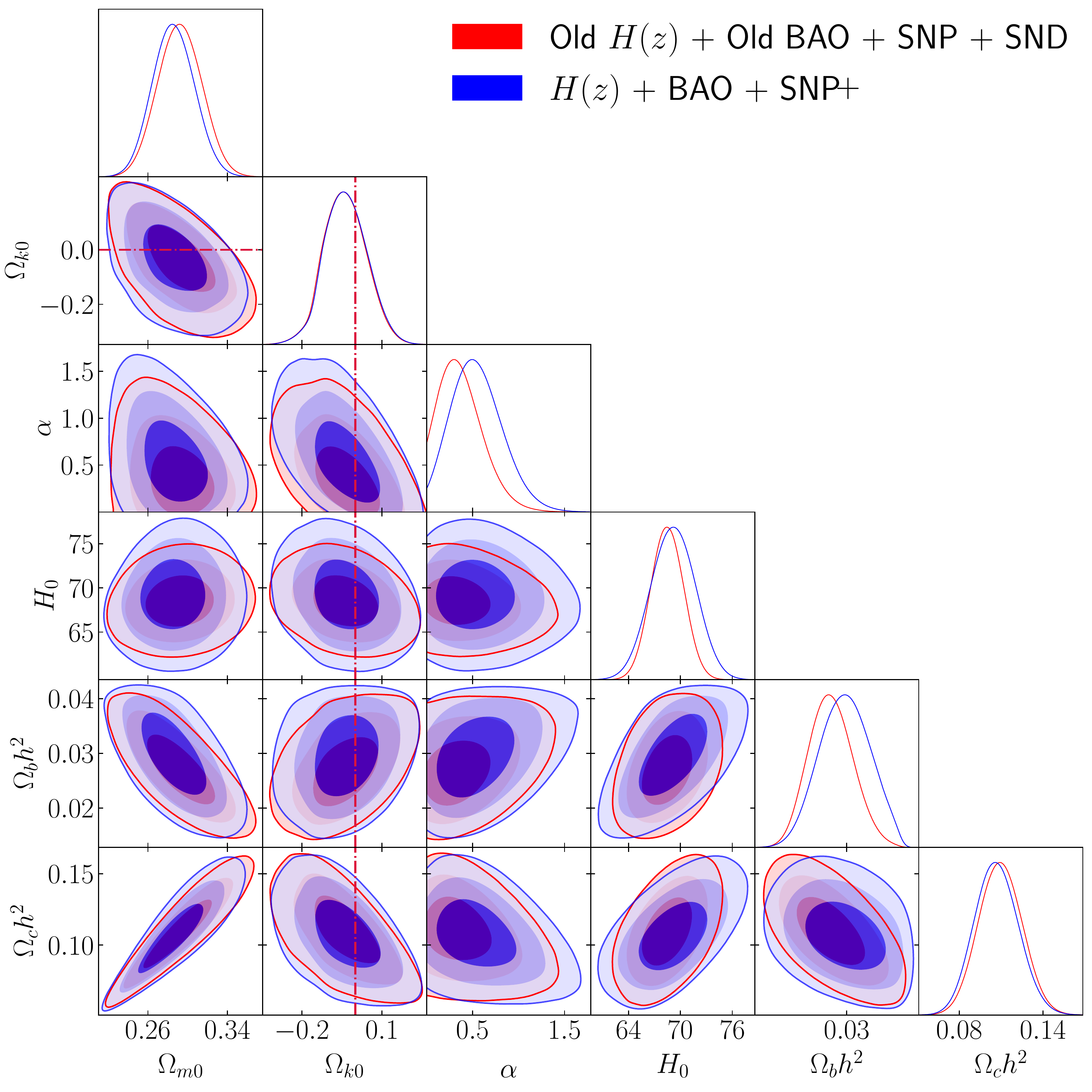}}\\
\caption{Same as Fig.\ \ref{fig1} but for Old $H(z)$ + Old BAO + SNP + SND (red) and $H(z)$ + BAO + SNP\plus\ (blue) data.}
\label{fig5}
\end{figure*}

For Old $H(z)$ + Old BAO + SNP + SND and $H(z)$ + BAO + SNP\plus\ data, the best-fitting parameter values, likelihood values, and information criteria values for all models are given in Table \ref{tab:BFP} and the marginalized posterior mean parameter values and uncertainties for all models are listed in Table \ref{tab:1d_BFP}. Figures \ref{fig4} and \ref{fig5} show the probability distributions and confidence regions of cosmological parameters, obtained from SNP\plus, $H(z)$ + BAO, $H(z)$ + BAO + SNP\plus, and Old $H(z)$ + Old BAO + SNP + SND data. Constraints from SNP\plus, $H(z)$, and BAO data are mutually consistent so these data can be used together to constrain cosmological parameters, as discussed below. Note that the Old $H(z)$ + Old BAO + SNP + SND data results are from Ref.\ \cite{CaoRatra2022} and are used here to compare to $H(z)$ + BAO + SNP\plus\ constraints.

$H(z)$ + BAO + SNP\plus\ (HzBSN) data constraints on \om\ range from $0.282\pm0.018$ (flat \pcdm) to $0.312\pm0.013$ (flat \lcdm), with a difference of $1.4\sigma$. In contrast, Old $H(z)$ + Old BAO + SNP + SND data favor values of \om\ higher by $\lesssim0.22\sigma$ (or lower by $0.42\sigma$ for flat \lcdm) and ranging from $0.287\pm0.017$ (flat \pcdm) to $0.304^{+0.014}_{-0.015}$ (flat \lcdm), with a difference of $0.75\sigma$.

HzBSN data constraints on $H_0$ range from $68.89\pm2.44$ \hunit\ (non-flat \lcdm) to $69.65\pm2.48$ \hunit\ (flat \lcdm), with a difference of $0.22\sigma$. These $H_0$ values are $0.24\sigma$ (non-flat \lcdm) and $0.44\sigma$ (flat \lcdm) higher than the median statistics estimate of $H_0=68\pm2.8$ \hunit\ \citep{chenratmed}, $0.31\sigma$ (non-flat \lcdm) and $0.05\sigma$ (flat \lcdm) lower than the TRGB and SN Ia estimate of $H_0=69.8\pm1.7$ \hunit\ \citep{Freedman2021}, and $1.6\sigma$ (non-flat \lcdm) and $1.3\sigma$ (flat \lcdm) lower than the Cepheids and SN Ia measurement of $73.04\pm1.04$ \hunit\ \cite{Riessetal2022}. The $H_0$ constraints from flat \lcdm\ is $0.90\sigma$ higher than the $H_0$ estimate of $67.36 \pm 0.54$ \hunit\ from \textit{Planck} 2018 TT,TE,EE+lowE+lensing CMB anisotropy data \cite{planck2018b}. In contrast, Old $H(z)$ + Old BAO + SNP + SND data favor values of $H_0$ lower by $\lesssim0.41\sigma$ (and with smaller error bars), ranging from $68.29\pm1.78$ \hunit\ (flat \pcdm) to $69.04\pm1.77$ \hunit\ (flat \lcdm), with a difference of $0.30\sigma$.

HzBSN data constraints on \ok\ are $0.087\pm0.063$, $-0.017\pm0.095$, and $-0.035^{+0.071}_{-0.085}$ for non-flat \lcdm, XCDM, and \pcdm, respectively. In contrast, Old $H(z)$ + Old BAO + SNP + SND data constraints on \ok\ are $0.040\pm0.070$, $-0.001\pm0.098$, and $-0.038^{+0.071}_{-0.085}$ for non-flat \lcdm, XCDM, and \pcdm, which are $0.50\sigma$ lower, $0.12\sigma$ higher, and $0.027\sigma$ lower than those from HzBSN data, respectively. For both data sets, non-flat \lcdm\ favors open spatial geometry, with HzBSN data being $1.4\sigma$ away from flat and Old $H(z)$ + Old BAO + SNP + SND data being within $1\sigma$ of flat, while closed spatial geometry is favored by non-flat XCDM and non-flat \pcdm, with flatness well within 1$\sigma$.

HzBSN data indicate a strong preference for dark energy dynamics. In particular, the central values of the XCDM equation of state parameter, $w_{\rm X}$, are found to be $>2\sigma$ and slightly $<2\sigma$ higher than $-1$ for flat and non-flat parametrizations respectively. Similarly, the central values of the parameter $\alpha$ in both the flat and non-flat \pcdm\ models are found to be $>2\sigma$ away from 0. Note that these constraints are skewed and non-Gaussian. Old $H(z)$ + Old BAO + SNP + SND data show somewhat less preference for dark energy dynamics. Specifically, for flat (non-flat) XCDM, $w_{\rm X}=-0.941\pm0.064$ ($w_{\rm X}=-0.948^{+0.098}_{-0.068}$), with central values being $0.92\sigma$ ($0.76\sigma$) higher than $w_{\rm X}=-1$, and for flat (non-flat) \pcdm, $\alpha=0.324^{+0.122}_{-0.264}$ ($\alpha=0.382^{+0.151}_{-0.299}$), with central values being $1.2\sigma$ ($1.3\sigma$) away from $\alpha=0$.

Relative to the Old $H(z)$ + Old BAO + SNP + SND data constraints of Ref.\ \cite{CaoRatra2022}, the most significant changes in the constraints from HzBSN data are that they more strongly favor dark energy dynamics and provide larger $H_0$ error bars.

\subsection{Constraints from QSO-AS + \hiig\ + \mii\ + \civ\ + A118 data}
 \label{subsec:comp3}

\begin{table*}
\centering
\resizebox*{2.05\columnwidth}{2.6\columnwidth}{%
\begin{threeparttable}
\caption{Unmarginalized best-fitting parameter values for all models from various combinations of data.}\label{tab:BFPs}
\begin{tabular}{lcccccccccccccccccccccccc}
\toprule
Model & Data set & $\Omega_{b}h^2$ & $\Omega_{c}h^2$ & \om & \ok & $w_{\mathrm{X}}$/$\alpha$\tnote{a} & $H_0$\tnote{b} & $l_{\rm m}$ & $\gamma_{\rm \textsc{m}}$ & $\beta_{\rm \textsc{m}}$ & $\sigma_{\rm int,\,\textsc{m}}$ & $\gamma_{\rm \textsc{c}}$ & $\beta_{\rm \textsc{c}}$ & $\sigma_{\rm int,\,\textsc{c}}$ & $\gamma_{\rm \textsc{a}}$ & $\beta_{\rm \textsc{a}}$ & $\sigma_{\rm int,\,\textsc{a}}$ & $-2\ln\mathcal{L}_{\mathrm{max}}$ & AIC & BIC & DIC & $\Delta \mathrm{AIC}$ & $\Delta \mathrm{BIC}$ & $\Delta \mathrm{DIC}$ \\
\midrule
 & QSO-AS + \hiig & 0.0332 & 0.0947 & 0.251 & -- & -- & 71.53 & 11.06 & -- & -- & -- & -- & -- & -- & -- & -- & -- & 786.45 & 794.45 & 809.28 & 792.69 & 0.00 & 0.00 & 0.00\\
 & \mii\ + \civ & -- & 0.0082 & 0.068 & -- & -- & -- & -- & 0.286 & 1.647 & 0.280 & 0.412 & 0.995 & 0.274 & -- & -- & -- & 50.94 & 64.94 & 84.21 & 69.10 & 0.00 & 0.00 & 0.00\\
Flat & A118 & -- & 0.2768 & 0.616 & -- & -- & -- & -- & -- & -- & -- & -- & -- & -- & 1.166 & 49.92 & 0.382 & 118.63 & 126.63 & 137.71 & 125.95 & 0.00 & 0.00 & 0.00\\
\lcdm & QHMCA\tnote{c} & 0.0376 & 0.0925 & 0.256 & -- & -- & 71.45 & 11.01 & 0.286 & 1.686 & 0.287 & 0.435 & 1.066 & 0.274 & 1.205 & 49.99 & 0.380 & 958.61 & 984.61 & 1040.28 & 984.21 & 0.00 & 0.00 & 0.00\\
 & HzBSN\tnote{d} & 0.0239 & 0.1256 & 0.312 & -- & -- & 69.35 & -- & -- & -- & -- & -- & -- & -- & -- & -- & -- & 1439.59 & 1445.59 & 1461.79 & 1446.05 & 0.00 & 0.00 & 0.00\\
 & OHzNBSNQHMA\tnote{e} & 0.0258 & 0.1207 & 0.300 & -- & -- & 70.06 & 10.93 & 0.296 & 1.671 & 0.286 & -- & -- & -- & 1.110 & 50.26 & 0.406 & 2031.30 & 2051.30 & 2105.13 & 2051.86 & 0.00 & 0.00 & 0.00\\
 & HzBSNQHMCA\tnote{f} & 0.0258 & 0.1279 & 0.309 & -- & -- & 70.64 & 10.81 & 0.306 & 1.687 & 0.273 & 0.423 & 1.095 & 0.268 & 1.219 & 49.93 & 0.383 & 2400.54 & 2426.54 & 2500.41 & 2427.47 & 0.00 & 0.00 & 0.00\\
 & HzBSNQMCA\tnote{g} & 0.0246 & 0.1291 & 0.316 & -- & -- & 69.94 & 10.88 & 0.289 & 1.699 & 0.273 & 0.446 & 1.047 & 0.287 & 1.165 & 50.08 & 0.380 & 1964.13 & 1990.13 & 2062.86 & 1990.91 & 0.00 & 0.00 & 0.00\\
\\
 & QSO-AS + \hiig & 0.0228 & 0.1145 & 0.260 & $-0.360$ & -- & 72.91 & 11.60 & -- & -- & -- & -- & -- & -- & -- & -- & -- & 784.18 & 794.18 & 812.71 & 793.24 & $-0.27$ & 3.43 & 0.55\\
 & \mii\ + \civ & -- & 0.0791 & 0.213 & $-0.678$ & -- & -- & -- & 0.293 & 1.642 & 0.279 & 0.512 & 1.018 & 0.269 & -- & -- & -- & 42.92 & 58.92 & 80.95 & 68.83 & $-6.01$ & $-3.26$ & $-0.27$\\
Non-flat & A118 & -- & 0.4633 & 0.997 & 1.553 & -- & -- & -- & -- & -- & -- & -- & -- & -- & 1.177 & 49.71 & 0.380 & 117.47 & 127.47 & 141.32 & 126.29 & 0.84 & 3.61 & 0.34\\
\lcdm & QHMCA\tnote{c} & 0.0301 & 0.0997 & 0.246 & $-0.229$ & -- & 72.83 & 11.49 & 0.311 & 1.668 & 0.274 & 0.432 & 1.042 & 0.271 & 1.172 & 50.10 & 0.387 & 957.08 & 985.08 & 1045.03 & 984.54 & 0.47 & 4.75 & 0.33\\
 & HzBSN\tnote{d} & 0.0276 & 0.1078 & 0.288 & 0.084 & -- & 68.69 & -- & -- & -- & -- & -- & -- & -- & -- & -- & -- & 1437.61 & 1445.61 & 1467.21 & 1446.04 & 0.02 & 5.42 & $-0.01$\\
 & OHzBSNQHMA\tnote{e} & 0.0261 & 0.1182 & 0.297 & 0.008 & -- & 69.90 & 10.94 & 0.301 & 1.674 & 0.281 & -- & -- & -- & 1.126 & 50.20 & 0.400 & 2031.26 & 2053.26 & 2112.48 & 2053.69 & 1.96 & 7.35 & 1.84\\
 & HzBSNQHMCA\tnote{f} & 0.0296 & 0.1100 & 0.286 & 0.088 & -- & 70.00 & 10.85 & 0.311 & 1.670 & 0.280 & 0.444 & 1.054 & 0.278 & 1.234 & 49.90 & 0.381 & 2398.84 & 2426.84 & 2506.39 & 2427.33 & 0.30 & 5.98 & $-0.14$\\
 & HzBSNQMCA\tnote{g} & 0.0268 & 0.1061 & 0.289 & 0.098 & -- & 68.01 & 11.20 & 0.292 & 1.694 & 0.274 & 0.430 & 1.049 & 0.277 & 1.167 & 50.13 & 0.384 & 1962.87 & 1990.87 & 2069.20 & 1991.12 & 0.74 & 6.34 & 0.21\\
\\
 & QSO-AS + \hiig & 0.0174 & 0.1308 & 0.285 & -- & $-1.280$ & 72.32 & 11.23 & -- & -- & -- & -- & -- & -- & -- & -- & -- & 786.05 & 796.05 & 814.58 & 795.03 & 1.60 & 5.30 & 2.34\\
 & \mii\ + \civ & -- & $-0.0212$ & 0.008 & -- & $-4.875$ & -- & -- & 0.248 & 1.399 & 0.262 & 0.337 & 0.757 & 0.225 & -- & -- & -- & 40.24 & 56.24 & 78.27 & 66.94 & $-8.70$ & $-5.95$ & $-2.16$\\
Flat & A118 & -- & $-0.0223$ & 0.006 & -- & $-0.203$ & -- & -- & -- & -- & -- & -- & -- & -- & 1.186 & 49.87 & 0.383 & 118.07 & 128.07 & 141.92 & 126.87 & 1.44 & 4.21 & 0.92\\
XCDM & QHMCA\tnote{c} & 0.0275 & 0.1323 & 0.305 & -- & $-1.607$ & 72.58 & 11.43 & 0.288 & 1.667 & 0.276 & 0.438 & 1.053 & 0.265 & 1.176 & 50.11 & 0.384 & 957.29 & 985.29 & 1045.24 & 984.99 & 0.68 & 4.96 & 0.78\\
 & HzBSN\tnote{d} & 0.0283 & 0.1092 & 0.290 & -- & $-0.883$ & 68.96 & -- & -- & -- & -- & -- & -- & -- & -- & -- & -- & 1434.63 & 1442.63 & 1464.22 & 1443.28 & $-2.96$ & 2.43 & $-2.77$\\
 & OHzBSNQHMA\tnote{e} & 0.0266 & 0.1162 & 0.296 & -- & $-0.975$ & 69.62 & 10.94 & 0.278 & 1.696 & 0.280 & -- & -- & -- & 1.118 & 50.24 & 0.406 & 2030.88 & 2052.88 & 2112.10 & 2053.30 & 1.58 & 6.97 & 1.44\\
 & HzBSNQHMCA\tnote{f} & 0.0298 & 0.1056 & 0.282 & -- & $-0.879$ & 69.45 & 10.89 & 0.279 & 1.708 & 0.286 & 0.435 & 1.068 & 0.283 & 1.195 & 50.01 & 0.384 & 2396.06 & 2424.06 & 2503.61 & 2425.97 & $-2.48$ & 3.20 & $-1.50$\\
 & HzBSNQMCA\tnote{g} & 0.0288 & 0.1135 & 0.296 & -- & $-0.892$ & 69.43 & 10.80 & 0.289 & 1.684 & 0.287 & 0.411 & 1.100 & 0.287 & 1.166 & 50.08 & 0.381 & 1960.06 & 1988.06 & 2066.39 & 1988.01 & $-2.07$ & 3.52 & $-2.90$\\
\\
 & QSO-AS + \hiig & 0.0300 & 0.0031 & 0.065 & $-0.560$ & $-0.651$ & 71.87 & 11.45 & -- & -- & -- & -- & -- & -- & -- & -- & -- & 781.18 & 793.18 & 815.43 & 799.59 & $-1.27$ & 6.15 & 6.90\\
 & \mii\ + \civ & -- & $-0.0149$ & 0.021 & $-0.034$ & $-5.000$ & -- & -- & 0.285 & 1.262 & 0.270 & 0.374 & 0.808 & 0.226 & -- & -- & -- & 32.43 & 50.43 & 75.21 & 68.46 & $-14.51$ & $-9.00$ & $-0.71$\\
Non-flat & A118 & -- & 0.4579 & 0.986 & 1.260 & $-1.127$ & -- & -- & -- & -- & -- & -- & -- & -- & 1.179 & 49.71 & 0.383 & 117.50 & 129.50 & 146.12 & 126.97 & 2.87 & 8.41 & 1.02\\
XCDM & QHMCA\tnote{c} & 0.0248 & 0.1329 & 0.293 & $-0.196$ & $-1.194$ & 73.47 & 11.37 & 0.300 & 1.673 & 0.278 & 0.422 & 1.062 & 0.289 & 1.168 & 50.09 & 0.394 & 956.99 & 986.99 & 1051.22 & 986.72 & 2.38 & 10.94 & 2.51\\
 & HzBSN\tnote{e} & 0.0278 & 0.1118 & 0.294 & $-0.032$ & $-0.865$ & 69.07 & -- & -- & -- & -- & -- & -- & -- & -- & -- & -- & 1434.46 & 1444.46 & 1471.45 & 1445.42 & $-1.13$ & 9.66 & $-0.63$\\
 & OHzBSNQHMA\tnote{e} & 0.0260 & 0.1158 & 0.295 & $-0.016$ & $-0.947$ & 69.53 & 10.94 & 0.277 & 1.697 & 0.288 & -- & -- & -- & 1.151 & 50.15 & 0.409 & 2030.78 & 2054.78 & 2119.38 & 2055.42 & 3.48 & 14.25 & 3.56\\
 & HzBSNQHMCA\tnote{f} & 0.0298 & 0.1093 & 0.286 & 0.007 & $-0.900$ & 69.87 & 10.80 & 0.306 & 1.692 & 0.274 & 0.443 & 1.062 & 0.276 & 1.222 & 49.93 & 0.387 & 2396.10 & 2426.10 & 2511.33 & 2427.35 & $-0.44$ & 10.92 & $-0.12$\\
 & HzBSNQMCA\tnote{g} & 0.0288 & 0.1079 & 0.287 & $-0.002$ & $-0.873$ & 69.13 & 10.83 & 0.299 & 1.689 & 0.283 & 0.431 & 1.035 & 0.273 & 1.160 & 50.10 & 0.384 & 1959.83 & 1989.83 & 2073.76 & 1989.54 & $-0.30$ & 10.89 & $-1.36$\\
\\
 & QSO-AS + \hiig & 0.0198 & 0.1066 & 0.249 & -- & 0.000 & 71.42 & 11.10 & -- & -- & -- & -- & -- & -- & -- & -- & -- & 786.46 & 796.46 & 815.00 & 796.31 & 2.01 & 5.72 & 3.62\\
 & \mii\ + \civ & -- & $-0.0078$ & 0.035 & -- & 0.014 & -- & -- & 0.265 & 1.662 & 0.283 & 0.399 & 0.984 & 0.263 & -- & -- & -- & 51.19 & 67.19 & 89.21 & 72.04 & 2.25 & 5.00 & 2.94\\
Flat & A118 & -- & 0.1226 & 0.301 & -- & 9.805 & -- & -- & -- & -- & -- & -- & -- & -- & 1.173 & 49.89 & 0.383 & 118.24 & 128.24 & 142.10 & 125.51 & 1.62 & 4.39 & $-0.44$\\
\pcdm & QHMCA\tnote{c} & 0.0227 & 0.1014 & 0.243 & -- & 0.009 & 71.72 & 11.15 & 0.299 & 1.685 & 0.272 & 0.419 & 1.070 & 0.280 & 1.222 & 49.97 & 0.392 & 958.98 & 986.98 & 1046.93 & 988.57 & 2.37 & 6.65 & 4.36\\
 & HzBSN\tnote{d} & 0.0288 & 0.1060 & 0.286 & -- & 0.402 & 68.84 & -- & -- & -- & -- & -- & -- & -- & -- & -- & -- & 1434.43 & 1442.43 & 1464.02 & 1442.92 & $-3.16$ & 2.23 & $-3.14$\\
 & OHzBSNQHMA\tnote{e} & 0.0274 & 0.1116 & 0.289 & -- & 0.150 & 69.51 & 10.97 & 0.292 & 1.685 & 0.280 & -- & -- & -- & 1.121 & 50.23 & 0.409 & 2030.52 & 2052.52 & 2111.74 & 2053.18 & 1.22 & 6.61 & 1.33\\
 & HzBSNQHMCA\tnote{f} & 0.0292 & 0.1152 & 0.294 & -- & 0.286 & 70.28 & 10.78 & 0.286 & 1.696 & 0.286 & 0.426 & 1.067 & 0.294 & 1.183 & 50.04 & 0.384 & 2396.50 & 2424.50 & 2504.05 & 2424.36 & $-2.04$ & 3.64 & $-3.11$\\
 & HzBSNQMCA\tnote{g} & 0.0302 & 0.1073 & 0.288 & -- & 0.474 & 69.31 & 10.77 & 0.297 & 1.691 & 0.271 & 0.442 & 1.056 & 0.275 & 1.189 & 50.01 & 0.390 & 1959.57 & 1987.57 & 2065.90 & 1987.57 & $-2.56$ & 3.04 & $-3.33$\\
\\
 & QSO-AS + \hiig & 0.0338 & 0.0979 & 0.251 & $-0.250$ & 0.000 & 72.53 & 11.47 & -- & -- & -- & -- & -- & -- & -- & -- & -- & 784.61 & 796.61 & 818.85 & 801.32 & 2.16 & 9.57 & 8.63\\
 & \mii\ + \civ & -- & 0.0611 & 0.176 & $-0.173$ & 0.115 & -- & -- & 0.289 & 1.659 & 0.271 & 0.421 & 1.034 & 0.264 & -- & -- & -- & 50.74 & 68.74 & 93.52 & 72.86 & 3.80 & 9.31 & 3.76\\
Non-flat & A118 & -- & 0.2763 & 0.615 & 0.383 & 6.632 & -- & -- & -- & -- & -- & -- & -- & -- & 1.180 & 49.87 & 0.381 & 118.00 & 130.00 & 146.62 & 126.28 & 3.37 & 8.91 & 0.33\\
\pcdm & QHMCA\tnote{c} & 0.0321 & 0.1029 & 0.259 & $-0.208$ & 0.101 & 72.42 & 11.12 & 0.304 & 1.676 & 0.277 & 0.451 & 1.039 & 0.273 & 1.155 & 50.11 & 0.386 & 957.66 & 987.66 & 1051.89 & 990.25 & 3.05 & 11.61 & 6.04\\
 & HzBSN\tnote{d} & 0.0282 & 0.1104 & 0.291 & $-0.045$ & 0.480 & 69.13 & -- & -- & -- & -- & -- & -- & -- & -- & -- & -- & 1434.23 & 1444.23 & 1471.22 & 1444.26 & $-1.36$ & 9.43 & $-1.79$\\
 & OHzBSNQHMA\tnote{e} & 0.0251 & 0.1207 & 0.300 & $-0.056$ & 0.195 & 69.84 & 10.91 & 0.279 & 1.699 & 0.284 & -- & -- & -- & 1.132 & 50.19 & 0.411 & 2030.76 & 2054.76 & 2119.36 & 2055.13 & 3.46 & 14.23 & 3.28\\
 & HzBSNQHMCA\tnote{f} & 0.0313 & 0.1001 & 0.275 & $-0.015$ & 0.545 & 69.33 & 10.85 & 0.287 & 1.694 & 0.272 & 0.424 & 1.059 & 0.281 & 1.197 & 50.00 & 0.389 & 2396.34 & 2426.34 & 2511.57 & 2425.80 & $-0.20$ & 11.16 & $-1.66$\\
 & HzBSNQMCA\tnote{g} & 0.0287 & 0.1113 & 0.291 & $-0.018$ & 0.380 & 69.47 & 10.86 & 0.288 & 1.706 & 0.283 & 0.448 & 1.031 & 0.287 & 1.182 & 50.04 & 0.389 & 1959.10 & 1989.10 & 2073.03 & 1989.11 & $-1.03$ & 10.16 & $-1.80$\\
\bottomrule
\end{tabular}
\begin{tablenotes}[flushleft]
\item [a] \wx\ corresponds to flat/non-flat XCDM and $\alpha$ corresponds to flat/non-flat \pcdm.
\item [b] \hunit.
\item [c] QSO-AS + \hiig\ + \mii\ + \civ\ + A118.
\item [d] $H(z)$ + BAO + SNP\plus.
\item [e] Old $H(z)$ + Old BAO + SNP + SND + QSO-AS + \hiig\ + \mii\ + A118.
\item [f] $H(z)$ + BAO + SNP\plus\ + QSO-AS + \hiig\ + \mii\ + \civ\ + A118.
\item [g] $H(z)$ + BAO + SNP\plus\ + QSO-AS + \mii\ + \civ\ + A118.
\end{tablenotes}
\end{threeparttable}%
}
\end{table*}

\begin{table*}
\centering
\resizebox*{2.05\columnwidth}{2.6\columnwidth}{%
\begin{threeparttable}
\caption{One-dimensional posterior mean parameter values and uncertainties ($\pm 1\sigma$ error bars or $2\sigma$ limits) for all models from various combinations of data.}\label{tab:1d_BFPs}
\begin{tabular}{lccccccccccccccccc}
\toprule
Model & Data set & $\Omega_{b}h^2$ & $\Omega_{c}h^2$ & \om & \ok & $w_{\mathrm{X}}$/$\alpha$\tnote{a} & $H_0$\tnote{b} & $l_{\rm m}$ & $\gamma_{\rm \textsc{m}}$ & $\beta_{\rm \textsc{m}}$ & $\sigma_{\rm int,\,\textsc{m}}$ & $\gamma_{\rm \textsc{c}}$ & $\beta_{\rm \textsc{c}}$ & $\sigma_{\rm int,\,\textsc{c}}$ & $\gamma_{\rm \textsc{a}}$ & $\beta_{\rm \textsc{a}}$ & $\sigma_{\rm int,\,\textsc{a}}$\\
\midrule
 & QSO-AS + \hiig & $0.0225\pm0.0113$ & $0.1076^{+0.0197}_{-0.0224}$ & $0.257^{+0.037}_{-0.047}$ & -- & -- & $71.52\pm1.79$ & $11.04\pm0.34$ & -- & -- & -- & -- & -- & -- & -- & -- & -- \\
 & \mii\ + \civ & -- & -- & $<0.444$\tnote{c} & -- & -- & -- & -- & $0.292\pm0.045$ & $1.691\pm0.061$ & $0.289^{+0.023}_{-0.030}$ & $0.440\pm0.042$ & $1.030\pm0.089$ & $0.305^{+0.037}_{-0.054}$ & -- & -- & -- \\
Flat & A118 & -- & -- & $0.598^{+0.292}_{-0.226}$ & -- & -- & -- & -- & -- & -- & -- & -- & -- & -- & $1.171\pm0.087$ & $49.93\pm0.25$ & $0.393^{+0.027}_{-0.032}$ \\
\lcdm & QHMCA\tnote{d} & $0.0225\pm0.0117$ & $0.1090^{+0.0193}_{-0.0222}$ & $0.260^{+0.036}_{-0.046}$ & -- & -- & $71.43\pm1.73$ & $11.03\pm0.34$ & $0.292\pm0.043$ & $1.690\pm0.055$ & $0.288^{+0.023}_{-0.029}$ & $0.439\pm0.038$ & $1.030^{+0.072}_{-0.063}$ & $0.301^{+0.035}_{-0.052}$ & $1.197\pm0.086$ & $50.02\pm0.24$ & $0.393^{+0.025}_{-0.031}$ \\
 & HzBSN\tnote{e} & $0.0243\pm0.0034$ & $0.1267^{+0.0080}_{-0.0089}$ & $0.312\pm0.013$ & -- & -- & $69.65\pm2.48$ & -- & -- & -- & -- & -- & -- & -- & -- & -- & -- \\
 & OHzBSNQHMA\tnote{f} & $0.0256\pm0.0020$ & $0.1201\pm0.0061$ & $0.300\pm0.012$ & -- & -- & $69.87\pm1.13$ & $10.96\pm0.25$ & $0.293\pm0.044$ & $1.684\pm0.055$ & $0.292^{+0.023}_{-0.029}$ & -- & -- & -- & $1.131\pm0.087$ & $50.20\pm0.24$ & $0.413^{+0.026}_{-0.033}$ \\
 & HzBSNQHMCA\tnote{g} & $0.0250\pm0.0021$ & $0.1260\pm0.0064$ & $0.308\pm0.012$ & -- & -- & $70.13\pm1.25$ & $10.86\pm0.26$ & $0.294\pm0.044$ & $1.691\pm0.054$ & $0.288^{+0.023}_{-0.029}$ & $0.442\pm0.039$ & $1.035^{+0.072}_{-0.063}$ & $0.302^{+0.035}_{-0.052}$ & $1.190\pm0.085$ & $50.02\pm0.24$ & $0.392^{+0.025}_{-0.031}$ \\
 & HzBSNQMCA\tnote{i} & $0.0241^{+0.0032}_{-0.0035}$ & $0.1263^{+0.0080}_{-0.0089}$ & $0.313\pm0.012$ & -- & -- & $69.50^{+2.45}_{-2.44}$ & $10.94^{+0.39}_{-0.44}$ & $0.294\pm0.044$ & $1.689\pm0.055$ & $0.288^{+0.023}_{-0.029}$ & $0.442\pm0.039$ & $1.032^{+0.075}_{-0.064}$ & $0.302^{+0.035}_{-0.052}$ & $1.190\pm0.085$ & $50.03\pm0.24$ & $0.392^{+0.025}_{-0.031}$ \\
\\
 & QSO-AS + \hiig & $0.0224\pm0.0111$ & $0.1122^{+0.0223}_{-0.0218}$ & $0.260^{+0.039}_{-0.045}$ & $-0.196^{+0.112}_{-0.295}$ & -- & $72.25\pm1.99$ & $11.35\pm0.49$ & -- & -- & -- & -- & -- & -- & -- & -- & -- \\
 & \mii\ + \civ & -- & -- & $0.473^{+0.187}_{-0.311}$ & $-0.818^{+0.391}_{-0.637}$ & -- & -- & -- & $0.314^{+0.048}_{-0.052}$ & $1.662\pm0.065$ & $0.285^{+0.023}_{-0.030}$ & $0.491^{+0.050}_{-0.064}$ & $1.073^{+0.093}_{-0.094}$ & $0.299^{+0.036}_{-0.053}$ & -- & -- & -- \\
Non-flat & A118 & -- & -- & $>0.267$ & $0.789^{+0.664}_{-0.775}$ & -- & -- & -- & -- & -- & -- & -- & -- & -- & $1.186\pm0.089$ & $49.82\pm0.26$ & $0.392^{+0.026}_{-0.032}$ \\
\lcdm & QHMCA\tnote{d} & $0.0224\pm0.0117$ & $0.1144^{+0.0213}_{-0.0212}$ & $0.266^{+0.036}_{-0.044}$ & $-0.157^{+0.109}_{-0.211}$ & -- & $71.97\pm1.85$ & $11.24\pm0.41$ & $0.292\pm0.043$ & $1.684\pm0.055$ & $0.288^{+0.023}_{-0.029}$ & $0.441\pm0.038$ & $1.028^{+0.071}_{-0.064}$ & $0.298^{+0.035}_{-0.052}$ & $1.182\pm0.087$ & $50.06\pm0.24$ & $0.395^{+0.026}_{-0.031}$ \\
 & HzBSN\tnote{e} & $0.0282^{+0.0046}_{-0.0050}$ & $0.1082\pm0.0152$ & $0.288\pm0.021$ & $0.087\pm0.063$ & -- & $68.89\pm2.44$ & -- & -- & -- & -- & -- & -- & -- & -- & -- & -- \\
 & OHzBSNQHMA\tnote{f} & $0.0265^{+0.0032}_{-0.0038}$ & $0.1168\pm0.0127$ & $0.295\pm0.019$ & $0.018\pm0.059$ & -- & $69.79\pm1.14$ & $10.96\pm0.25$ & $0.293\pm0.044$ & $1.685\pm0.055$ & $0.292^{+0.023}_{-0.029}$ & -- & -- & -- & $1.131\pm0.086$ & $50.20\pm0.24$ & $0.413^{+0.027}_{-0.033}$ \\
 & HzBSNQHMCA\tnote{g} & $0.0291^{+0.0036}_{-0.0041}$ & $0.1115\pm0.0126$ & $0.288\pm0.019$ & $0.074\pm0.056$ & -- & $70.02\pm1.25$ & $10.87\pm0.26$ & $0.293\pm0.044$ & $1.692\pm0.054$ & $0.288^{+0.023}_{-0.029}$ & $0.440\pm0.038$ & $1.033^{+0.073}_{-0.062}$ & $0.303^{+0.035}_{-0.052}$ & $1.197^{+0.084}_{-0.085}$ & $50.01\pm0.24$ & $0.392^{+0.025}_{-0.031}$ \\
 & HzBSNQMCA\tnote{i} & $0.0273^{+0.0041}_{-0.0047}$ & $0.1112^{+0.0138}_{-0.0151}$ & $0.293\pm0.020$ & $0.073\pm0.059$ & -- & $68.88\pm2.45$ & $11.04^{+0.39}_{-0.45}$ & $0.294\pm0.044$ & $1.687\pm0.055$ & $0.288^{+0.022}_{-0.029}$ & $0.441\pm0.039$ & $1.027^{+0.075}_{-0.065}$ & $0.303^{+0.035}_{-0.052}$ & $1.195\pm0.085$ & $50.03\pm0.24$ & $0.392^{+0.025}_{-0.031}$ \\
\\
 & QSO-AS + \hiig & $0.0224\pm0.0112$ & $0.1391^{+0.0333}_{-0.0256}$ & $0.305^{+0.056}_{-0.047}$ & -- & $-1.683^{+0.712}_{-0.387}$ & $72.92^{+2.15}_{-2.40}$ & $11.31\pm0.43$ & -- & -- & -- & -- & -- & -- & -- & -- & -- \\
 & \mii\ + \civ & -- & -- & $<0.563$ & -- & $<-1.509$ & -- & -- & $0.282^{+0.042}_{-0.046}$ & $1.557^{+0.117}_{-0.101}$ & $0.283^{+0.023}_{-0.029}$ & $0.405\pm0.047$ & $0.900^{+0.122}_{-0.121}$ & $0.282^{+0.038}_{-0.054}$ & -- & -- & -- \\
Flat & A118 & -- & -- & $0.557^{+0.277}_{-0.274}$ & -- & $-2.521^{+2.330}_{-2.370}$ & -- & -- & --  & -- & -- & -- & -- & -- & $1.167\pm0.088$ & $50.01^{+0.27}_{-0.31}$ & $0.393^{+0.026}_{-0.032}$ \\
XCDM & QHMCA\tnote{d} & $0.0225\pm0.0117$ & $0.1457^{+0.0291}_{-0.0242}$ & $0.313\pm0.045$ & -- & $-1.929^{+0.825}_{-0.426}$ & $73.51^{+2.15}_{-2.52}$ & $11.41\pm0.43$ & $0.295\pm0.044$ & $1.677\pm0.056$ & $0.287^{+0.023}_{-0.029}$ & $0.439\pm0.038$ & $1.029^{+0.071}_{-0.064}$ & $0.297^{+0.035}_{-0.052}$ & $1.183\pm0.086$ & $50.07\pm0.24$ & $0.392^{+0.025}_{-0.031}$ \\
 & HzBSN\tnote{e} & $0.0287\pm0.0044$ & $0.1097\pm0.0117$ & $0.290\pm0.016$ & -- & $-0.886\pm0.053$ & $69.15\pm2.52$ & -- & -- & -- & -- & -- & -- & -- & -- & -- & -- \\
 & OHzBSNQHMA\tnote{f} & $0.0271^{+0.0027}_{-0.0031}$ & $0.1147^{+0.0098}_{-0.0097}$ & $0.294\pm0.015$ & -- & $-0.959\pm0.059$ & $69.66\pm1.16$ & $10.94\pm0.25$ & $0.293\pm0.044$ & $1.685\pm0.055$ & $0.292^{+0.023}_{-0.029}$ & -- & -- & -- & $1.131\pm0.087$ & $50.20\pm0.24$ & $0.413^{+0.027}_{-0.033}$ \\
 & HzBSNQHMCA\tnote{g} & $0.0294^{+0.0030}_{-0.0035}$ & $0.1106^{+0.0100}_{-0.0102}$ & $0.288\pm0.015$ & -- & $-0.895\pm0.051$ & $69.84\pm1.26$ & $10.80\pm0.26$ & $0.294\pm0.044$ & $1.692\pm0.054$ & $0.288^{+0.023}_{-0.029}$ & $0.442\pm0.039$ & $1.034^{+0.074}_{-0.062}$ & $0.303^{+0.035}_{-0.052}$ & $1.193\pm0.085$ & $50.01\pm0.24$ & $0.392^{+0.025}_{-0.031}$ \\
 & HzBSNQMCA\tnote{i} & $0.0284^{+0.0040}_{-0.0045}$ & $0.1101\pm0.0112$ & $0.292\pm0.016$ & -- & $-0.889\pm0.052$ & $69.05\pm2.40$ & $10.90^{+0.38}_{-0.43}$ & $0.294\pm0.044$ & $1.690\pm0.055$ & $0.288^{+0.023}_{-0.029}$ & $0.442\pm0.039$ & $1.031^{+0.075}_{-0.064}$ & $0.303^{+0.035}_{-0.052}$ & $1.192\pm0.084$ & $50.02\pm0.24$ & $0.392^{+0.025}_{-0.031}$ \\
\\
 & QSO-AS + \hiig & $0.0224\pm0.0114$ & $0.1122^{+0.0473}_{-0.0326}$ & $0.258^{+0.086}_{-0.057}$ & $0.018^{+0.345}_{-0.383}$ & $-1.670^{+1.063}_{-0.245}$ & $72.34\pm2.16$ & $11.20\pm0.49$ & -- & -- & -- & -- & -- & -- & -- & -- & -- \\
 & \mii\ + \civ & -- & -- & $0.338^{+0.101}_{-0.299}$ & $-0.410^{+0.368}_{-0.222}$ & $<-1.124$ & -- & -- & $0.319^{+0.048}_{-0.054}$ & $1.526^{+0.132}_{-0.108}$ & $0.280^{+0.023}_{-0.030}$ & $0.456^{+0.047}_{-0.056}$ & $0.966^{+0.119}_{-0.110}$ & $0.282^{+0.037}_{-0.053}$ & -- & -- & -- \\
Non-flat & A118 & -- & -- & $>0.226$ & $0.690^{+0.512}_{-0.798}$ & $-2.342^{+2.067}_{-1.106}$ & -- & -- & -- & -- & -- & -- & -- & -- & $1.185\pm0.090$ & $49.82\pm0.28$ & $0.392^{+0.026}_{-0.032}$ \\
XCDM & QHMCA\tnote{d} & $0.0226\pm0.0118$ & $0.1327^{+0.0335}_{-0.0277}$ & $0.291^{+0.056}_{-0.049}$ & $0.039^{+0.219}_{-0.229}$ & $-2.066^{+1.287}_{-0.446}$ & $73.13^{+2.14}_{-2.46}$ & $11.30\pm0.46$ & $0.294\pm0.044$ & $1.680\pm0.056$ & $0.287^{+0.023}_{-0.030}$ & $0.439\pm0.038$ & $1.030^{+0.071}_{-0.065}$ & $0.298^{+0.035}_{-0.052}$ & $1.187\pm0.087$ & $50.06\pm0.24$ & $0.393^{+0.025}_{-0.031}$ \\
 & HzBSN\tnote{e} & $0.0284\pm0.0047$ & $0.1115^{+0.0151}_{-0.0165}$ & $0.293\pm0.021$ & $-0.017\pm0.095$ & $-0.884^{+0.082}_{-0.058}$ & $69.23\pm2.53$ & -- & -- & -- & -- & -- & -- & -- & -- & -- & -- \\
 & OHzBSNQHMA\tnote{f} & $0.0269^{+0.0033}_{-0.0039}$ & $0.1155^{+0.0128}_{-0.0127}$ & $0.295\pm0.019$ & $-0.009^{+0.077}_{-0.083}$ & $-0.959^{+0.090}_{-0.063}$ & $69.65\pm1.16$ & $10.93\pm0.26$ & $0.293\pm0.044$ & $1.685\pm0.055$ & $0.292^{+0.023}_{-0.029}$ & -- & -- & -- & $1.130\pm0.087$ & $50.20\pm0.24$ & $0.413^{+0.027}_{-0.033}$ \\
 & HzBSNQHMCA\tnote{g} & $0.0293^{+0.0036}_{-0.0041}$ & $0.1108^{+0.0124}_{-0.0125}$ & $0.289\pm0.019$ & $-0.004\pm0.078$ & $-0.897^{+0.075}_{-0.055}$ & $69.81\pm1.26$ & $10.80\pm0.27$ & $0.294\pm0.044$ & $1.692\pm0.054$ & $0.288^{+0.023}_{-0.030}$ & $0.442\pm0.039$ & $1.034^{+0.074}_{-0.063}$ & $0.303^{+0.035}_{-0.052}$ & $1.192\pm0.085$ & $50.01\pm0.24$ & $0.392^{+0.025}_{-0.031}$ \\
 & HzBSNQMCA\tnote{i} & $0.0279^{+0.0042}_{-0.0047}$ & $0.1134^{+0.0139}_{-0.0153}$ & $0.296\pm0.020$ & $-0.032\pm0.085$ & $-0.877^{+0.075}_{-0.053}$ & $69.26\pm2.45$ & $10.84^{+0.41}_{-0.45}$ & $0.294\pm0.043$ & $1.691\pm0.055$ & $0.289^{+0.023}_{-0.030}$ & $0.443\pm0.039$ & $1.032^{+0.077}_{-0.064}$ & $0.303^{+0.036}_{-0.053}$ & $1.189\pm0.085$ & $50.02\pm0.24$ & $0.392^{+0.025}_{-0.031}$ \\
\\
 & QSO-AS + \hiig & $0.0217^{+0.0081}_{-0.0138}$ & $0.0543^{+0.0225}_{-0.0471}$ & $0.154^{+0.053}_{-0.086}$ & -- & $<6.506$ & $70.64\pm1.80$ & $10.81\pm0.34$ & -- & -- & -- & -- & -- & -- & -- & -- & -- \\
 & \mii\ + \civ & -- & -- & $<0.537$\tnote{c} & -- & $<6.202$\tnote{c} & -- & -- & $0.299\pm0.046$ & $1.717^{+0.059}_{-0.053}$ & $0.289^{+0.023}_{-0.030}$ & $0.449\pm0.043$ & $1.069^{+0.091}_{-0.074}$ & $0.312^{+0.037}_{-0.054}$ & -- & -- & -- \\
Flat & A118 & -- & -- & $0.535^{+0.293}_{-0.287}$ & -- & -- & --  & -- & -- & -- & -- & -- & -- & -- & $1.171\pm0.088$ & $49.88\pm0.24$ & $0.392^{+0.026}_{-0.032}$ \\
\pcdm & QHMCA\tnote{d} & $0.0219^{+0.0082}_{-0.0165}$ & $0.0647^{+0.0378}_{-0.0408}$ & $0.175^{+0.074}_{-0.080}$ & -- & $<6.630$ & $70.66\pm1.75$ & $10.83\pm0.35$ & $0.292\pm0.044$ & $1.695\pm0.054$ & $0.288^{+0.023}_{-0.029}$ & $0.439\pm0.039$ & $1.033^{+0.074}_{-0.064}$ & $0.303^{+0.035}_{-0.053}$ & $1.204\pm0.086$ & $50.00\pm0.24$ & $0.393^{+0.025}_{-0.031}$ \\
 & HzBSN\tnote{e} & $0.0300^{+0.0047}_{-0.0046}$ & $0.1040\pm0.0129$ & $0.282\pm0.018$ & -- & $0.475^{+0.189}_{-0.265}$ & $69.01\pm2.43$ & -- & -- & -- & -- & -- & -- & -- & -- & -- & -- \\
 & OHzBSNQHMA\tnote{f} & $0.0286^{+0.0025}_{-0.0033}$ & $0.1089^{+0.0103}_{-0.0083}$ & $0.286\pm0.015$ & -- & $0.249^{+0.069}_{-0.239}$ & $69.50\pm1.14$ & $10.92\pm0.25$ & $0.293\pm0.044$ & $1.686\pm0.054$ & $0.292^{+0.023}_{-0.029}$ & -- & -- & -- & $1.132\pm0.086$ & $50.20\pm0.24$ & $0.413^{+0.027}_{-0.033}$ \\
 & HzBSNQHMCA\tnote{g} & $0.0307^{+0.0032}_{-0.0038}$ & $0.1057^{+0.0118}_{-0.0107}$ & $0.281\pm0.017$ & -- & $0.423^{+0.168}_{-0.246}$ & $69.79^{+1.25}_{-1.26}$ & $10.79\pm0.26$ & $0.294\pm0.044$ & $1.692\pm0.054$ & $0.288^{+0.023}_{-0.029}$ & $0.442\pm0.039$ & $1.034^{+0.075}_{-0.062}$ & $0.303^{+0.035}_{-0.052}$ & $1.193\pm0.085$ & $50.01\pm0.24$ & $0.392^{+0.025}_{-0.031}$ \\
 & HzBSNQMCA\tnote{i} & $0.0296^{+0.0043}_{-0.0047}$ & $0.1048\pm0.0122$ & $0.284\pm0.017$ & -- & $0.450^{+0.174}_{-0.253}$ & $68.92\pm2.35$ & $10.90^{+0.37}_{-0.43}$ & $0.294\pm0.043$ & $1.690\pm0.055$ & $0.288^{+0.023}_{-0.029}$ & $0.442\pm0.039$ & $1.031^{+0.075}_{-0.064}$ & $0.303^{+0.035}_{-0.052}$ & $1.192\pm0.085$ & $50.02\pm0.24$ & $0.392^{+0.025}_{-0.031}$ \\
\\
 & QSO-AS + \hiig & $0.0219^{+0.0093}_{-0.0130}$ & $0.0576^{+0.0268}_{-0.0424}$ & $0.163^{+0.058}_{-0.081}$ & $0.181^{+0.180}_{-0.339}$ & $<7.875$ & $70.21\pm1.83$ & $10.70^{+0.36}_{-0.41}$ & -- & -- & -- & -- & -- & -- & -- & -- & -- \\
 & \mii\ + \civ & -- & -- & $<0.536$\tnote{c} & $0.088^{+0.384}_{-0.364}$ & $<6.162$\tnote{c} & -- & -- & $0.299\pm0.046$ & $1.719\pm0.055$ & $0.290^{+0.023}_{-0.030}$ & $0.450\pm0.043$ & $1.072^{+0.088}_{-0.076}$ & $0.312^{+0.037}_{-0.054}$ & -- & -- & -- \\%
Non-flat & A118 & -- & -- & $0.516^{+0.215}_{-0.288}$ & $0.064^{+0.293}_{-0.282}$ & $5.209^{+3.855}_{-2.462}$ & -- & -- & --  & -- & -- & -- & -- & -- & $1.174\pm0.089$ & $49.88\pm0.24$ & $0.392^{+0.026}_{-0.032}$ \\
\pcdm & QHMCA\tnote{d} & $0.0221^{+0.0117}_{-0.0165}$ & $0.0742^{+0.0405}_{-0.0329}$ & $0.194^{+0.077}_{-0.063}$ & $0.018^{+0.083}_{-0.232}$ & $<6.688$ & $70.63\pm1.85$ & $10.84\pm0.39$ & $0.292\pm0.044$ & $1.694\pm0.055$ & $0.288^{+0.023}_{-0.029}$ & $0.439\pm0.039$ & $1.033^{+0.075}_{-0.064}$ & $0.303^{+0.035}_{-0.053}$ & $1.201\pm0.086$ & $50.00\pm0.24$ & $0.393^{+0.026}_{-0.031}$ \\
 & HzBSN\tnote{e} & $0.0296^{+0.0048}_{-0.0047}$ & $0.1067^{+0.0153}_{-0.0154}$ & $0.286\pm0.021$ & $-0.035^{+0.071}_{-0.085}$ & $0.550^{+0.231}_{-0.314}$ & $69.15\pm2.53$ & -- & -- & -- & -- & -- & -- & -- & -- & -- & -- \\
 & OHzBSNQHMA\tnote{f} & $0.0277^{+0.0034}_{-0.0040}$ & $0.1126\pm0.0128$ & $0.291\pm0.019$ & $-0.040^{+0.064}_{-0.072}$ & $0.316^{+0.101}_{-0.292}$ & $69.52\pm1.15$ & $10.89\pm0.25$ & $0.294\pm0.044$ & $1.685\pm0.055$ & $0.292^{+0.023}_{-0.030}$ & -- & -- & -- & $1.128\pm0.087$ & $50.20\pm0.24$ & $0.413^{+0.027}_{-0.033}$ \\
 & HzBSNQHMCA\tnote{g} & $0.0303^{+0.0038}_{-0.0041}$ & $0.1068\pm0.0127$ & $0.283\pm0.019$ & $-0.021^{+0.067}_{-0.074}$ & $0.468^{+0.200}_{-0.292}$ & $69.76\pm1.25$ & $10.78\pm0.26$ & $0.294\pm0.044$ & $1.692\pm0.054$ & $0.288^{+0.023}_{-0.029}$ & $0.442\pm0.039$ & $1.035^{+0.074}_{-0.063}$ & $0.303^{+0.035}_{-0.052}$ & $1.191\pm0.084$ & $50.01\pm0.24$ & $0.392^{+0.025}_{-0.031}$ \\
 & HzBSNQMCA\tnote{i} & $0.0292^{+0.0045}_{-0.0048}$ & $0.1083^{+0.0144}_{-0.0145}$ & $0.288\pm0.020$ & $-0.044^{+0.067}_{-0.081}$ & $0.543^{+0.226}_{-0.308}$ & $69.18^{+2.43}_{-2.44}$ & $10.83^{+0.40}_{-0.44}$ & $0.294\pm0.044$ & $1.691\pm0.055$ & $0.288^{+0.023}_{-0.029}$ & $0.443\pm0.039$ & $1.033^{+0.076}_{-0.063}$ & $0.303^{+0.035}_{-0.052}$ & $1.189\pm0.085$ & $50.02\pm0.24$ & $0.392^{+0.025}_{-0.031}$ \\
\bottomrule
\end{tabular}
\begin{tablenotes}[flushleft]
\item [a] \wx\ corresponds to flat/non-flat XCDM and $\alpha$ corresponds to flat/non-flat \pcdm.
\item [b] \hunit.
\item [c] This is the 1$\sigma$ limit. The $2\sigma$ limit is set by the prior, and is not shown here.
\item [d] QSO-AS + \hiig\ + \mii\ + \civ\ + A118.
\item [e] $H(z)$ + BAO + SNP\plus.
\item [f] Old $H(z)$ + Old BAO + SNP + SND + QSO-AS + \hiig\ + \mii\ + A118.
\item [g] $H(z)$ + BAO + SNP\plus\ + QSO-AS + \hiig\ + \mii\ + \civ\ + A118.
\item [i] $H(z)$ + BAO + SNP\plus\ + QSO-AS + \mii\ + \civ\ + A118.
\end{tablenotes}
\end{threeparttable}%
}
\end{table*}

\begin{figure*}
\centering
 \subfloat[]{%
    \includegraphics[width=0.45\textwidth,height=0.35\textwidth]{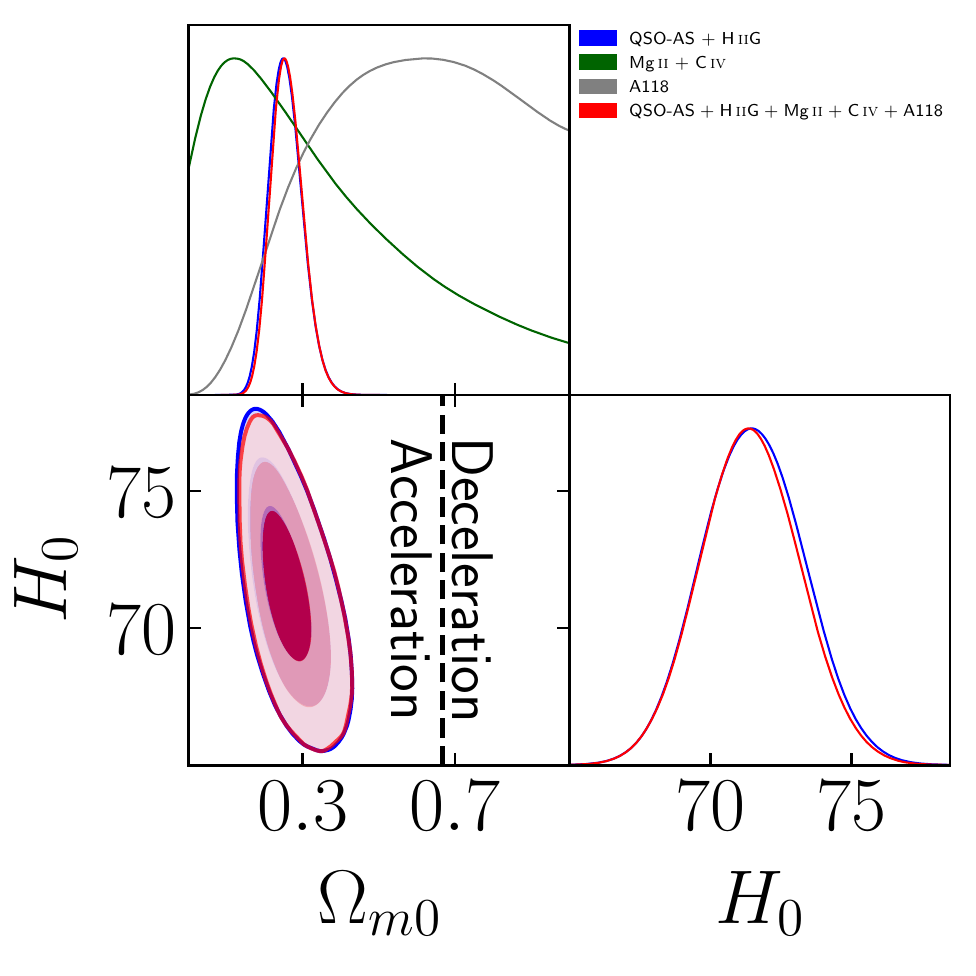}}
 \subfloat[]{%
    \includegraphics[width=0.45\textwidth,height=0.35\textwidth]{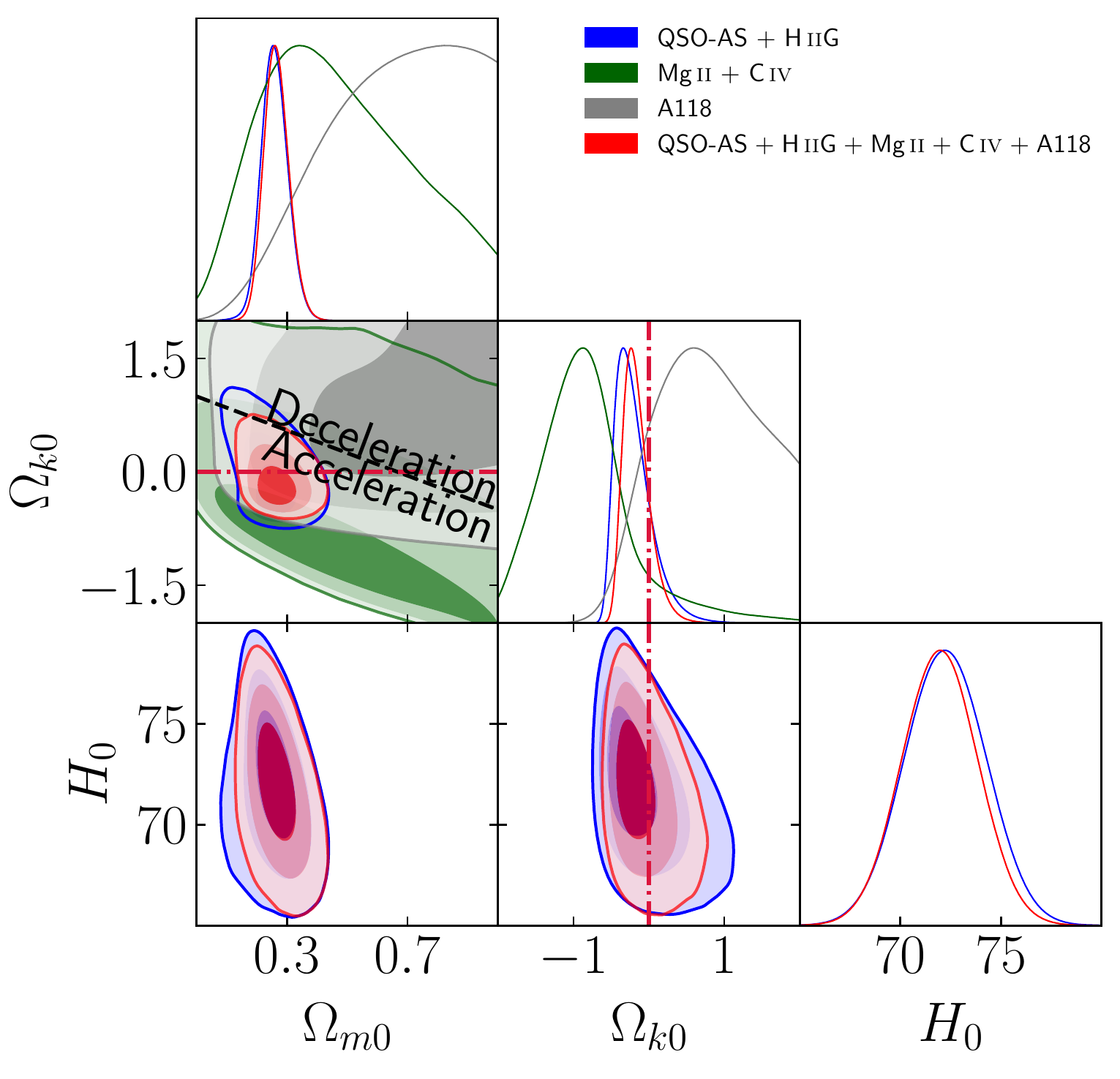}}\\
 \subfloat[]{%
    \includegraphics[width=0.45\textwidth,height=0.35\textwidth]{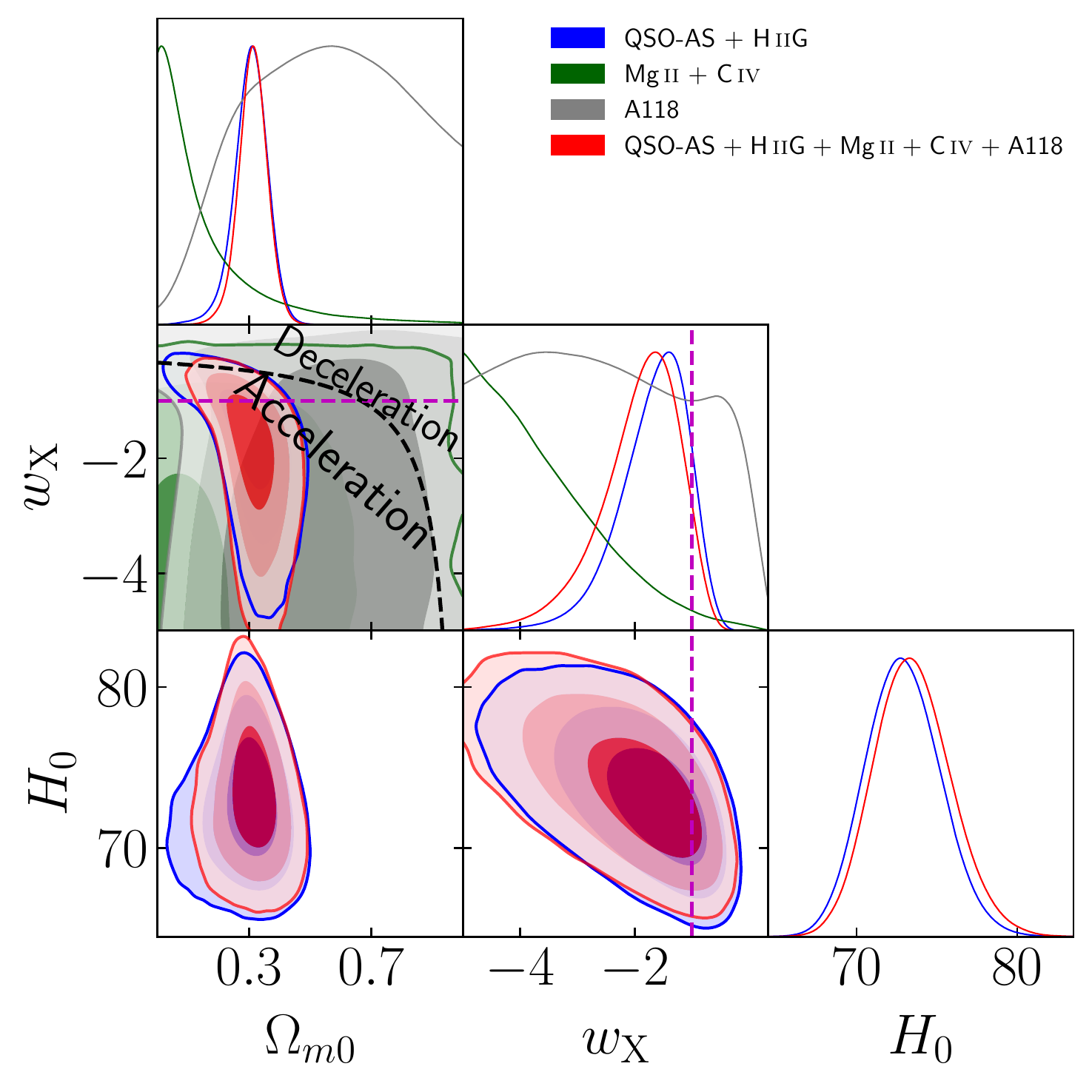}}
 \subfloat[]{%
    \includegraphics[width=0.45\textwidth,height=0.35\textwidth]{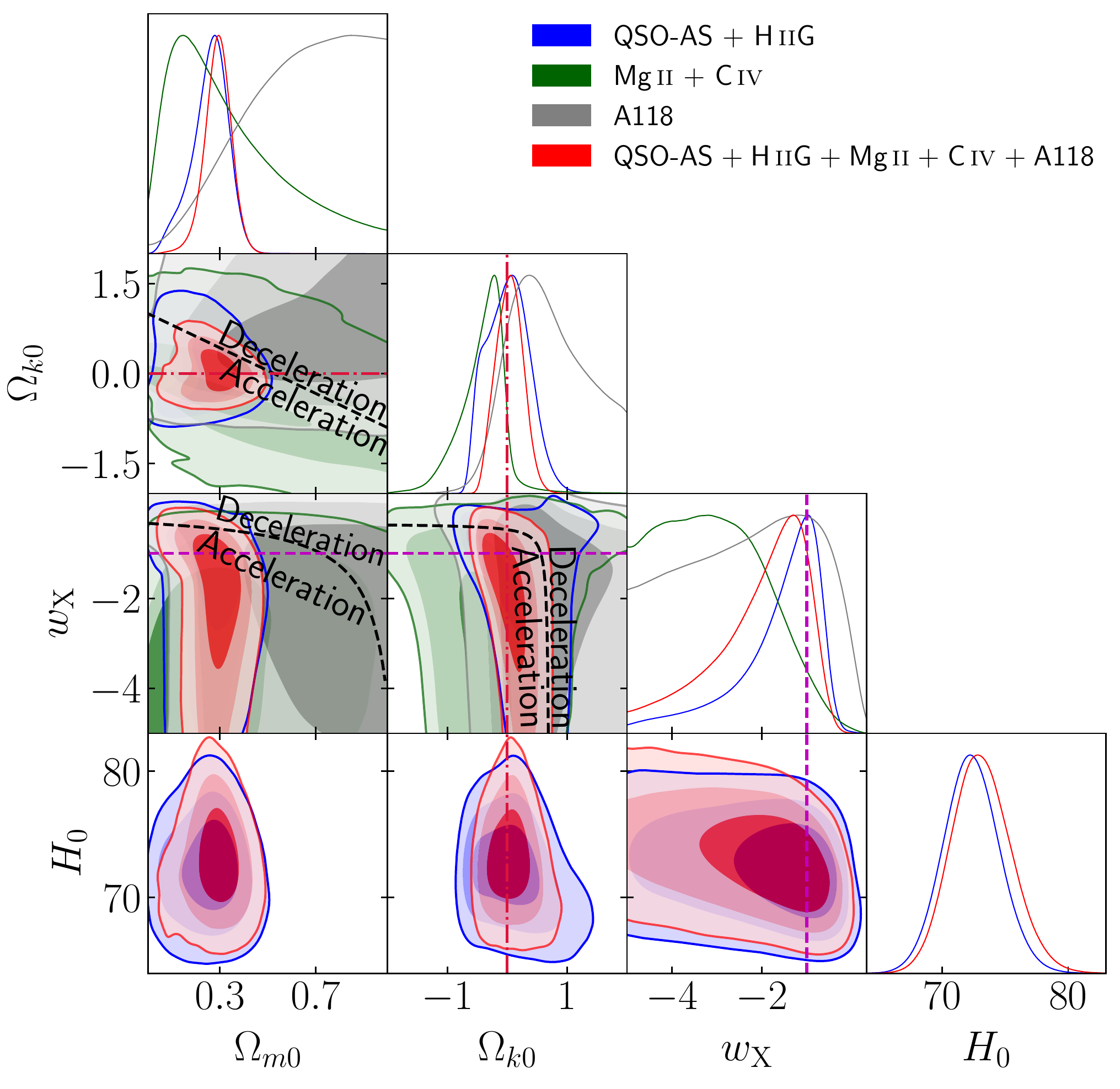}}\\
 \subfloat[]{%
    \includegraphics[width=0.45\textwidth,height=0.35\textwidth]{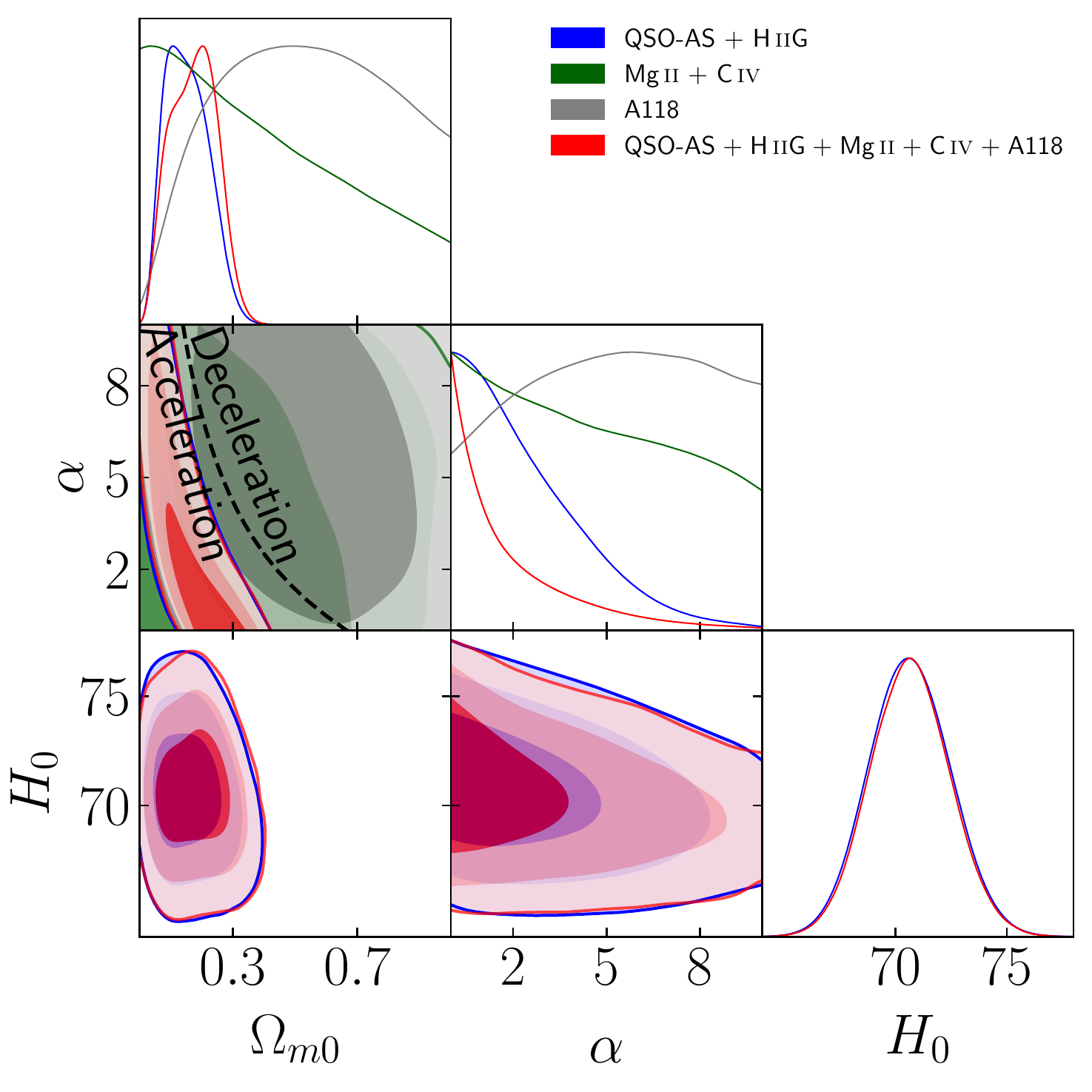}}
 \subfloat[]{%
    \includegraphics[width=0.45\textwidth,height=0.35\textwidth]{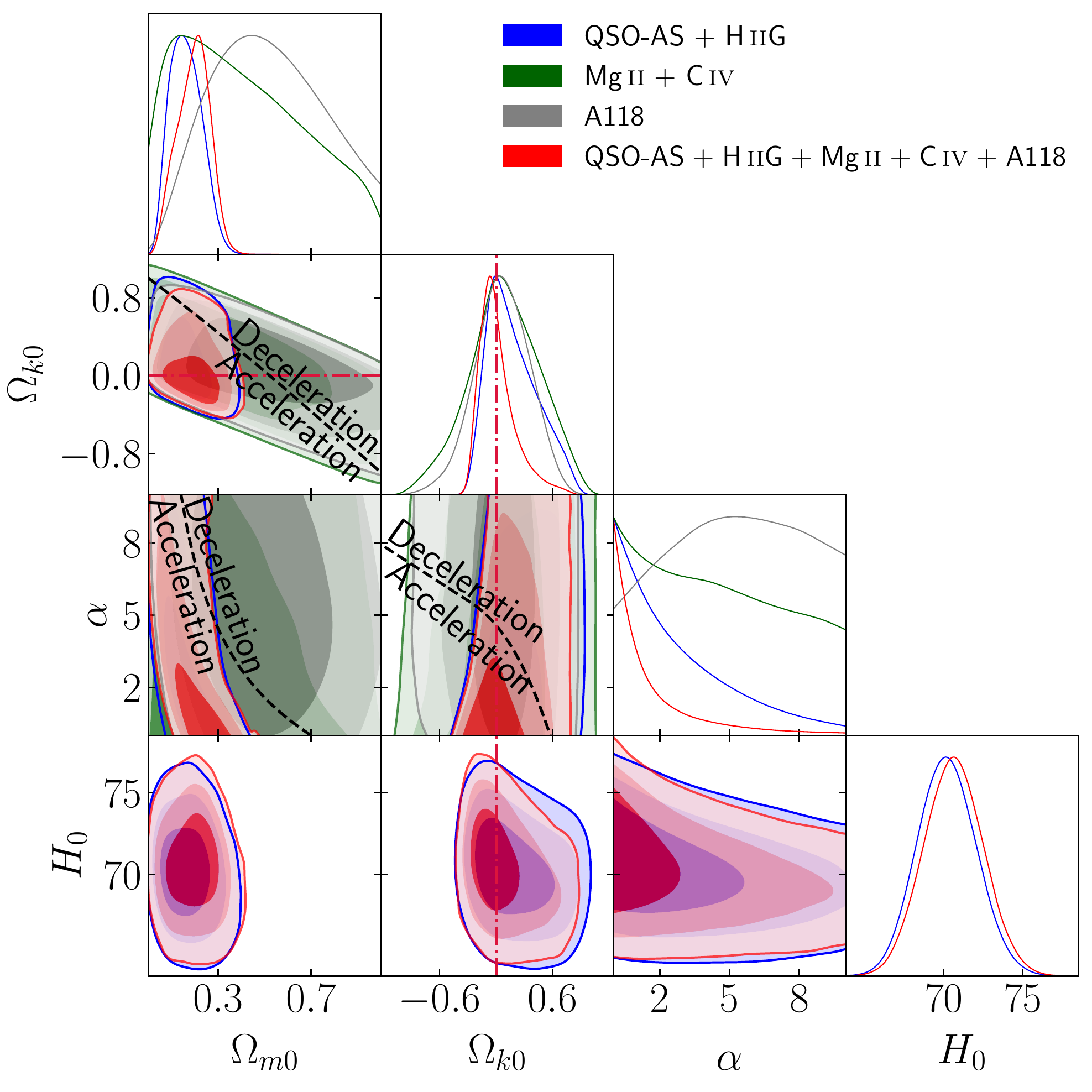}}\\
\caption{Same as Fig.\ \ref{fig1} but for QSO-AS + \hiig\ (blue), \mii\ + \civ\ (green), A118 (gray), and QSO-AS + \hiig\ + \mii\ + \civ\ + A118 (red) data. The black dashed zero-acceleration lines, computed for the third cosmological parameter set to the $H(z)$ + BAO data best-fitting values listed in Table \ref{tab:BFP} in panels (d) and (f), divides the parameter space into regions associated with currently-accelerating (below left) and currently-decelerating (above right) cosmological expansion.}
\label{fig6}
\end{figure*}

Constraints from QSO-AS + \hiig, \mii\ + \civ, and A118 data are listed in Tables \ref{tab:BFPs} and \ref{tab:1d_BFPs} and their confidence regions are shown in Fig.\ \ref{fig6}. Note that A118 results here are corrected with respect to the ones reported in Ref.\ \cite{CaoDainottiRatra2022} and used in Ref.\ \cite{CaoRatra2022}. Since the changes are insignificant we do not discuss these results in detail here. QSO-AS + \hiig, \mii\ + \civ, and A118 data constraints are mutually consistent, so these data can be used together to constrain cosmological parameters, as discussed next.

Joint QSO-AS + \hiig\ + \mii\ + \civ\ + A118 (QHMCA) data constraints on \om\ range from $0.175^{+0.074}_{-0.080}$ (flat \pcdm) to $0.313\pm0.045$ (flat XCDM), with a difference of $1.6\sigma$. 

QHMCA data constraints on $H_0$ range from $70.63\pm1.85$ \hunit\ (non-flat \pcdm) to $73.51^{+2.15}_{-2.52}$ \hunit\ (flat XCDM), with a difference of $0.92\sigma$. These $H_0$ values  are $0.78\sigma$ (non-flat \pcdm) and $1.5\sigma$ (flat XCDM) higher than the median statistics estimate of $H_0=68\pm2.8$ \hunit\ \citep{chenratmed} (also see Refs.\ \citep{gott_etal_2001, Calabreseetal2012}), $0.33\sigma$ (non-flat \pcdm) and $1.2\sigma$ (flat XCDM) higher than the TRGB and SN Ia estimate of $H_0=69.8\pm1.7$ \hunit\ \citep{Freedman2021}, and $1.1\sigma$ (non-flat \pcdm) lower and $0.17\sigma$ (flat XCDM) higher than the Cepheids and SN Ia measurement of $73.04\pm1.04$ \hunit\ \cite{Riessetal2022}. The $H_0$ constraints from flat \lcdm\ here, $71.43\pm1.73$ \hunit\, is $2.2\sigma$ higher than the $H_0$ estimate of $67.36 \pm 0.54$ \hunit\ from \textit{Planck} 2018 TT,TE,EE+lowE+lensing CMB anisotropy data \cite{planck2018b}.

QHMCA data constraints on \ok\ are $-0.157^{+0.109}_{-0.211}$, $0.039^{+0.219}_{-0.229}$, and $0.018^{+0.083}_{-0.282}$ for non-flat \lcdm, XCDM, and \pcdm, respectively. The non-flat \lcdm\ result favors closed spatial geometry, being $1.4\sigma$ away from flat, while open spatial geometry is favored in both the non-flat XCDM and non-flat \pcdm\ models, but with flatness well within 1$\sigma$.

QHMCA data indicate a preference for dark energy dynamics. In particular, the central values of the XCDM equation of state parameter, $w_{\rm X}$, are found to be $1.1\sigma$ and $0.83\sigma$ lower than $-1$ for the flat and non-flat parametrizations respectively. $2\sigma$ upper limits of $\alpha$ in both the flat and non-flat \pcdm\ models suggest $\alpha=0$ is within $1\sigma$.

QSO-AS + \hiig\ + \mii\ + \civ\ + A118 data constraints derived here are very consistent with the QSO-AS + \hiig\ + \mii\ + A118 data constraints derived in Ref.\ \cite{CaoRatra2022}.

\subsection{Constraints from $H(z)$ + BAO + SNP\plus\ + QSO-AS + \hiig\ + \mii\ + \civ\ + A118 and Old $H(z)$ + Old BAO + SNP + SND + QSO-AS + \hiig\ + \mii\ + A118 data}
 \label{subsec:comp4}

\begin{figure*}
\centering
 \subfloat[]{%
    \includegraphics[width=0.45\textwidth,height=0.35\textwidth]{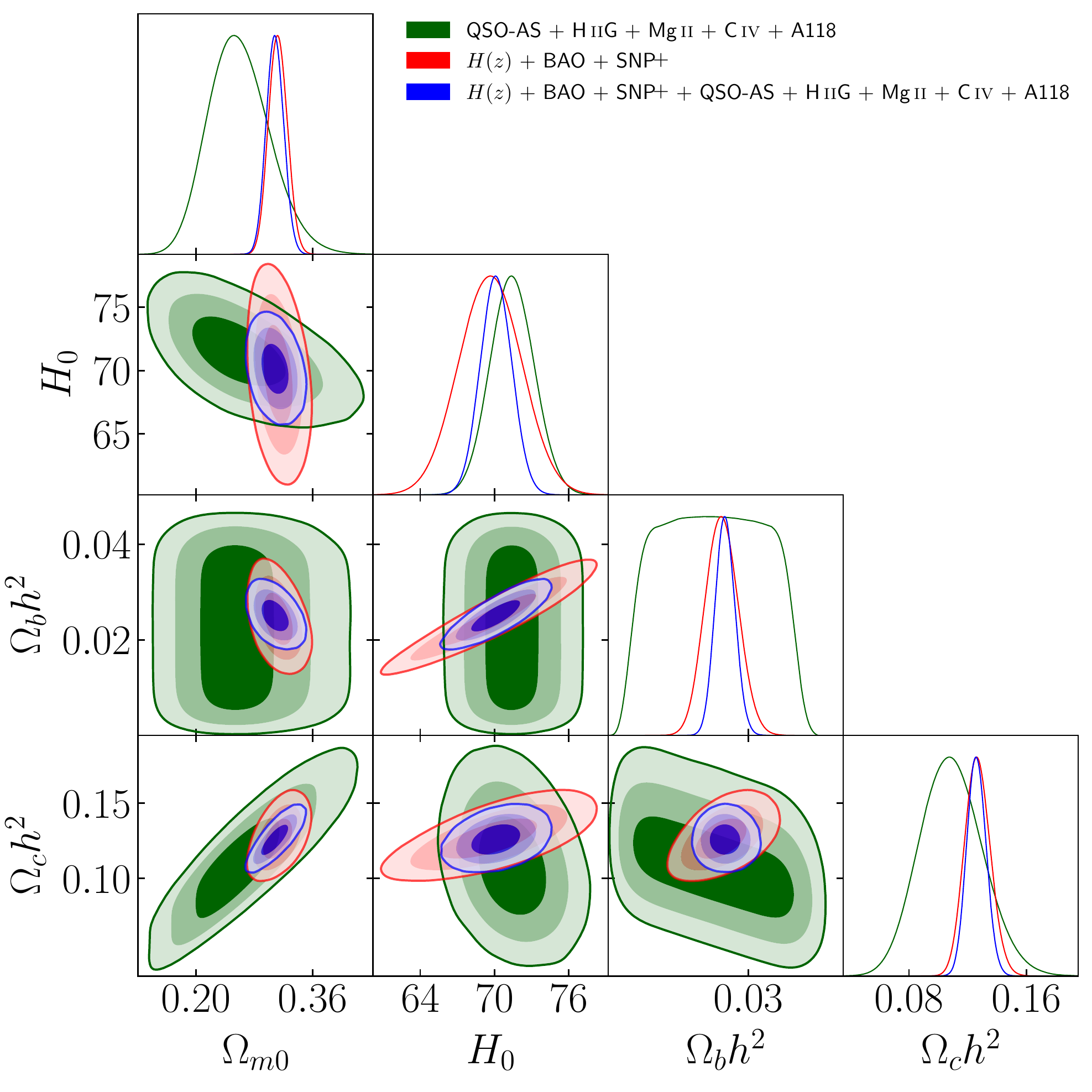}}
 \subfloat[]{%
    \includegraphics[width=0.45\textwidth,height=0.35\textwidth]{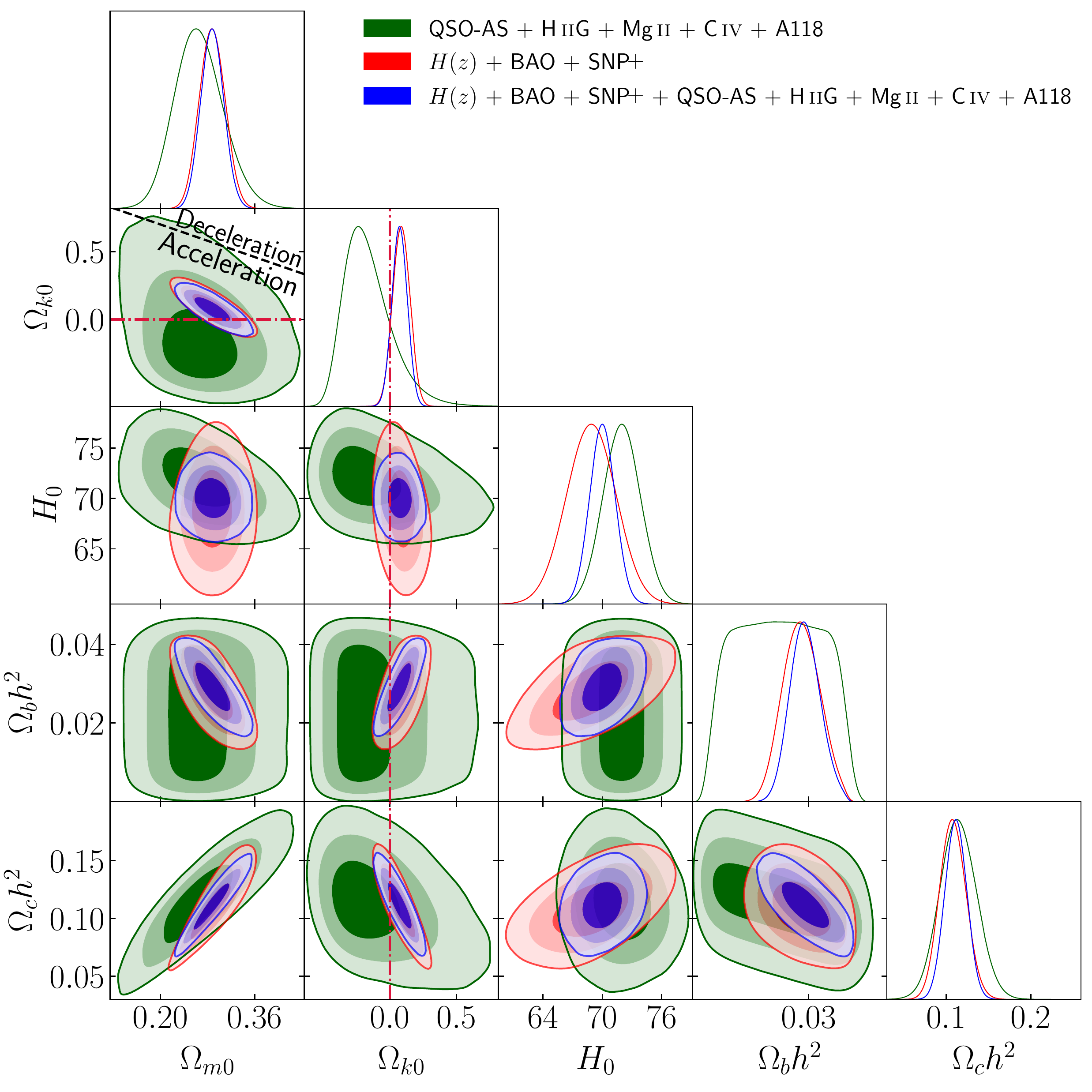}}\\
 \subfloat[]{%
    \includegraphics[width=0.45\textwidth,height=0.35\textwidth]{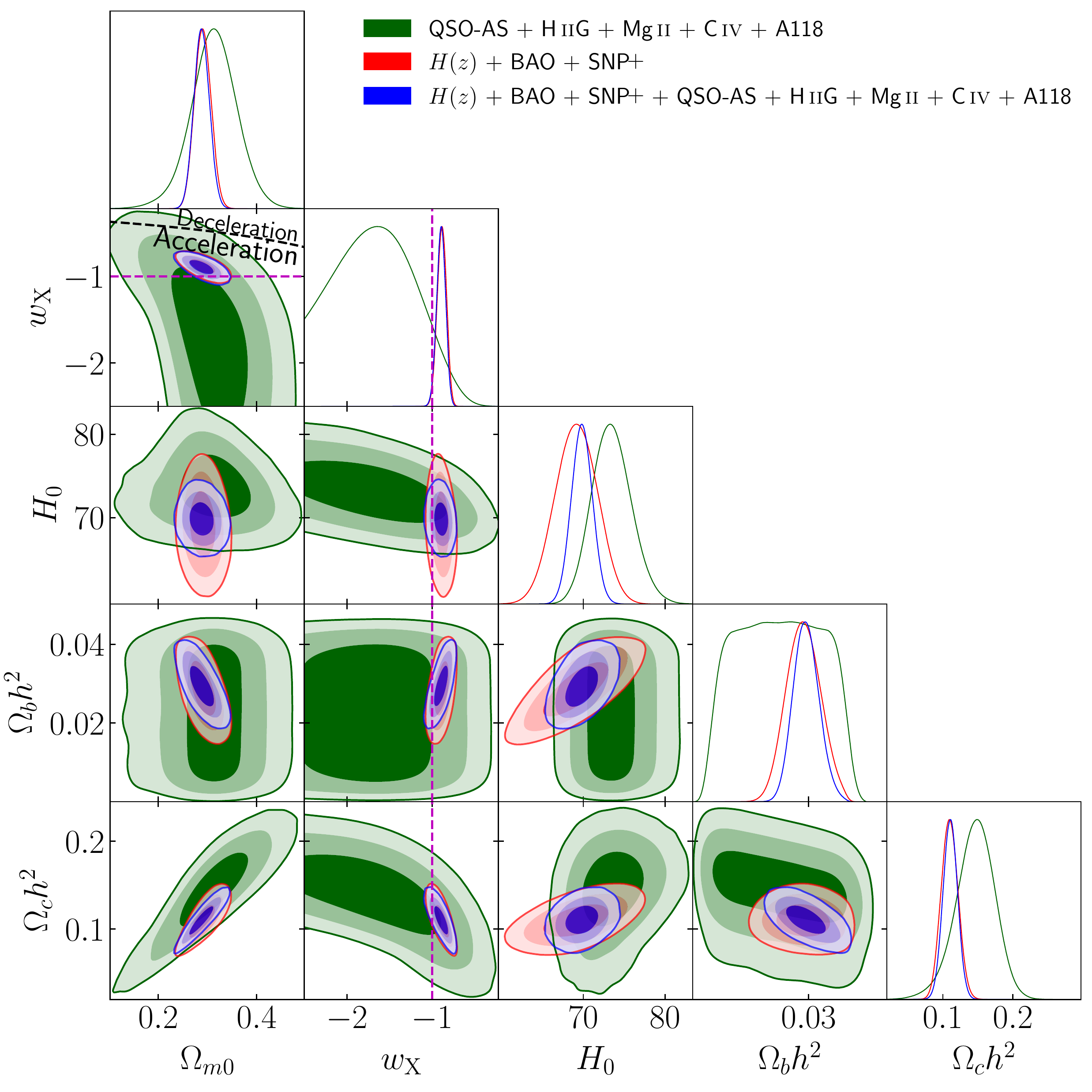}}
 \subfloat[]{%
    \includegraphics[width=0.45\textwidth,height=0.35\textwidth]{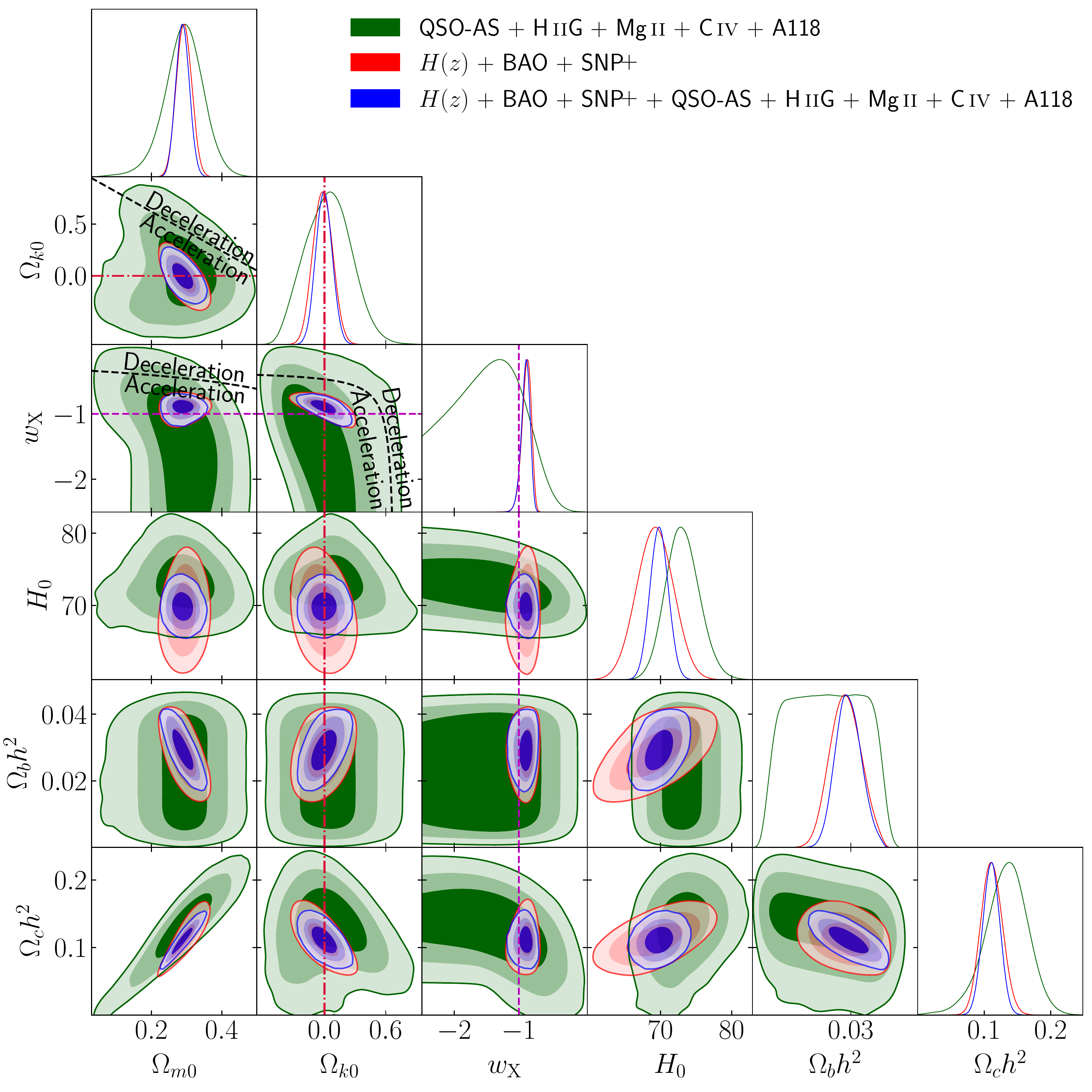}}\\
 \subfloat[]{%
    \includegraphics[width=0.45\textwidth,height=0.35\textwidth]{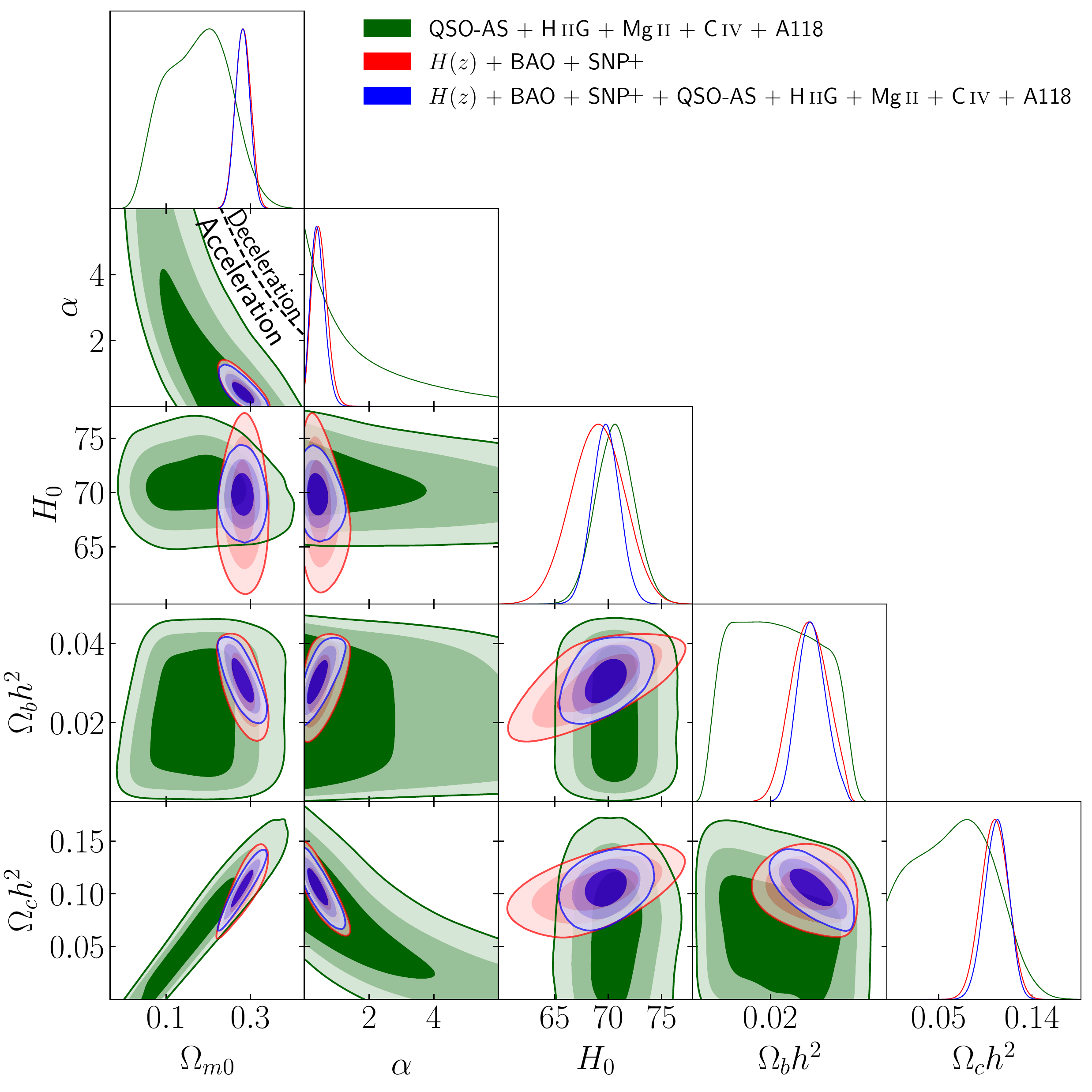}}
 \subfloat[]{%
    \includegraphics[width=0.45\textwidth,height=0.35\textwidth]{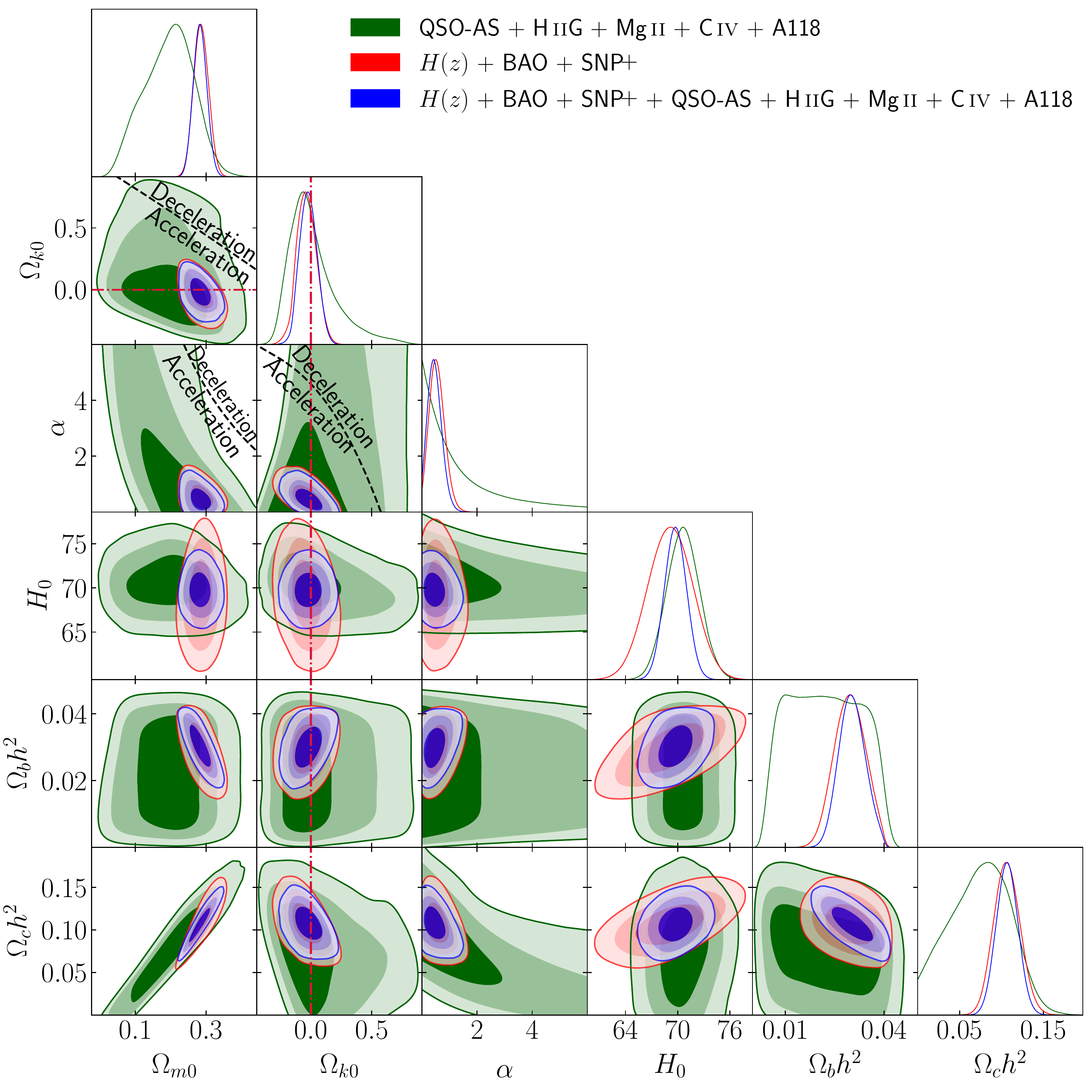}}\\
\caption{Same as Fig.\ \ref{fig1} but for QSO-AS + \hiig\ + \mii\ + \civ\ + A118 (green), $H(z)$ + BAO + SNP\plus\ (red), and $H(z)$ + BAO + SNP\plus\ + QSO-AS + \hiig\ + \mii\ + \civ\ + A118  (blue) data. The black dashed zero-acceleration lines in panels (b)--(f), computed for the third cosmological parameter set to the $H(z)$ + BAO data best-fitting values listed in Table \ref{tab:BFP} in panels (d) and (f), divides the parameter space into regions associated with currently-accelerating (below or below left) and currently-decelerating (above or above right) cosmological expansion.}
\label{fig7}
\end{figure*}

\begin{figure*}
\centering
 \subfloat[]{%
    \includegraphics[width=0.45\textwidth,height=0.35\textwidth]{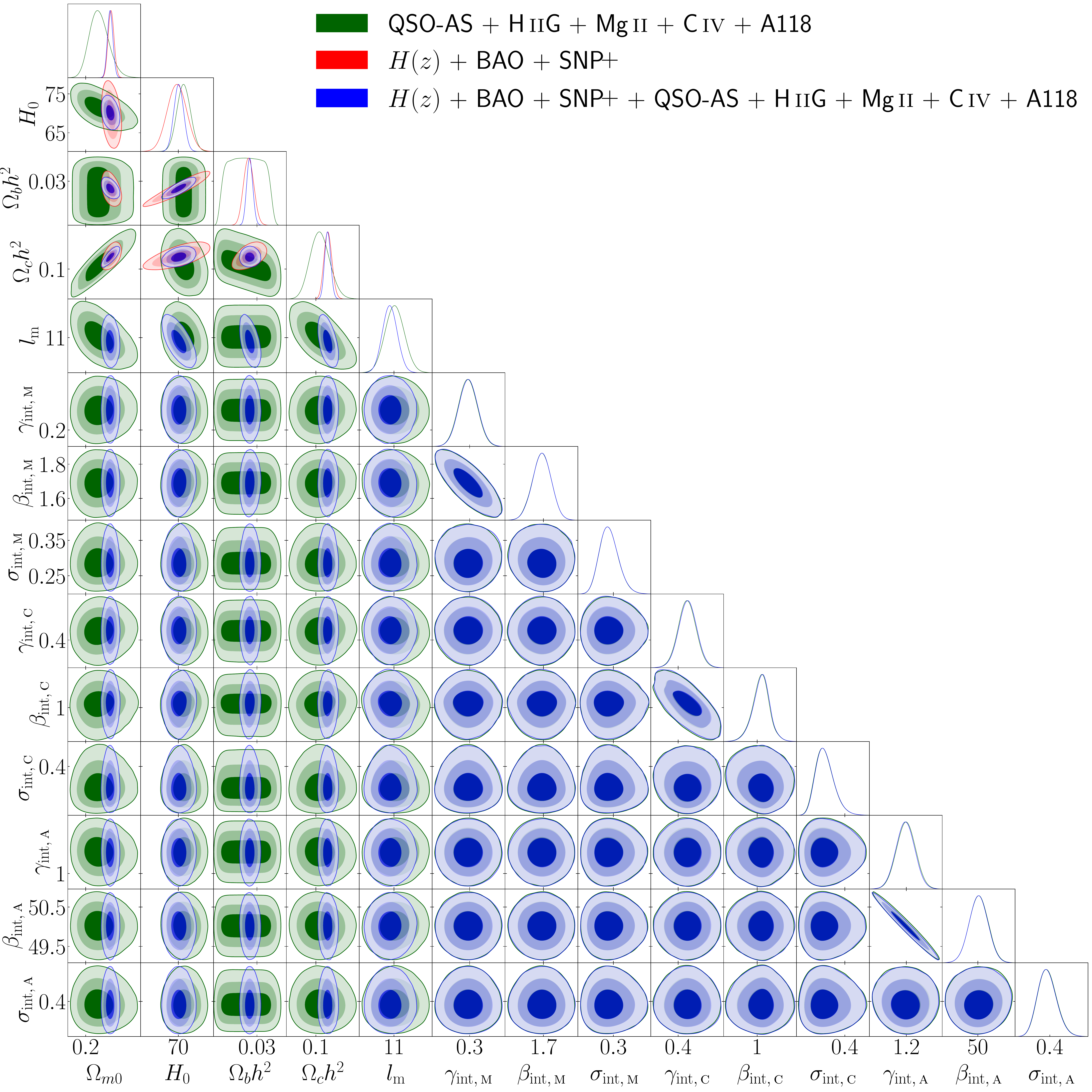}}
 \subfloat[]{%
    \includegraphics[width=0.45\textwidth,height=0.35\textwidth]{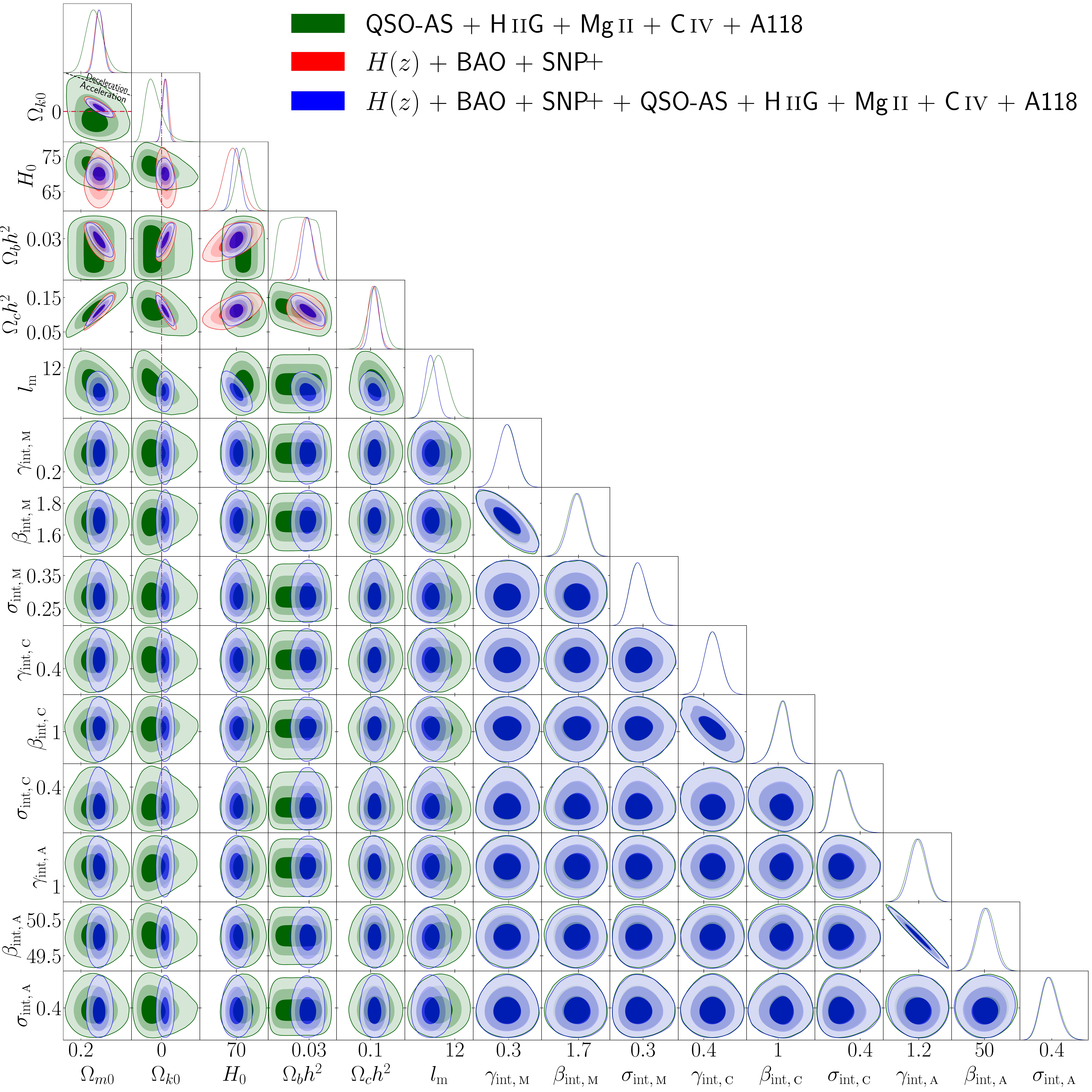}}\\
 \subfloat[]{%
    \includegraphics[width=0.45\textwidth,height=0.35\textwidth]{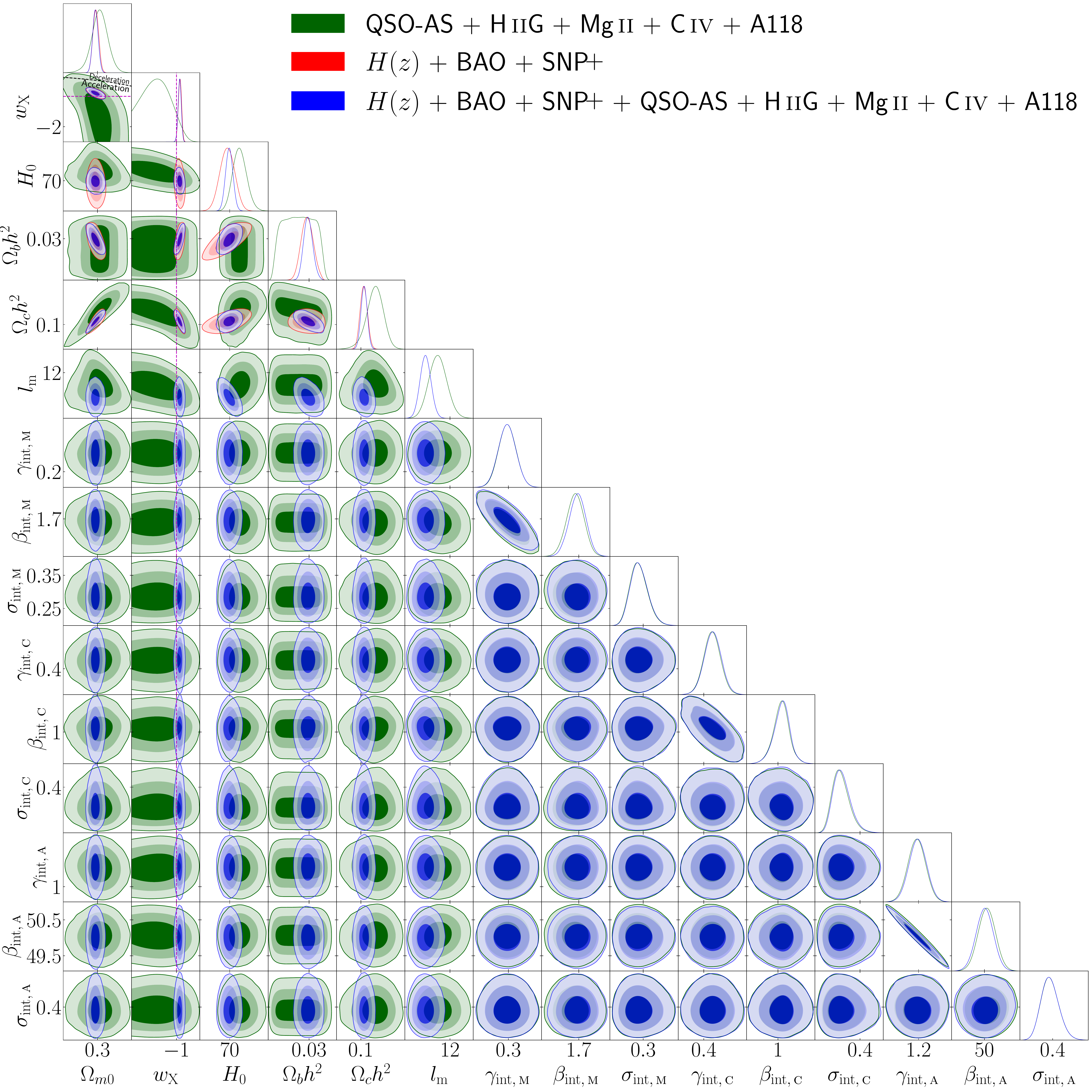}}
 \subfloat[]{%
    \includegraphics[width=0.45\textwidth,height=0.35\textwidth]{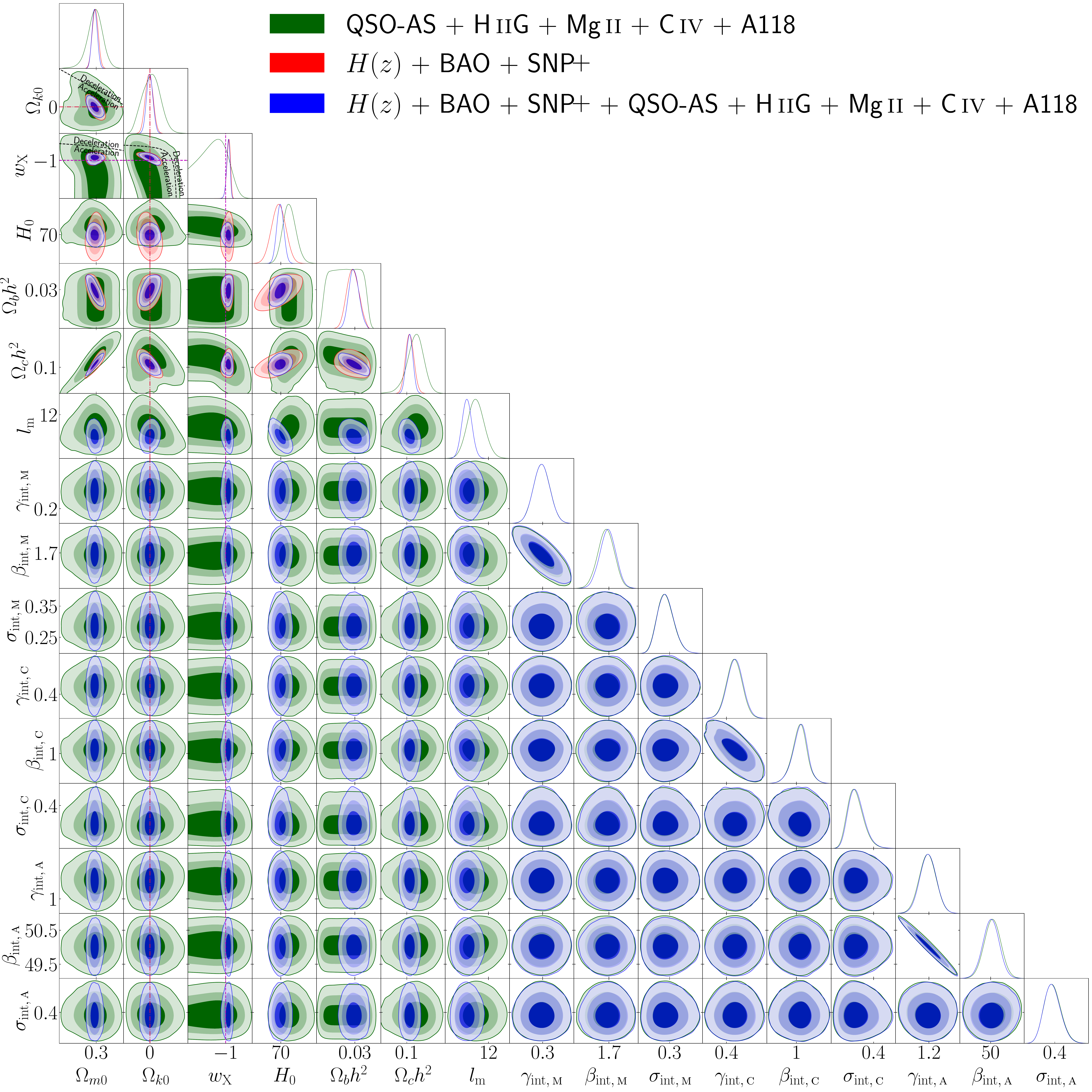}}\\
 \subfloat[]{%
    \includegraphics[width=0.45\textwidth,height=0.35\textwidth]{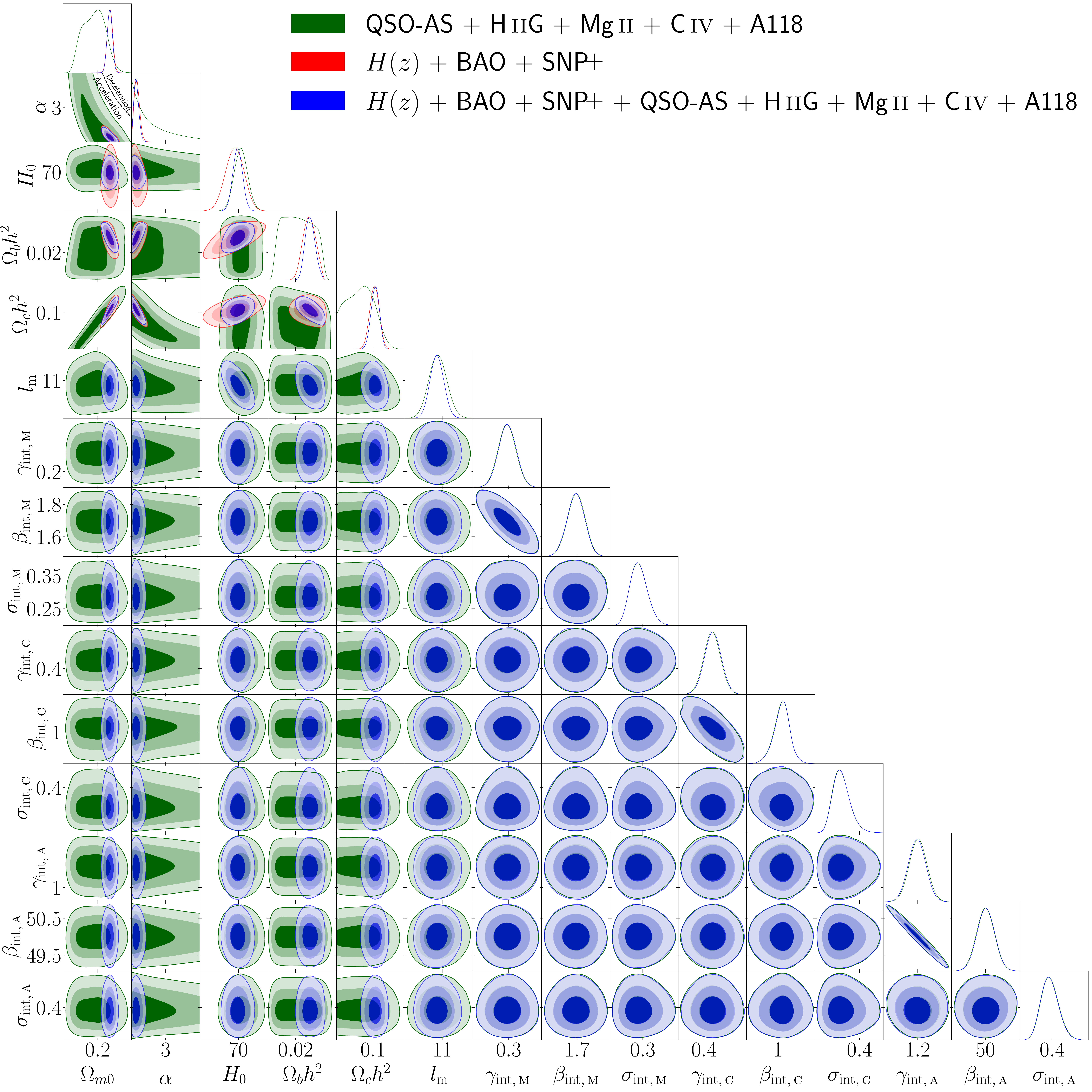}}
 \subfloat[]{%
    \includegraphics[width=0.45\textwidth,height=0.35\textwidth]{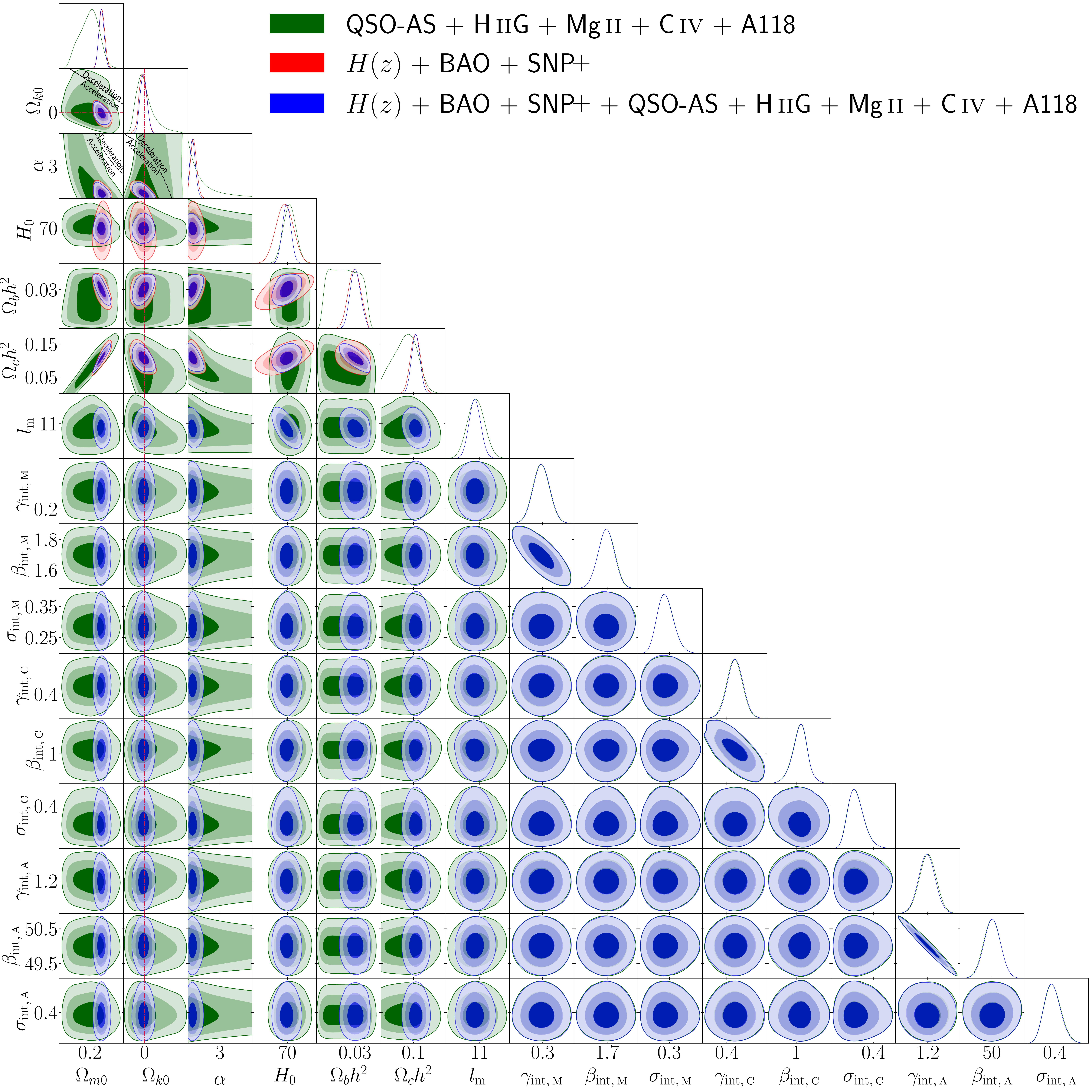}}\\
\caption{Same as Fig.\ \ref{fig7} but including non-cosmological parameters.}
\label{fig8}
\end{figure*}

\begin{figure*}
\centering
 \subfloat[]{%
    \includegraphics[width=0.45\textwidth,height=0.35\textwidth]{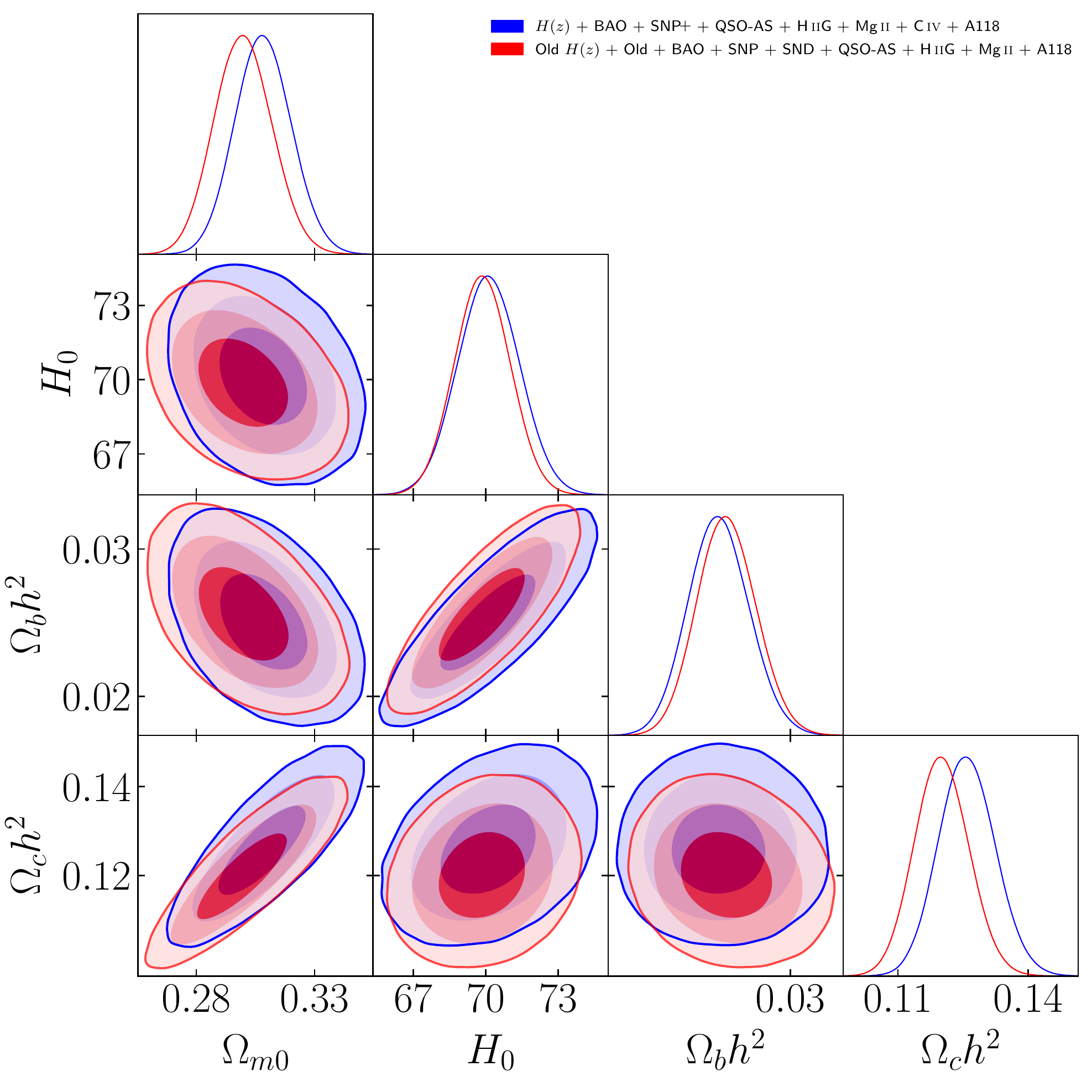}}
 \subfloat[]{%
    \includegraphics[width=0.45\textwidth,height=0.35\textwidth]{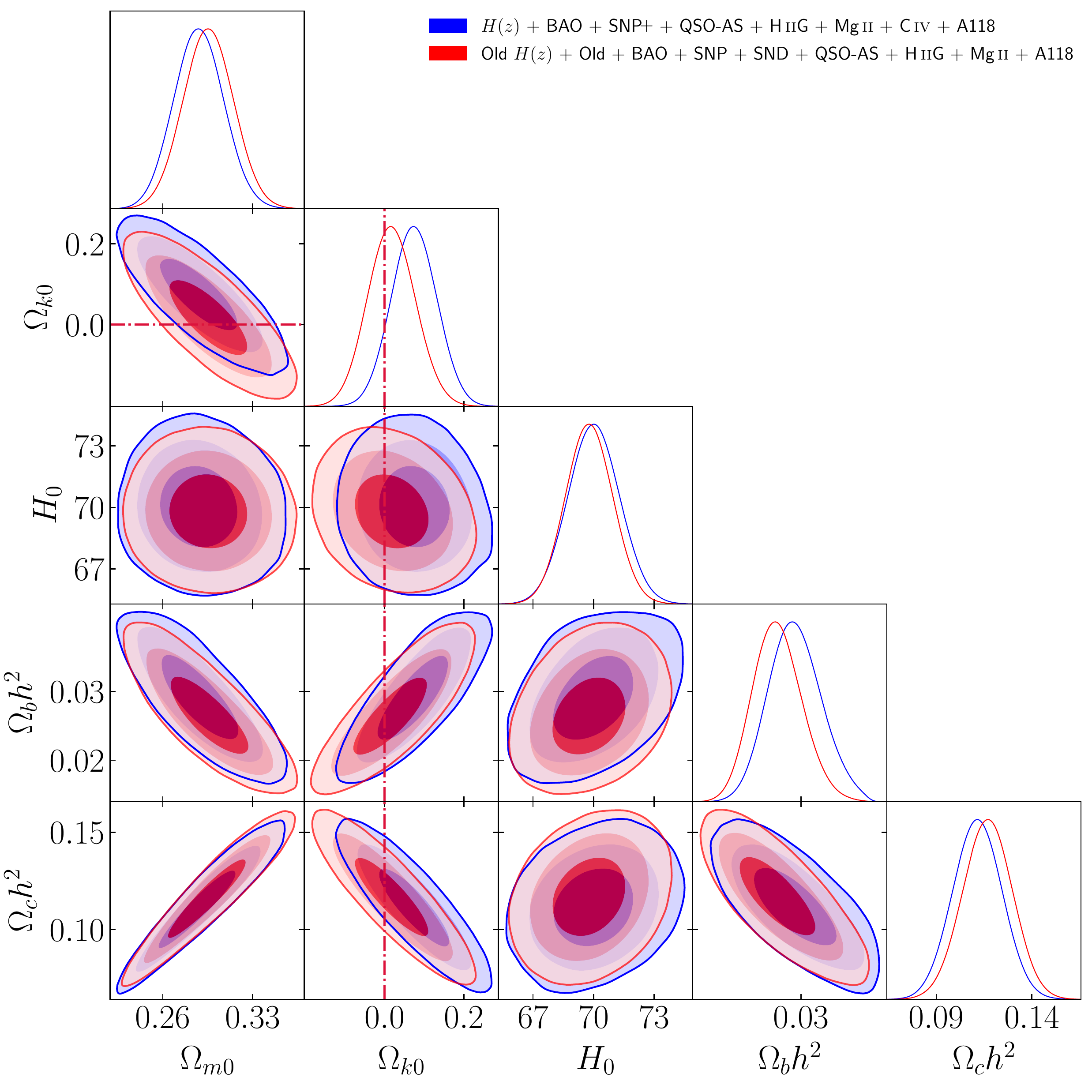}}\\
 \subfloat[]{%
    \includegraphics[width=0.45\textwidth,height=0.35\textwidth]{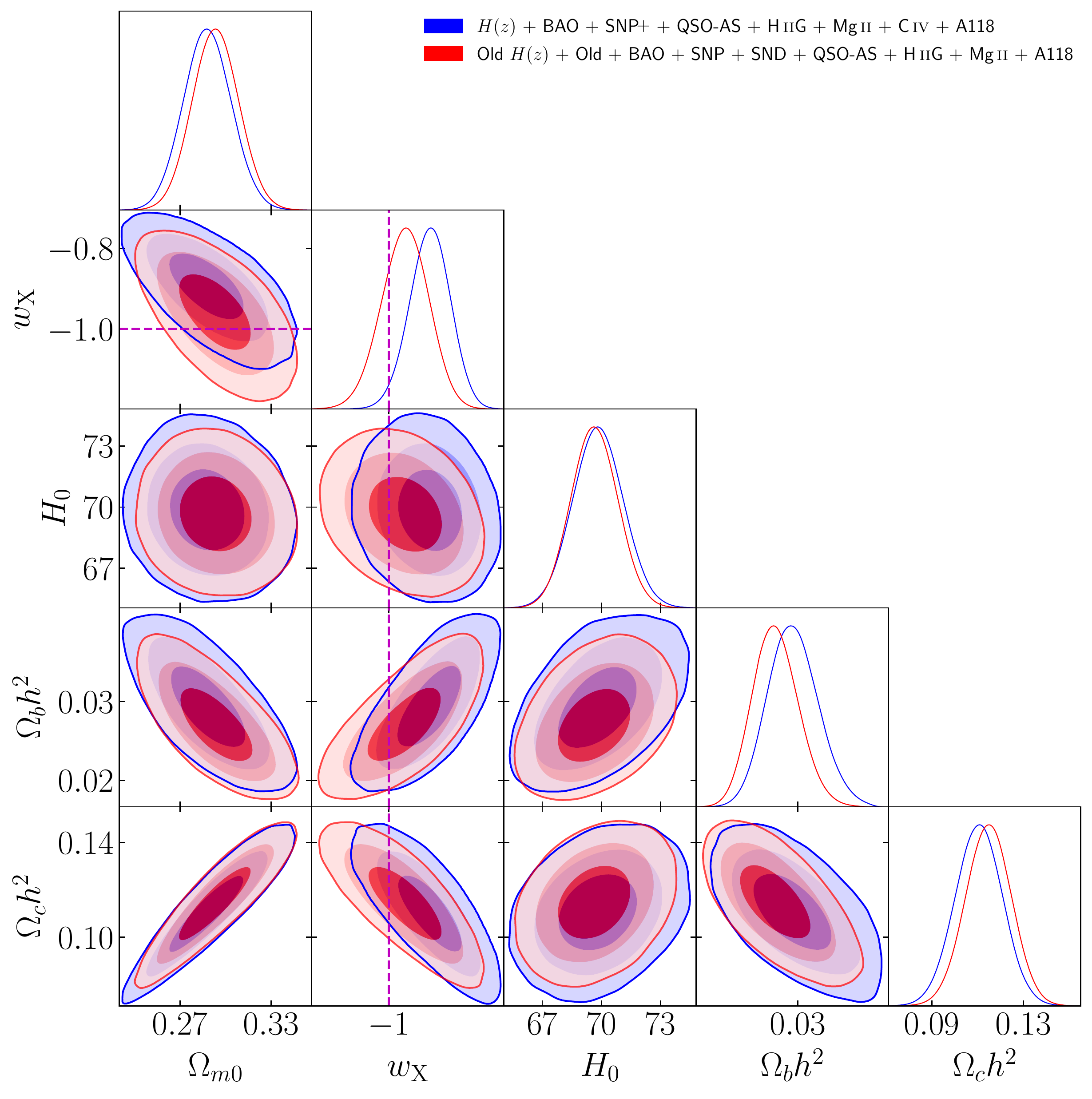}}
 \subfloat[]{%
    \includegraphics[width=0.45\textwidth,height=0.35\textwidth]{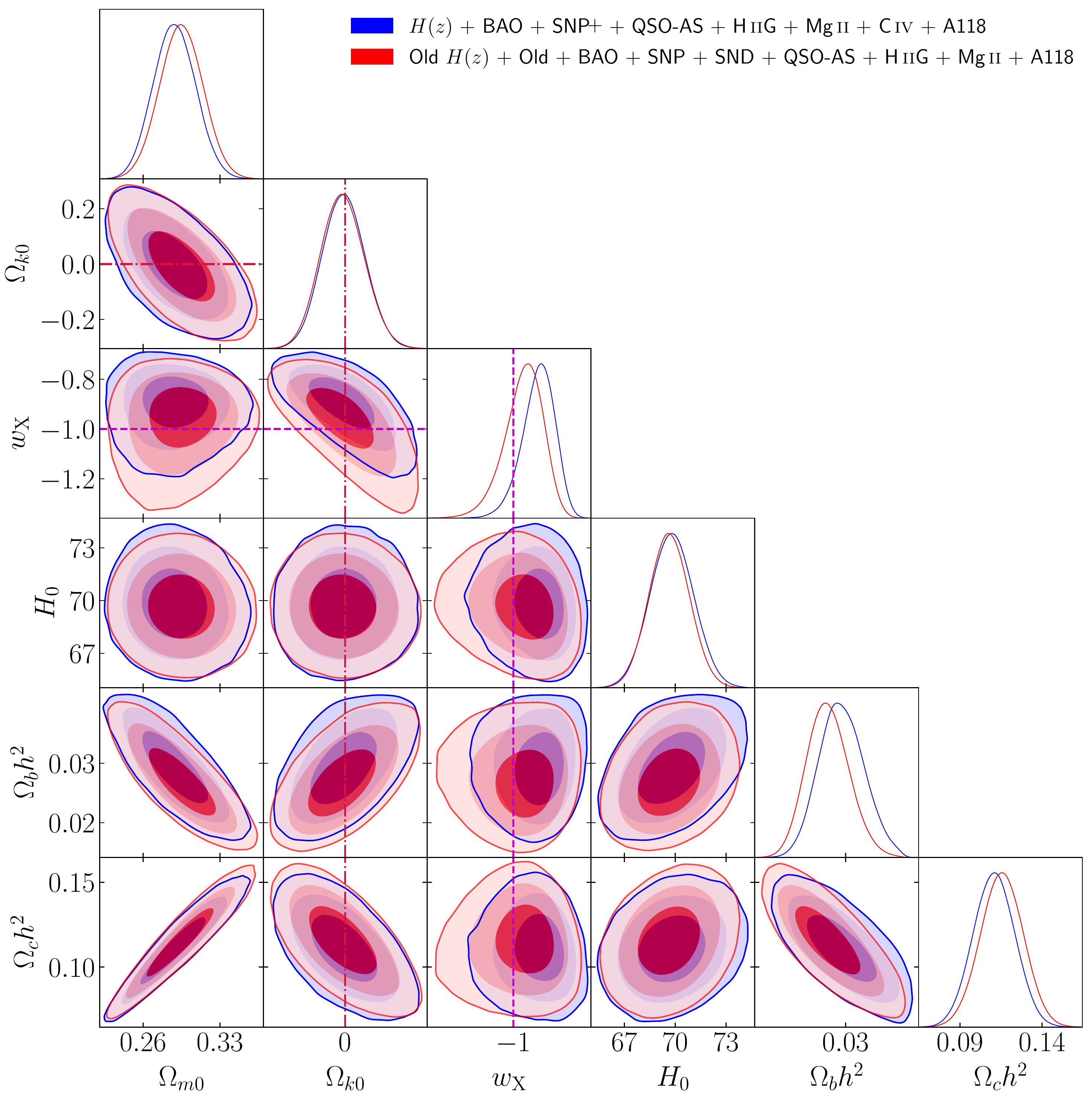}}\\
 \subfloat[]{%
    \includegraphics[width=0.45\textwidth,height=0.35\textwidth]{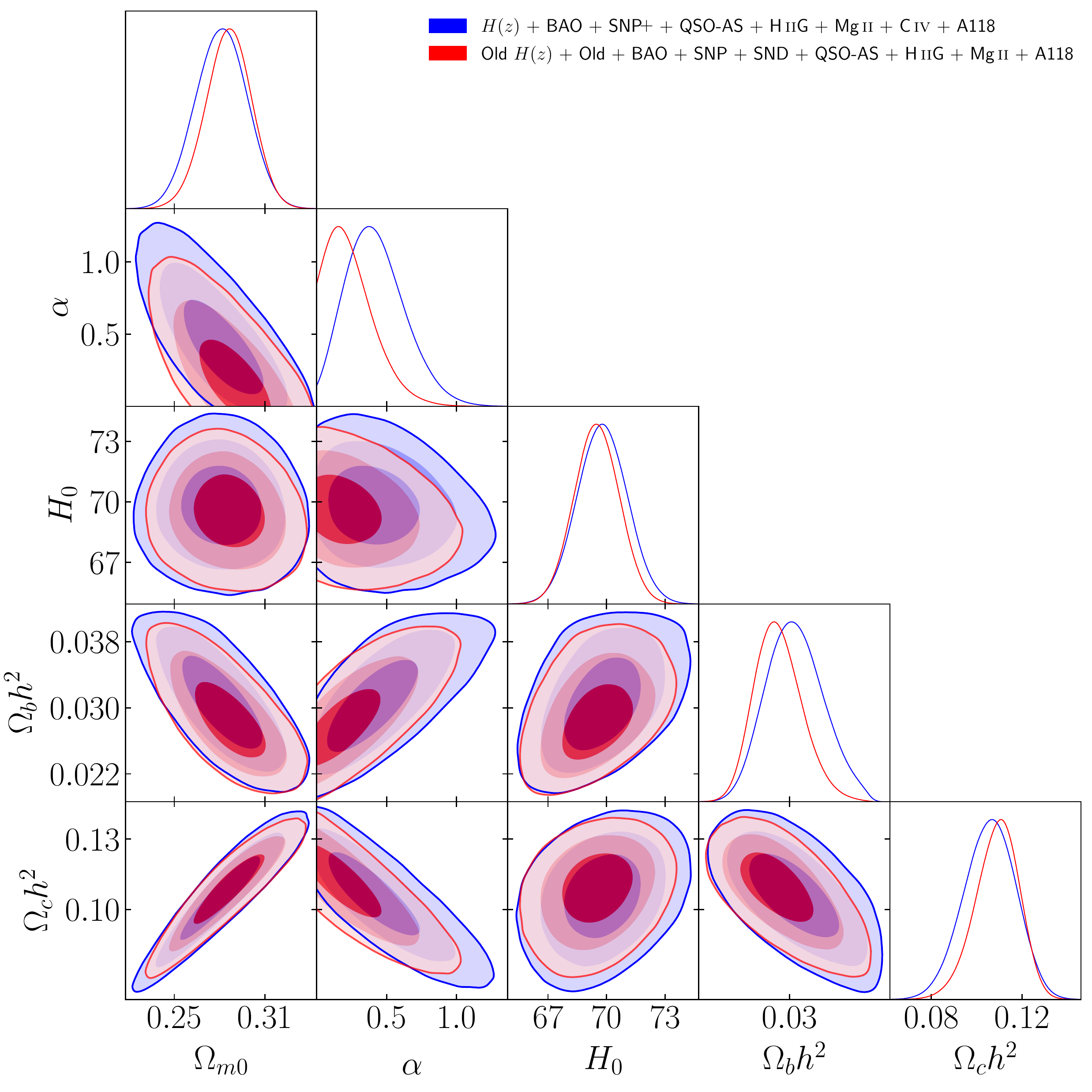}}
 \subfloat[]{%
    \includegraphics[width=0.45\textwidth,height=0.35\textwidth]{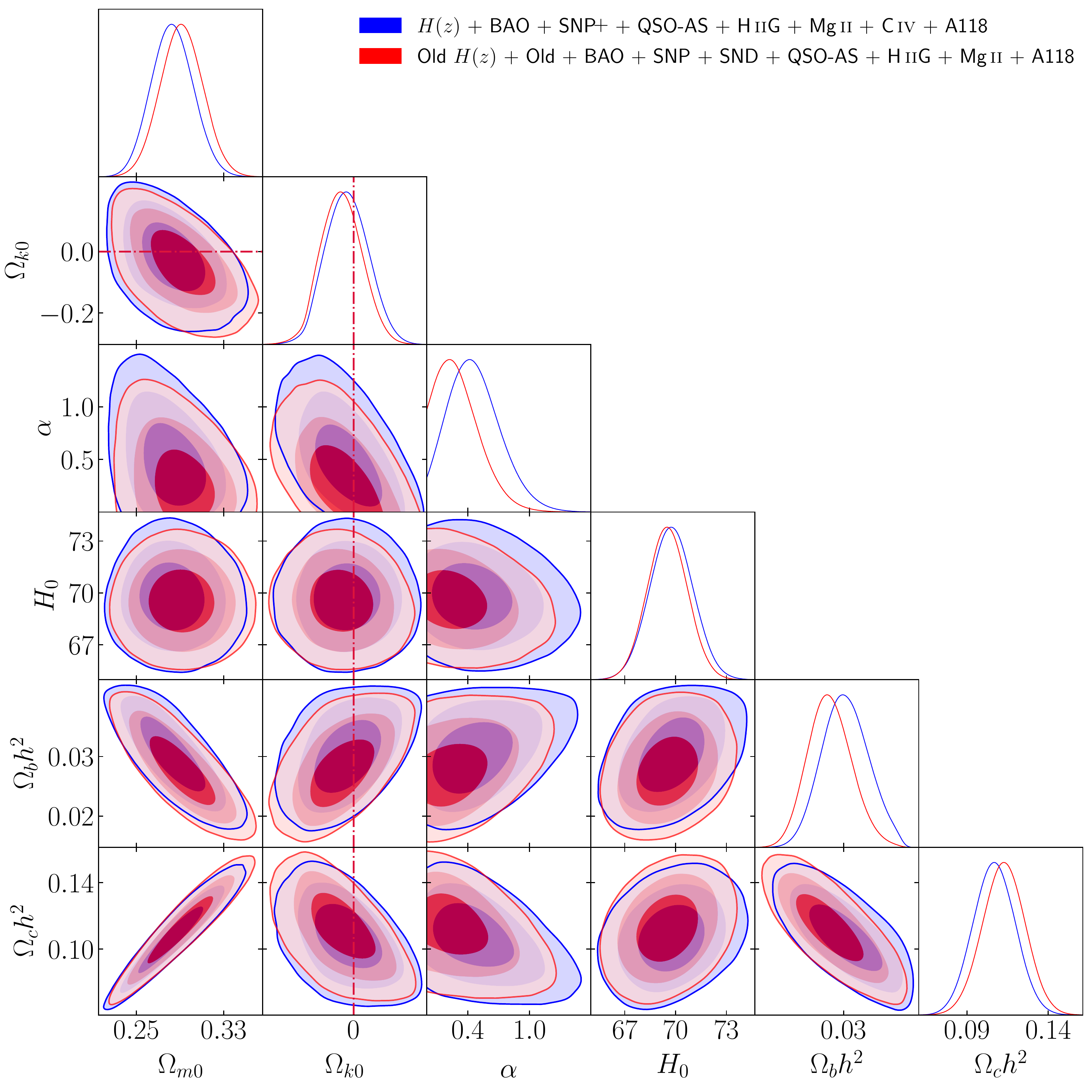}}\\
\caption{Same as Fig.\ \ref{fig1} but for Old $H(z)$ + Old BAO + QSO-AS + \hiig\ + \mii\ + A118 (red) and $H(z)$ + BAO + SNP\plus\ + QSO-AS + \hiig\ + \mii\ + \civ\ + A118 (blue) data.}
\label{fig9}
\end{figure*}

\begin{figure*}
\centering
 \subfloat[]{%
    \includegraphics[width=0.45\textwidth,height=0.35\textwidth]{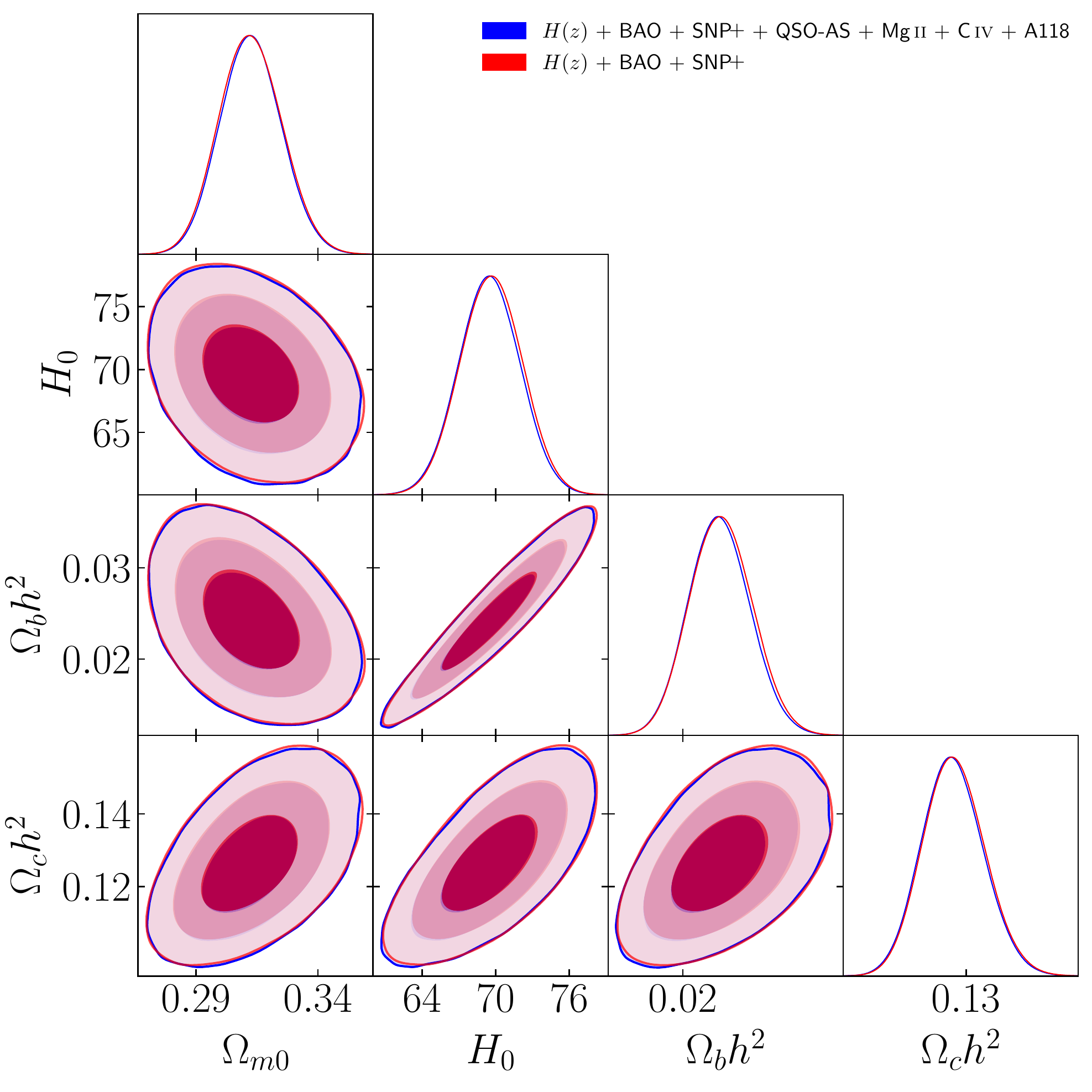}}
 \subfloat[]{%
    \includegraphics[width=0.45\textwidth,height=0.35\textwidth]{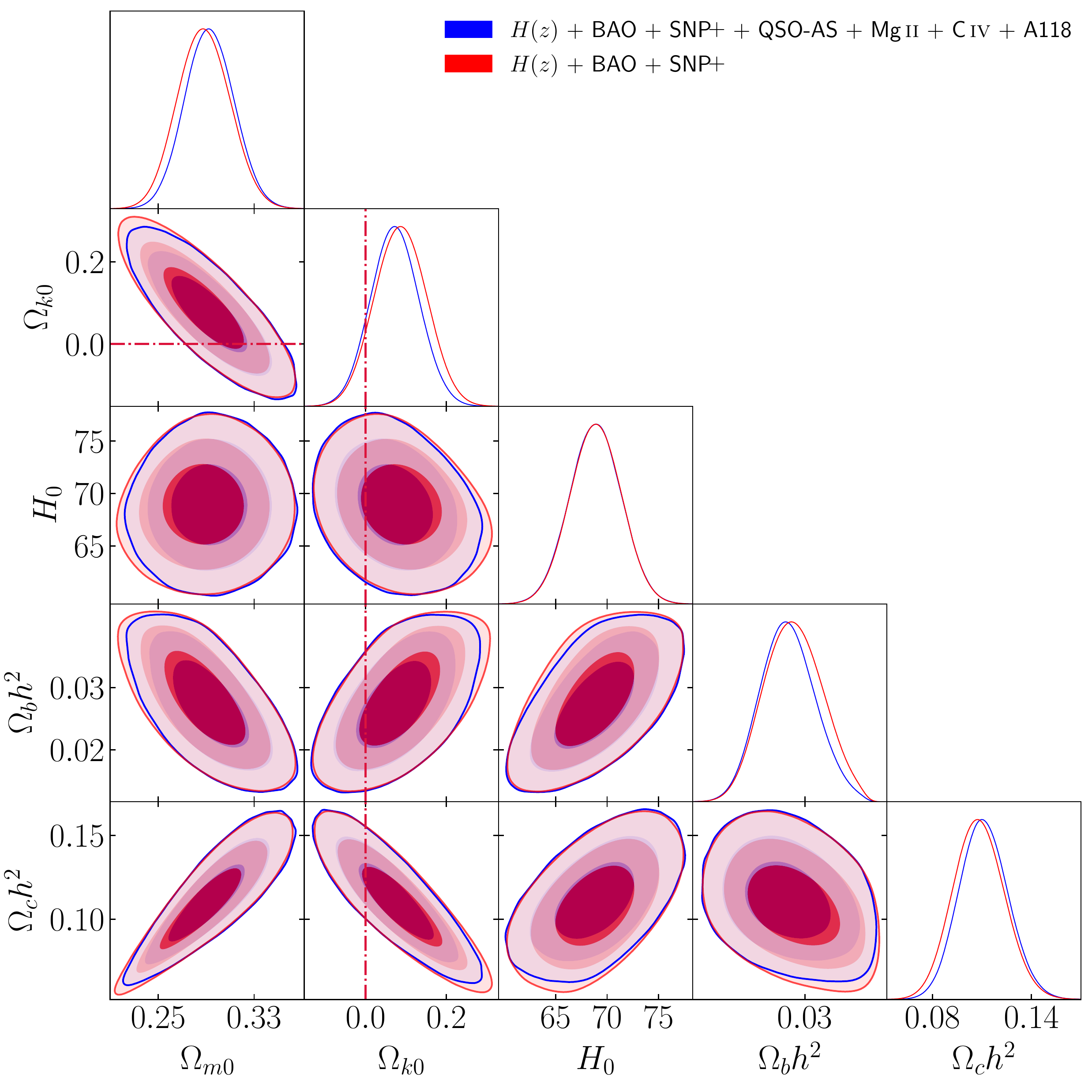}}\\
 \subfloat[]{%
    \includegraphics[width=0.45\textwidth,height=0.35\textwidth]{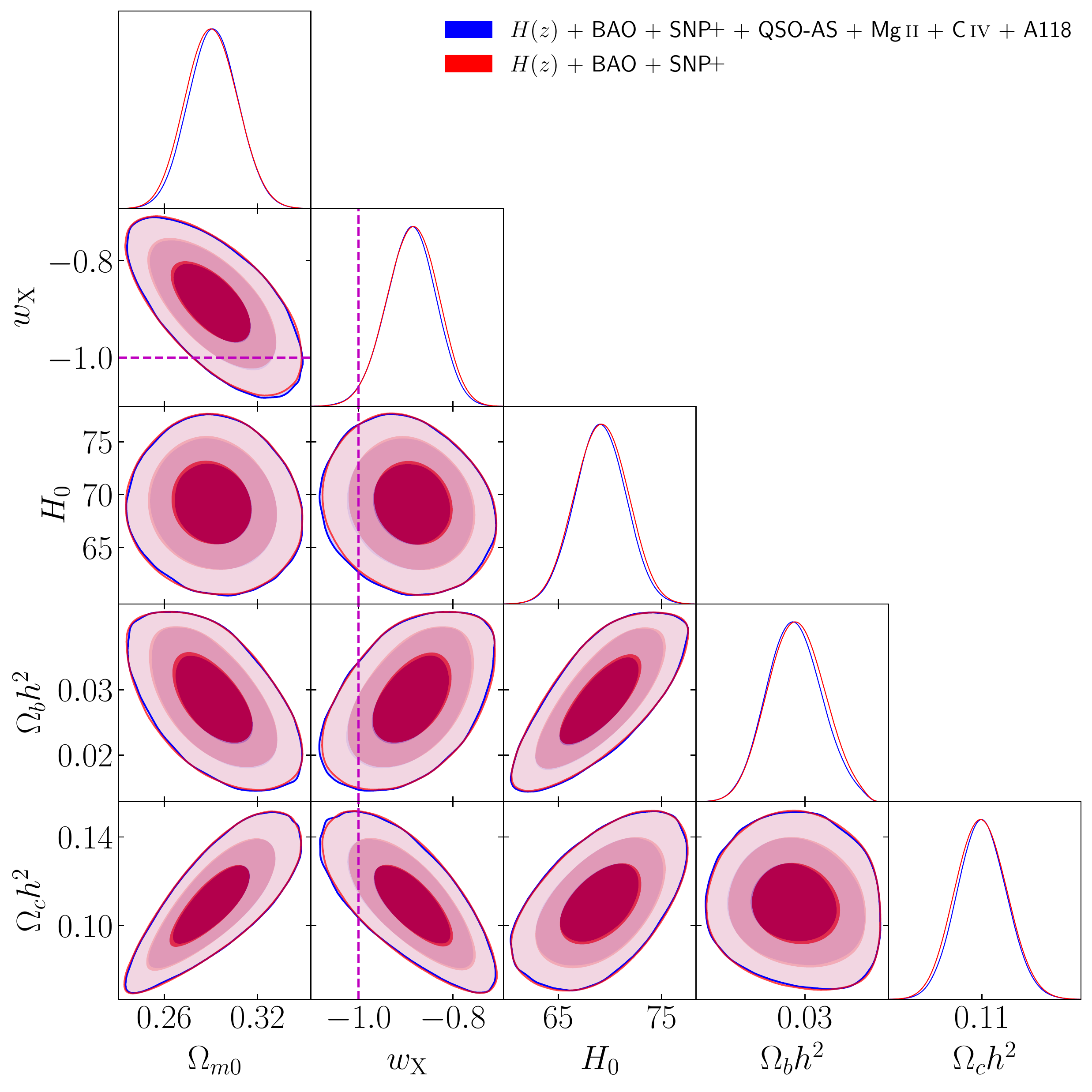}}
 \subfloat[]{%
    \includegraphics[width=0.45\textwidth,height=0.35\textwidth]{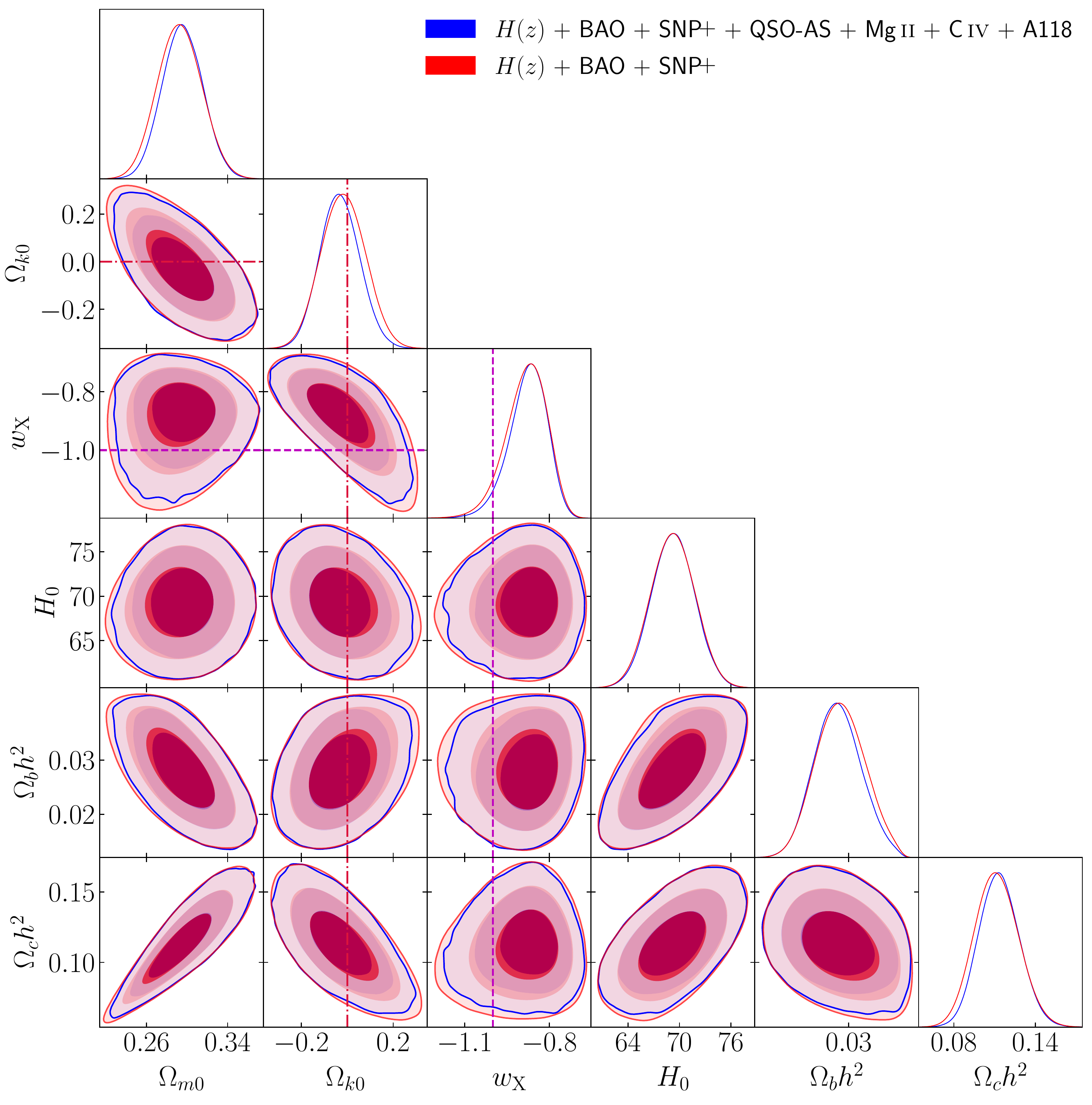}}\\
 \subfloat[]{%
    \includegraphics[width=0.45\textwidth,height=0.35\textwidth]{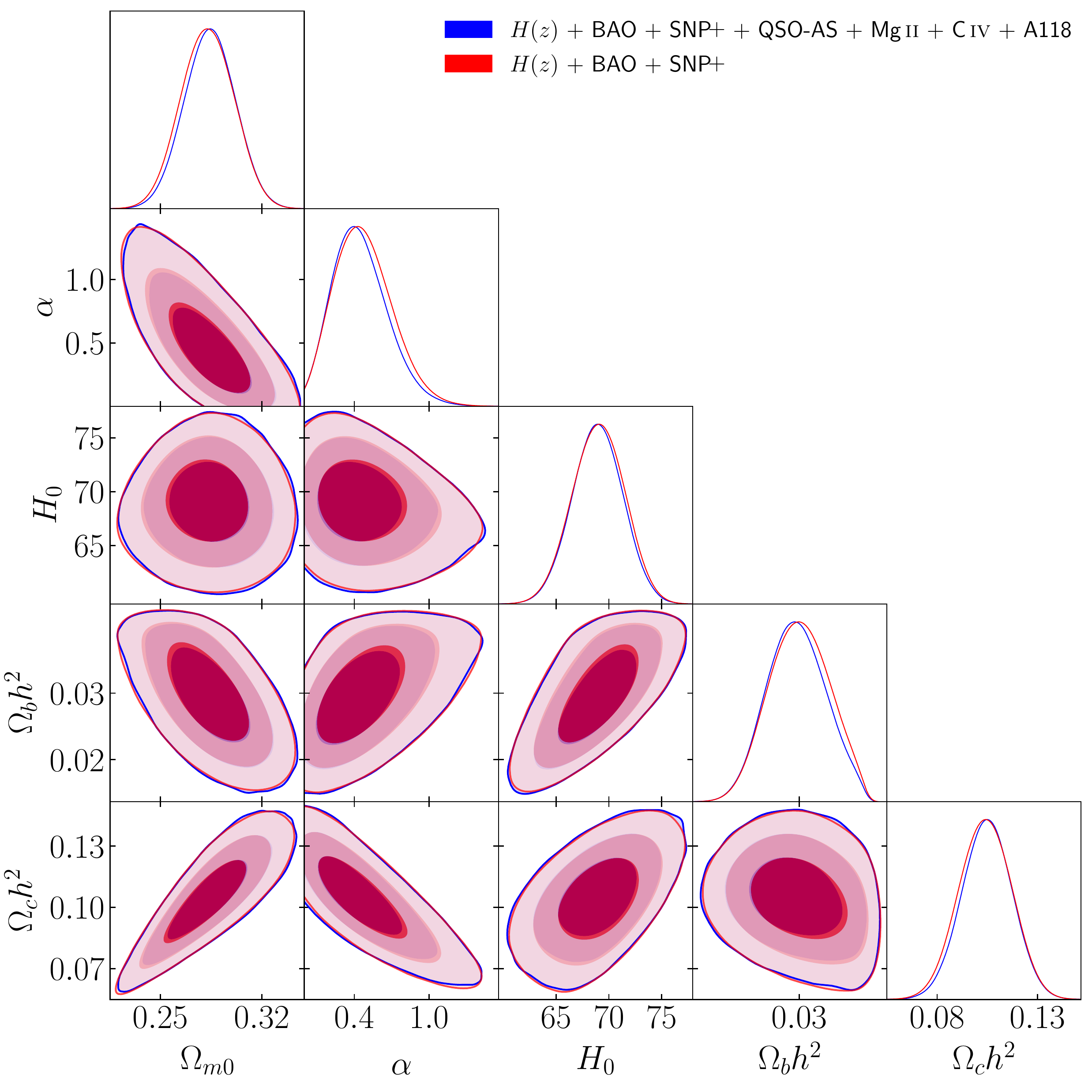}}
 \subfloat[]{%
    \includegraphics[width=0.45\textwidth,height=0.35\textwidth]{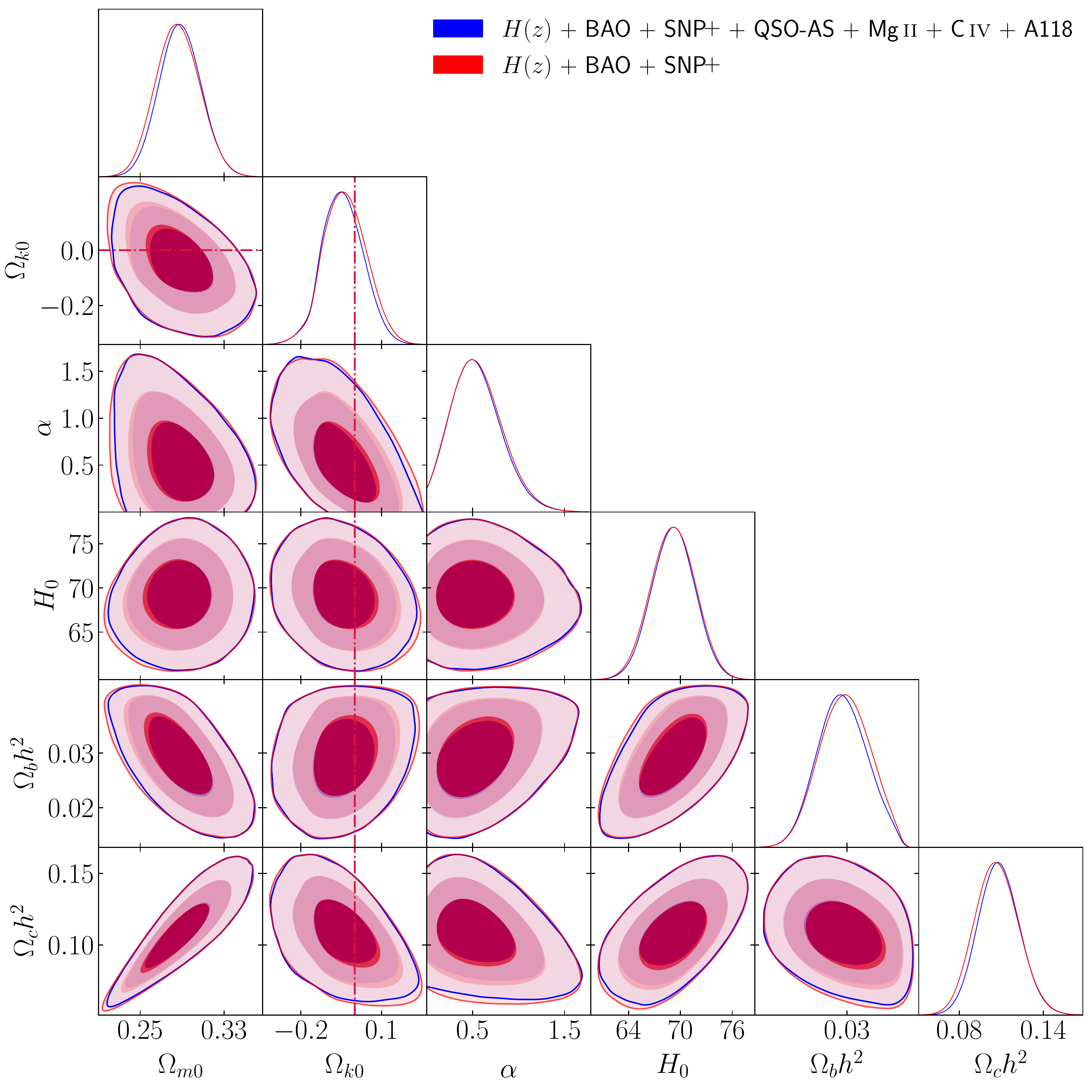}}\\
\caption{Same as Fig.\ \ref{fig1} but for $H(z)$ + BAO + SNP\plus\ + QSO-AS + \mii\ + \civ\ + A118 (blue) and $H(z)$ + BAO + SNP\plus\ (red) data.}
\label{fig10}
\end{figure*}

For $H(z)$ + BAO + SNP\plus\ + QSO-AS + \hiig\ + \mii\ + \civ\ + A118 (HzBSNQHMCA) data and Old $H(z)$ + Old BAO + SNP + SND + QSO-AS + \hiig\ + \mii\ + A118 (OHzBSNQHMA) data, the best-fitting parameter values, likelihood values, and information criteria values for all models are given in Table \ref{tab:BFPs} and the marginalized posterior mean parameter values and uncertainties for all models are listed in Table \ref{tab:1d_BFPs}. The OHzBSNQHMA data results are from Ref.\ \cite{CaoRatra2022} and are used here to compare to HzBSNQHMCA data constraints. Note that there is a typo in table 5 of Ref.\ \cite{CaoRatra2022}, where in the flat XCDM OHzBSNQHMA case, \obhs\ should be $0.1147^{+0.0098}_{-0.0097}$ instead of $0.1449^{+0.0098}_{-0.0097}$. Constraints derived without using \hiig\ data are discussed in the Appendix.

Figures \ref{fig7} and \ref{fig8} show that $H(z)$ + BAO + SNP\plus\ and QSO-AS + \hiig\ + \mii\ + \civ\ + A118 data constraints are mutually consistent so these data can be used together to more restrictively constrain cosmological parameter values. Figure \ref{fig9} compares the probability distributions and confidence regions of cosmological parameters, obtained from HzBSNQHMCA and OHzBSNQHMA data.

HzBSNQHMCA data constraints on \om\ range from $0.281\pm0.017$ (flat \pcdm) to $0.308\pm0.012$ (flat \lcdm), with a difference of $1.3\sigma$. This difference is somewhat larger than the $0.73\sigma$ difference for the OHzBSNQHMA data set, \cite{CaoRatra2022}.

HzBSNQHMCA data constraints on \obhs\ range from $0.0250\pm0.0021$ (flat \lcdm) to $0.0307^{+0.0032}_{-0.0038}$ (flat \pcdm), with a difference of $1.3\sigma$. The constraints on \ochs\ range from $0.1057^{+0.0118}_{-0.0107}$ (flat \pcdm) to $0.1260\pm0.0064$ (flat \lcdm), with a difference of $1.5\sigma$.

HzBSNQHMCA data constraints on $H_0$ range from $69.76\pm1.25$ \hunit\ (non-flat \pcdm) to $70.13\pm1.25$ \hunit\ (flat \lcdm), with a difference of $0.21\sigma$. This difference is slightly smaller than the $0.23\sigma$ difference for the OHzBSNQHMA data set, \cite{CaoRatra2022}. These $H_0$ values are $0.57\sigma$ (non-flat \pcdm) and $0.69\sigma$ (flat \lcdm) higher than the median statistics estimate of $H_0=68\pm2.8$ \hunit\ \citep{chenratmed}, $0.02\sigma$ (non-flat \pcdm) lower and $0.16\sigma$ (flat \lcdm) higher than the TRGB and SN Ia estimate of $H_0=69.8\pm1.7$ \hunit\ \citep{Freedman2021}, and $2.0\sigma$ (non-flat \pcdm) and $1.8\sigma$ (flat \lcdm) lower than the Cepheids and SN Ia measurement of $73.04\pm1.04$ \hunit\ \cite{Riessetal2022}. The $H_0$ constraints from flat \lcdm\ is $2.0\sigma$ higher than the $H_0$ estimate of $67.36 \pm 0.54$ \hunit\ from \textit{Planck} 2018 TT,TE,EE+lowE+lensing CMB anisotropy data \cite{planck2018b}.

HzBSNQHMCA data constraints on \ok\ are $0.074\pm0.056$, $-0.004\pm0.078$, and $-0.021^{+0.067}_{-0.074}$ for non-flat \lcdm, XCDM, and \pcdm, respectively. Non-flat \lcdm\ favors open spatial geometry, being $1.3\sigma$ away from flat, while closed spatial geometry is favored by non-flat XCDM and non-flat \pcdm, with flatness within 1$\sigma$. It is interesting that for the non-CMB data compilation of Ref.\ \cite{deCruzPerezetal2022} the two non-flat \lcdm\ models, with two different primordial power spectra, favor closed geometry at $\sim 0.7\sigma$, but that non-CMB data compilation includes growth factor measurements and so those constraints also depend on the primordial power spectrum assumed, unlike the constraints we derive here.

HzBSNQHMCA data indicate a strong preference for dark energy dynamics. For flat (non-flat) XCDM ($1\sigma$ and $2\sigma$), $w_{\rm X}=-0.895\pm0.051{}^{+0.099}_{-0.105}$ ($w_{\rm X}=-0.897^{+0.075}_{-0.055}{}^{+0.127}_{-0.139}$), with central values being $2.0\sigma$ ($1.6\sigma$) higher than $w_{\rm X}=-1$ (\lcdm). For flat (non-flat) \pcdm\ ($1\sigma$ and $2\sigma$), $\alpha=0.423^{+0.168}_{-0.246}{}^{+0.405}_{-0.393}$ ($\alpha=0.468^{+0.200}_{-0.292}{}^{+0.454}$), with central values being $>2\sigma$ ($1.6\sigma$) away from $\alpha=0$ (\lcdm).

HzBSNQHMCA data constraints are in good agreement with OHzBSNQHMA data constraints. Specifically, the \obhs\ constraints from the former are between $0.21\sigma$ lower (flat \lcdm) and $0.52\sigma$ higher (flat XCDM); the \ochs\ constraints from the former are between $0.32\sigma$ lower (non-flat \pcdm) and $0.67\sigma$ higher (flat \lcdm); the \om\ constraints from the former are between $0.30\sigma$ lower (non-flat \pcdm) and $0.47\sigma$ higher (flat \lcdm); and the $H_0$ constraints from the former are between $0.093\sigma$ higher (non-flat XCDM) and $0.17\sigma$ higher (flat \pcdm).

In non-flat \lcdm, XCDM, and \pcdm, the \ok\ constraints from HzBSNQHMCA data are $0.69\sigma$, $0.046\sigma$, and $0.19\sigma$ higher than those from OHzBSNQHMA data, respectively. In both data sets, open spatial geometry is favored by non-flat \lcdm, and closed spatial geometry is favored by non-flat XCDM and non-flat \pcdm.

In flat and non-flat XCDM and \pcdm, the \wx\ and $\alpha$ constraints from HzBSNQHMCA data are $0.82\sigma$, $0.59\sigma$, $0.68\sigma$, and $0.49\sigma$ higher than those from OHzBSNQHMA data, respectively. The former, new, data indicate stronger evidence for dark energy dynamics than do the latter, old data of Ref.\ \cite{CaoRatra2022}.

Overall, the changes in constraints found here, using updated and more data and improved analyses, compared to those found in Ref.\  \cite{CaoRatra2022}, are not large. This lends hope to the belief that the constraints we have derived here are more than somewhat reliable. We note however some trends from the above numerical values of the differences, and from Fig.\ \ref{fig9}: In all six models the new HzBSNQHMCA data results here favor slightly larger values of $H_0$; in the four dynamical dark energy models they favor slightly more dark energy dynamics; and in five of the six models they favor slightly larger \obhs\ and slightly smaller \ochs\ and \om\ values, with the exception being the opposite behavior of the flat \lcdm\ model.  

\subsection{Model Comparison}
 \label{subsec:comp}

From the AIC, BIC, and DIC values listed in Tables \ref{tab:BFP} and \ref{tab:BFPs}, 
we find the following results:
\begin{itemize}
    \item {\bf AIC.} Flat \lcdm\ is favored the most by $H(z)$, SNP\plus, A118, and QHMCA data; flat \pcdm\ is favored the most by $H(z)$ + BAO and $H(z)$ + BAO + SNP\plus\ data; non-flat XCDM is favored the most by QSO-AS + \hiig\ and \mii\ + \civ\ data; and non-flat XCDM is favored the most by HzBSNQHMCA data.
    
    The evidence against the rest of the models/parametrizations is either only weak or positive, except for the \mii\ + \civ\ data, where the evidence against flat XCDM is positive, against non-flat \lcdm\ is strong, and against other models/parametrizations are very strong.
    
    \item {\bf BIC.} $H(z)$ + BAO data favor flat \pcdm\ the most, \mii\ + \civ\ favor non-flat XCDM the most, and the other data combinations favor flat \lcdm\ the most.

    $H(z)$ + BAO and $H(z)$ + BAO + SNP\plus\ data provide only weak or positive evidence against other models/parametrizations.

    $H(z)$, QSO-AS + \hiig, and A118 data provide positive or strong (non-flat XCDM and non-flat \pcdm) evidence against other models/parametrizations.

    \mii\ + \civ\ data provide positive evidence against non-flat \lcdm\ and flat XCDM, strong evidence against flat \lcdm, and very strong evidence against flat and non-flat \pcdm.

    SNP\plus\ data provide strong or very strong (non-flat XCDM and non-flat \pcdm) evidence against other models/parametrizations.
    
    QHMCA data provide positive evidence against non-flat \lcdm\ and flat XCDM, strong evidence against flat \pcdm, and very strong evidence against non-flat XCDM and non-flat \pcdm.
    
    HzBSNQHMCA data provide positive or very strong (non-flat XCDM and non-flat \pcdm) evidence against other models/parametrizations.

    \item {\bf DIC.} \mii\ + \civ\ data favor flat XCDM the most, $H(z)$ + BAO, $H(z)$ + BAO + SNP\plus, A118, and HzBSNQHMCA data favor flat \pcdm\ the most, and the other data combinations favor flat \lcdm\ the most.
    
    There is strong evidence against non-flat XCDM and non-flat \pcdm\ from QSO-AS + \hiig\ data, strong evidence against non-flat \pcdm\ from QHMCA data, and weak or positive evidence against the others from the remaining data sets.
\end{itemize}

Based on the more reliable DIC values, we conclude that HzBSNQHMCA data do not provide strong evidence against any of the considered cosmological models and parametrizations, and that this is also the case based on DIC values for the more standard HzBSN data.

\section{Conclusion}
\label{sec:conclusion}

We have used a large compilation of available lower-redshift, non-CMB, expansion-rate data sets to derive cosmological constraints. By analyzing 32 $H(z)$, 12 BAO, 1590 Pantheon\plus\ SN Ia (SNP\plus), 120 QSO-AS, 181 \hiig, 78 \mq, 38 \cq, and 118 A118 GRB measurements, we find that the results from each individual data set are mutually consistent and so these data can be jointly used to study comological models. Additionally, we compare these new data set constraints to their older counterparts and do not find large differences.

The $H(z)$ + BAO + SNP\plus\ + QSO-AS + \hiig\ + \mii\ + \civ\ + A118 (HzBSNQHMCA) data combination results in a fairly precise summary value of $\Om=0.288\pm0.017$ (very similar to the $\Om=0.295\pm0.017$ we found in Ref.\ \cite{CaoRatra2022}), which is in agreement with many recent measurements, e.g.\ Ref.\ \cite{Hang:2020gwn}, and summary values of $\obh=0.0294\pm0.0036$ and $\obh=0.1107\pm0.0113$. Our summary value of the Hubble constant, $H_0=69.8\pm1.3$ \hunit\ (very similar to the $H_0=69.7\pm1.2$ \hunit\ we found in Ref.\ \cite{CaoRatra2022}), is more in line with the values reported in Refs.\ \cite{Freedman2021} and \cite{chenratmed} than with the result of Ref.\ \cite{Riessetal2022}. Specifically, our summary central value of $H_0$ matches that of Ref.\ \cite{Freedman2021}, and is $0.58\sigma$ higher and $1.9\sigma$ lower than the values reported in Refs.\ \cite{chenratmed} and \cite{Riessetal2022}, respectively. Similar to the approach outlined in Refs.\ \cite{CaoRyanRatra2021, CaoRyanRatra2022, CaoRatra2022}, the summary central value is the average of the two (of six model) central mean values, while the uncertainties are the quadrature sum of the systematic uncertainty, defined as half of the difference between the two central mean values, and the statistical uncertainty, defined as the average of the error bars of the two central results. We note that from model to model the \om\ and \obhs\ values range over $1.3\sigma$, while those of \ochs\ range over $1.5\sigma$, unlike the $H_0$ values which range over only $0.21\sigma$.

Our flat \lcdm\ HzBSNQHMCA constraints, $\obh=0.0250\pm0.0021$, $\och=0.1260\pm0.0064$, $H_0=70.13\pm1.25$ \hunit, and $\Om=0.308\pm0.012$, are $1.2\sigma$, $0.92\sigma$, and $2.0\sigma$ higher, and $0.52\sigma$ lower than those from \textit{Planck} TT,TE,EE+lowE+lensing CMB anisotropy data, $\obh=0.02237\pm0.00015$, $\och=0.1200\pm0.0012$, $H_0=67.36\pm0.54$ \hunit, and $\Om=0.3153\pm0.0073$, \citep{planck2018b}, where our uncertainties are 14, 5.3, 2.3, and 1.6 times larger than those from \textit{Planck} data, respectively. Our summary values of \obhs, \ochs, $H_0$, and \om, are $2.0\sigma$ higher, $0.82\sigma$ lower, $1.7\sigma$ higher, and $1.5\sigma$ lower than those from flat \lcdm\ \textit{Planck} data, where our summary values uncertainties are 24, 9.4, 2.4, and 2.3 times larger than those from \textit{Planck} data, respectively.

Our estimated error bar for $H_0$ is slightly larger than that of Ref.\ \cite{Riessetal2022}, but is still much (2.4 times) larger than the error bar from the flat \lcdm\ model \textit{Planck} value \citep{planck2018b}. Our measured summary value for $H_0 = 69.8 \pm 1.3$ \hunit\ falls between the results from the flat \lcdm\ model \textit{Planck} value, \citep{planck2018b}, and the Cepheids and SN Ia measurement of Ref.\ \cite{Riessetal2022}, differing by about 2$\sigma$ from both. (As discussed in the Appendix, excluding \hiig\ data results in $H_0$ values that are $\sim 0.9 \sigma$ higher than the \textit{Planck} flat \lcdm\ model value and $\sim 1.3-1.6\sigma$ lower than the Cepheids and SN Ia local expansion rate value of Ref.\ \cite{Riessetal2022}.) Our $H_0$ value is reasonably consistent with the slightly less-constraining flat \lcdm\ model Atacama Cosmology Telescope (ACT) and South Pole Telescope (SPT) CMB anisotropy values, $H_0 = 67.9 \pm 1.5$ \hunit\ and $H_0 = 68.8 \pm 1.5$ \hunit, \citep{ACT:2020gnv, SPT-3G:2021eoc}, respectively. Our measured summary value of $H_0$ also agrees well with the slightly less-constraining TRGB and SN Ia measurement of Ref.\ \cite{Freedman2021}. These agreements might mean that $H_0=69.8\pm1.3$ \hunit\ is the current most reasonable value for the Hubble constant.

HzBSNQHMCA data show at most mild evidence for non-flat geometry (the strongest being 1.3$\sigma$ evidence for open geometry in the non-flat \lcdm\ model), but indicate more significant evidence for dark energy dynamics, from $1.6\sigma$ in the non-flat models to $2\sigma$ or larger in the flat dynamical dark energy models.

The DIC analysis shows that the HzBSNQHMCA data combination supports flat \pcdm\ the most, but it does not provide strong evidence against models with constant dark energy or a small spatial curvature energy density (the evidence against them is either weak or positive).

We look forward to a future where the quality and quantity of lower-redshift, non-CMB and non-distance-ladder, expansion-rate data, such as those utilized in this study, are significantly improved to the level where they can measure cosmological parameter values with error bars comparable to those obtained from \textit{Planck} CMB anisotropy data.

\begin{acknowledgments}
We thank Dillon Brout and Javier de Cruz P\'{e}rez for useful discussions about Pantheon\plus\ data, Ricardo Ch\'{a}vez and Adam Riess for useful discussions about \hiig\ data, and Michele Moresco and Adam Riess for encouraging us to account for $H(z)$ correlations. We also acknowledge valuable comments from Jim Peebles and Adam Riess. This research was supported in part by DOE grant DE-SC0011840. The computations for this project were performed on the Beocat Research Cluster at Kansas State University, which is funded in part by NSF grants CNS-1006860, EPS-1006860, EPS-0919443, ACI-1440548, CHE-1726332, and NIH P20GM113109.
\end{acknowledgments}

\appendix

\section{Constraints from $H(z)$ + BAO + SNP\plus\ + QSO-AS + \mii\ + \civ\ + A118 data}

For $H(z)$ + BAO + SNP\plus\ + QSO-AS + \mii\ + \civ\ + A118 (HzBSNQMCA) data, the best-fitting parameter values, likelihood values, and information criteria values for all models are given in Table \ref{tab:BFPs} and the marginalized posterior mean parameter values and uncertainties for all models are listed in Table \ref{tab:1d_BFPs}. These results are independent of both CMB data and local distance-ladder data, such as Cepheid or TRGB distances.

The results presented in Figure \ref{fig10} suggest that the differences between the HzBSN and HzBSNQMCA data constraints are relatively small. Furthermore, the comparison of the one-dimensional marginalized constraints from the two data sets supports this conclusion. Specifically, the central values of \om\ derived from HzBSNQMCA data are only $0.057-0.17\sigma$ higher than those obtained from HzBSN data, with error bars $4.8-7.7\%$ smaller, except that in the flat XCDM case the error bars are similar.

The central values of \ok\ derived from HzBSNQMCA data are $0.16\sigma$, $0.12\sigma$, and $0.083\sigma$ lower than those obtained from HzBSN data, with error bars $6.3\%$, $11\%$, and $5.1\%$ smaller, for non-flat \lcdm, XCDM, and \pcdm\ respectively.

The central values of $H_0$ derived from HzBSNQMCA data are from $0.043\sigma$ lower to $0.0085\sigma$ (non-flat XCDM and \pcdm) higher than those obtained from HzBSN data, with error bars $1.6-4.8\%$ smaller, except that in the non-flat \lcdm\ case the error bar is $0.41\%$ larger.

Similar to HzBSN $H_0$ constraints, HzBSNQMCA data $H_0$ constraints range from $68.88\pm2.45$ \hunit\ (non-flat \lcdm) to $69.50^{+2.45}_{-2.44}$ \hunit\ (flat \lcdm), with a difference of $0.18\sigma$. These $H_0$ values are $0.24\sigma$ (non-flat \lcdm) and $0.40\sigma$ (flat \lcdm) higher than the median statistics estimate of $H_0=68\pm2.8$ \hunit\ \citep{chenratmed}, $0.31\sigma$ (non-flat \lcdm) and $0.10\sigma$ (flat \lcdm) lower than the TRGB and SN Ia estimate of $H_0=69.8\pm1.7$ \hunit\ \citep{Freedman2021}, and $1.6\sigma$ (non-flat \lcdm) and $1.3\sigma$ (flat \lcdm) lower than the Cepheids and SN Ia measurement of $73.04\pm1.04$ \hunit\ \cite{Riessetal2022}. The $H_0$ constraints from flat \lcdm\ is $0.86\sigma$ higher than the $H_0$ estimate of $67.36 \pm 0.54$ \hunit\ from \textit{Planck} 2018 TT,TE,EE+lowE+lensing CMB anisotropy data \cite{planck2018b}.

The central values of \wx\ derived from HzBSNQMCA data are $0.040\sigma$ lower and $0.072\sigma$ higher than those obtained from HzBSN data, with error bars $1.9\%$ and $8.6\%$ smaller, for flat and non-flat XCDM respectively. The central values of $\alpha$ derived from HzBSNQMCA data are $0.079\sigma$ and $0.018\sigma$ lower than those obtained from HzBSN data, with error bars $5.9\%$ and $2.0\%$ smaller, for flat and non-flat \pcdm\ respectively.


\bibliography{apssamp}

\end{document}